\documentclass[12pt]{book}
\usepackage{amsmath,amssymb}
\usepackage{graphics,epsfig,pst-all}
\usepackage{graphicx}
\usepackage{bm}
\usepackage[spanish]{babel}
\usepackage{makeidx}
\makeindex

\title{Introducci\'on a la Teor\'{\i}a de los Monopolos Magn\'eticos}

\author{Mauricio Vargas Villegas \\
Grupo PROTOS \\
Facultad de Ciencias Naturales y Matem\'aticas \\
mauricio.vargas@unibague.edu.co \\
Universidad de Ibagu\'e \\
Ibagué, Colombia}
\date{}
\begin{document}

\maketitle
\tableofcontents

\chapter{Resumen}
\'Este documento tiene como objetivo hacer una introducci\'on a la teor\'{\i}a de los monopolos magn\'eticos cl\'asicos y cu\'anticos, introduciendo la electrodin\'amica desde el punto de vista conceptual intentando hasta donde sea posible no descuidar la formulaci\'on matem\'atica con el fin de explicar tanto f\'{\i}sica como matem\'aticamente las ecuaciones de Maxwell; posteriormente se har\'a una breve pero profunda exposici\'on de las bases de la relatividad especial que tiene como objetivo usarla para que junto con la electrodin\'amica se pueda formular la teor\'{\i}a covariante de la electrodin\'amica y analizar su invarianza de Lorentz. El documento busca modelar el campo electromagn\'etico del monopolo magn\'etico con el objetivo de compararlo con el campo electromagn\'etico de los polos el\'ectricos est\'aticos, y en movimiento. A manera introductoria se puede observar que el campo de los monopolos es interesante pues en la electrodin\'amica cl\'asica se ha supuesto la no existencia de los polos magn\'eticos libres (PM). Cl\'asicamente los monopolos se definen mediante una rotaci\'on de $\pi/2$, lo que se llama una transformaci\'on de dualidad. Dirac \cite{Dirac1}, demostr\'o que el polo magn\'etico ser\'{\i}a necesario para cuantizar la carga el\'ectrica, pero se encontr\'o un problema muy serio con una singularidad para $\theta=(0,\;\pi)$ que se llam\'o la cuerda de Dirac. Aunque nunca hubo evidencia experimental del monopolo y Dirac al final de su vida puso en duda la existencia de \'este, parece ser que la falla est\'a en elementos err\'oneos de la teor\'{\i}a de Dirac. Adicionalmentelos monopolos relacionan muy cercanamente la f\'{\i}sica de part\'{\i}culas y la cosmolog\'{\i}a, pues a muy altas energ\'{\i}as se restablece el grupo de unificaci\'on, con cuyo enfriamiento la simetr\'{\i}a fue espont\'aneamente rota. Tambi\'en la teor\'{\i}a puede definir un nuevo tipo de interacciones a un orden tal que se podr\'{\i}a cambiar el n\'umero bari\'onico llevando a la predicci\'on de que el prot\'on ser\'{\i}a inestable. Finalmente en la teor\'{\i}a de unificaci\'on habr\'{\i}an part\'{\i}culas estables que llevan cargas magn\'eticas o monopolos magn\'eticos. Una de las teor\'{\i}as postuladas para la unificaci\'on es la teor\'{\i}a de las cuerdas, aunque \'esta a\'un est\'a en desarrollo es muy matem\'atica y en este momento, no parece que sea posible su verificaci\'on experimental lo que la hace una teor\'{\i}a que sea de mucho agrado para los f\'{\i}cos.
\par
Las cr\'{\i}ticas a los monopolos son muy interesantes \cite{Hadjesfandiari}, \cite{Comay1}, \cite{Comay2}, \cite{Hagen} pero solo desde el punto de vista de la simetr\'{\i}a de las ecuaciones de Maxwell, es innegable la importancia del monopolo magn\'etico. La dificultad para estudiar experimentalmente al PM es que se requerir\'{\i}an aceleradores con una energ\'{\i}a del orden de $10^{15}$ GeV, los cuales a\'un no se han construido.
\chapter{Introducci\'on a la Electrodin\'amica \label{IntroED}}
Las ecuaciones de Maxwell en unidades Gaussianas $\mu_0=\varepsilon_0=1$ (para convertirlas al sistema MKS \'o SI, se debe tener en cuenta la equivalencia $\varepsilon_0=\frac{10^7}{4\pi c^2}$ y $\mu_0=4\pi\times10^{-7}$)son:
\begin{align}
&\bm{\nabla}\centerdot\bm{D}=4\pi\rho\quad \bm{\nabla}\times\bm{H}-\frac{1}{c}\frac{\partial\bm{D}}{\partial t}=\frac{4\pi}{c}\bm{J} \notag \\
&\bm{\nabla}\centerdot\bm{B}=0\qquad \bm{\nabla}\times\bm{E}+\frac{1}{c}\frac{\partial\bm{B}}{\partial t}=0 \label{EcMaxwell}
\end{align}
en las cuales la Ley de Gauss, $\bm{\nabla}\centerdot\bm{D}=\bm{\nabla}\centerdot\bm{E}=4\pi\rho$, significa que la densidad de flujo el\'ectrico diverge desde un foco (densidad de carga el\'ectrica positiva) o hacia un sumidero (densidad de carga el\'ectrica negativa), con lo que se pueden graficar las l\'{\i}neas de campo el\'ectrico de forma radial. As\'{\i} en una regi\'on del espacio tridimensional se construye una superficie cerrada imaginaria, denominada superficie Gaussiana, con el objetivo de analizar el flujo de las l\'{\i}neas de campo. Si se considera el espacio en un volumen $\tau$, la Ley de Gauss toma la forma,
\begin{equation}
\int_\tau(\bm{\nabla}\centerdot\bm{D})d\tau=4\pi\int_\tau\rho d\tau,
\end{equation}
se puede observar que hay una relaci\'on directa entre ese volumen y una superficie cerrada por medio del Teorema de la Divergencia o de Gauss, quedando
\begin{equation}
\oint_S\bm{D}\centerdot d\bm{S}=4\pi q.
\end{equation}
Si el campo es constante, para una carga puntual, que es sim\'etrica con una superficie esf\'erica se llega a la conocida Ley de Coulomb,
\begin{equation}
\bm{D}=\frac{q}{r^3}\hat{\bm{r}}
\end{equation}
que el sistema MKS es
\begin{equation}
\bm{E}=\frac{q}{4\pi\varepsilon_0}\frac{\hat{\bm{r}}}{r^3}.
\end{equation}
donde se us\'o la relaci\'on $\bm{D}=\varepsilon_0\bm{E}$, que relaciona los dos campos vectoriales el\'ectricos (el campo densidad de flujo el\'ectrico $\bm{D}$ y el campo intensidad de campo el\'ectrico $\bm{E}$).
\newline
Teniendo en cuenta la identidad del c\'alculo vectorial $\bm{\nabla}r^{-1}=-r^{-3}\bm{r}$, se puede reescribir la Ley de Coulomb quedando
\begin{equation}
\bm{E}=-\bm{\nabla}\left(\frac{q}{4\pi\varepsilon_0r}\right)=-\bm{\nabla}\Phi \label{PotencEscalar}
\end{equation}
donde $\Phi$ es una funci\'on escalar, o campo escalar denominada potencial el\'ectrico, que define el trabajo por unidad de carga para mover \'esta de un punto a otro en contra del campo el\'ectrico.
\par
La Ley de Ampere-Maxwell, $\bm{\nabla}\times\bm{H}-\frac{1}{c}\frac{\partial\bm{D}}{\partial t}=\frac{4\pi}{c}\bm{J}$, dice que si existe una densidad de corriente el\'ectrica, habr\'a una circulaci\'on de la intensidad de campo magn\'etico alrededor de \'esta, pero para poder conservar la continuidad de corriente el\'ectrica, es necesario que la densidad de flujo el\'ectrico sea temporalmente variante. Lo anterior se puede observar analizando que la divergencia de un rotacional siempre es cero, pues el rotacional produce una circulaci\'on, que no posee divergencia. Por tanto si existe variaci\'on temporal del campo el\'ectrico existir\'a un campo magn\'etico sin necesidad alguna de fuentes. As\'{\i},
\begin{equation}
-\frac{1}{c}\bm{\nabla}\centerdot\left(\frac{\partial\bm{D}}{\partial t}\right)=\frac{4\pi}{c}\bm{\nabla}\centerdot\bm{J}
\end{equation}
lo que se reduce a
\begin{equation}
-\frac{\partial}{\partial t}\bm{\nabla}\centerdot\bm{D}=4\pi\bm{\nabla}\centerdot\bm{J}
\end{equation}
y mediante el uso de la Ley de Gauss, finalmente se llega a la ecuaci\'on de continuidad,
\begin{equation}
-\frac{\partial\rho}{\partial t}=\bm{\nabla}\centerdot\bm{J}
\end{equation}
que no se verificar\'{\i}a si no hubiera variaci\'on de la densidad de flujo el\'ectrico. Adicionalmente debido a la circulaci\'on del campo magn\'etico, \'este est\'a definiendo una trayectoria cerrada o una superficie abierta. Si la Ley de Ampere-Maxwell se integra sobre una superficie abierta, se llega a
\begin{equation}
\int_S(\bm{\nabla}\times\bm{H})\centerdot d\bm{S}-\frac{1}{c}\frac{\partial}{\partial t}\int_S\bm{D}\centerdot d\bm{S}=\frac{4\pi}{c}\int_S\bm{J}\centerdot d\bm{S}
\end{equation}
en donde se puede observar que $\int_S\bm{J}\centerdot d\bm{S}$ es una corriente el\'ectrica, que es originada por un movimiento de cargas puntuales en una direcci\'on determinada y local o sea mediante un vector densidad de corriente el\'ectrica. Hay que hacer claridad que en \'este punto se definen las tres corrientes el\'ectricas de la electrodin\'amica; la corriente de conducci\'on que se origina en la Ley de Ohm, $\bm{J}=\sigma\bm{E}$, donde $\sigma$ es la conductividad; la corriente de convecci\'on, $\bm{J}=\rho\bm{v}_d$, donde $\rho$ es la densidad volum\'etrica de carga y $\bm{v}_d$ es la velocidad de deriva o de arrastre que depende del tipo de conductor por el que se mueven las cargas libres, y la corriente de desplazamiento de Maxwell, $\bm{J}_d=\frac{1}{c}\frac{\partial\bm{D}}{\partial t}$, que se origina con la variaci\'on temporal de la densidad de flujo el\'ectrico.
\newline
Usando el Teorema de Stokes, la Ley de Ampere-Maxwell integral es
\begin{equation}
\oint_C\bm{H}\centerdot d\bm{L}=\frac{4\pi}{c}I
\end{equation}
donde la corriente $I$ contiene tanto la Ley de Ohm como la corriente de desplazamiento de Maxwell.
\par
La ley de Faraday-Henry, $\bm{\nabla}\times\bm{E}+\frac{1}{c}\frac{\partial\bm{B}}{\partial t}=0$, informa que si la densidad de flujo magn\'etico es temporalmente variante, se originar\'a una contra-circulaci\'on de la intensidad de campo el\'ectrico alrededor de \'esta. \'Esta es la base de la Ley de Lenz y del origen de las corrientes par\'asitas, o corrientes de Eddy. Por tanto una variaci\'on temporal del campo el\'ectrico produce una ciculaci\'on lev\'ogira del campo magn\'etico alrededor de \'esta. Realizando una integraci\'on sobre una superficie abierta se llega
\begin{equation}
\int_S(\bm{\nabla}\times\bm{E})\centerdot d\bm{S}=-\frac{1}{c}\frac{\partial}{\partial t}\int_S\bm{B}\centerdot d\bm{S}=-\frac{1}{c}\frac{\partial\Phi_m}{\partial t} \label{LeyF-H}
\end{equation}
donde utilizando el teorema de Stokes se verifica $\int_S(\bm{\nabla}\times\bm{E})\centerdot d\bm{S}=\oint_C\bm{E})\centerdot d\bm{L}=f\varepsilon m$ y por tanto,
\begin{equation}
f\mathcal{E}m=-\frac{1}{c}\frac{\partial\Phi_m}{\partial t}.
\end{equation}
donde $f\varepsilon m$ es la fuerza electromotr\'{\i}z, que no tiene nada que ver con una fuerza pues tiene unidades de voltios, pero se sigue llamando as\'{\i}, aunque no sea correcto, debido a 'razones hist\'oricas'. Es de observar que la derivada temporal del fujo magn\'etico en la Ley de Faraday-Henry (ec. (\ref{LeyF-H})) afecta tanto al campo como a la secci\'on areal; por tanto,
\begin{equation}
\frac{1}{c}\frac{\partial}{\partial t}\int_S\bm{B}\centerdot d\bm{S}=\frac{1}{c}\int_S\frac{\partial\bm{B}}{\partial t}\centerdot d\bm{S}+\frac{1}{c}\int_S\bm{B}\centerdot d\frac{\partial\bm{S}}{\partial t}
\end{equation}
donde en el primer t\'ermino del segundo  miembro la secci\'on areal es constante y el campo var\'{\i}a temporalmente y en el segundo t\'ermino el campo es constante y la secci\'on areal var\'{\i}a respecto al tiempo.
\begin{center}
\fbox{\parbox{10cm}{\textbf{Problema:} Describa un sistema f\'{\i}sico en el cual est\'en presentes los t\'erminos del segundo miembro de la Ley de Faraday-Henry.}}
\end{center}
\par
Finalmente, la llamada ley de Gauss del magnetismo, $\bm{\nabla}\centerdot\bm{B}=\bm{\nabla}\centerdot\bm{H}=0\text{  pues }\mu_0=1$, muestra que el campo magn\'etico no es divergente y por tanto los polos magn\'eticos (fuentes magn\'eticas) libres no existen. Si se analiza en todo el espacio, con un volumen $\tau$,
\begin{equation}
\int_{\tau}(\bm{\nabla}\centerdot\bm{B})d\tau=0
\end{equation}
usando el teorema de la divergencia, dicha ley se escribe en su forma integral como
\begin{equation}
\oint_S\bm{B}\centerdot d\bm{S}=0
\end{equation}
\subsection{Potenciales Vectoriales}
Si se consideran detenidamente las ecuaciones de Maxwell,
\begin{align}
&1)\;\bm{\nabla}\centerdot\bm{D}=4\pi\rho, \notag \\
&2)\;\bm{\nabla}\times\bm{H}-\frac{1}{c}\frac{\partial\bm{D}}{\partial t}=\frac{4\pi}{c}\bm{J}, \notag \\
&3)\;\bm{\nabla}\centerdot\bm{B}=0, \notag \\
&4)\;\bm{\nabla}\times\bm{E}+\frac{1}{c}\frac{\partial\bm{B}}{\partial t}=0 \label{EMx}
\end{align}
se puede observar en la ley de Gauss del magnetismo (ausencia de monopolos magn\'eticos libres, Ec. 3) de (\ref{EMx})) el hecho de que la divergencia del campo magn\'etico es nula, entonces dicho campo est\'a definido por el rotacional de otro campo al que se da el nombre de potencial magn\'etico,
\begin{equation}
\bm{\nabla}\centerdot\bm{B}=\bm{\nabla}\centerdot\bm{\nabla}\times\bm{A}=0
\end{equation}
\'esto es $\bm{B}=\bm{\nabla}\times\bm{A}$. Usando de la Ley de Biot-Savart,
\begin{equation}
\bm{B}(\bm{x})=\frac{1}{c}\int\frac{\bm{J}(\bm{x}')\times(\bm{x}-\bm{x}')}{\vert\bm{x}-\bm{x}'\vert^3}d^3x' \label{LeyBiotSavart}
\end{equation}
se nota claramente una analog\'{\i}a con la Ley de Coulomb,
\begin{equation}
\bm{E}(\bm{x})=\int\frac{\rho(\bm{x}')(\bm{x}-\bm{x}')}{\vert\bm{x}-\bm{x}'\vert^3}d^3x' \label{LeyCoulomb}
\end{equation}
lo que lleva a dilucidar la forma matem\'atica del potencial vectorial. \'Este es
\begin{equation}
\bm{A}(\bm{x})=\frac{1}{c}\int\frac{\bm{J}(\bm{x}')}{\vert\bm{x}-\bm{x}'\vert}d^3x' \label{PotencVect}
\end{equation}
pero dicha definici\'on llevar\'{\i}a a ajustar al potencial en un punto espec\'{\i}fico del espacio y no en cualquier punto. Por tanto a la ecuaci\'on (\ref{PotencVect}) se le adiciona el gradiente de una funci\'on escalar $\Psi$, $\bm{\nabla}\Psi(\bm{x})$ de forma tal que se calibre al potencial vectorial de la forma $\bm{A}\to\bm{A}+\bm{\nabla}\Psi$, lo que se llama realizar una transformaci\'on de gauge en la teor\'{\i}a, y lo que se busca con \'esto es demostrar que la teor\'{\i}a es invariante de gauge. Tambi\'en para el potencial escalar se verifica una transformaci\'on de gauge an\'aloga, $\Phi\to \Phi-\frac{\partial\Psi}{\partial t}$. Si se reemplaza $\bm{B}=\bm{\nabla}\times\bm{A}$ y, adicionalmente se usa la ec. (\ref{PotencEscalar}), en la ecuaci\'on 2) de (\ref{EMx}), se obtiene,
\begin{equation}
\bm{\nabla}\times(\bm{\nabla}\times\bm{A})-\frac{1}{c}\frac{\partial(-\bm{\nabla}\Phi)}{\partial t}=\frac{4\pi}{c}\bm{J}
\end{equation}
donde usando identidades vectoriales se puede modificar a
\begin{equation}
\bm{\nabla}(\bm{\nabla}\centerdot\bm{A})-\bm{\nabla}^2\bm{A}+\frac{1}{c}\frac{\partial(\bm{\nabla}\Phi)}{\partial t}=\frac{4\pi}{c}\bm{J}
\end{equation}
pero se observa que hay algo que no funciona, pues si hay campos variantes con el tiempo, la ecuaci\'on 4) de (\ref{EMx}) queda
\begin{equation}
\bm{\nabla}\times\bm{E}+\frac{1}{c}\frac{\partial(\bm{\nabla}\times\bm{A})}{\partial t}=\bm{\nabla}\times\left(\bm{E}+\frac{1}{c}\frac{\partial\bm{A}}{\partial t}\right)=0
\end{equation}
por lo que el campo el\'ectrico queda definido no como la ec. (\ref{PotencEscalar}), sino como $\bm{E}+\frac{1}{c}\frac{\partial\bm{A}}{\partial t}=-\bm{\nabla}\Phi$, y por tanto $\bm{E}=-\bm{\nabla}\Phi-\frac{1}{c}\frac{\partial\bm{A}}{\partial t}$. Realizando \'estos reemplazos en la Ley de Faraday-Henry, se obtiene una ecuaci\'on que posee la forma de una ecuaci\'on de onda, pero con potenciales acoplados:
\begin{align}
\bm{\nabla}(\bm{\nabla}\centerdot\bm{A})-\bm{\nabla}^2\bm{A}+\frac{1}{c}\frac{\partial(\bm{\nabla}\Phi+\frac{1}{c}\frac{\partial\bm{A}}{\partial t})}{\partial t} &=\frac{4\pi}{c}\bm{J} \notag \\
\bm{\nabla}(\bm{\nabla}\centerdot\bm{A})-\bm{\nabla}^2\bm{A}+\frac{1}{c}\bm{\nabla}\frac{\partial \Phi}{\partial t}+\frac{1}{c^2}\frac{\partial^2\bm{A}}{\partial t^2} &=\frac{4\pi}{c}\bm{J} \notag \\
\bm{\nabla}\left(\bm{\nabla}\centerdot\bm{A}+\frac{1}{c}\frac{\partial \Phi}{\partial t}\right)-\bm{\nabla}^2\bm{A}+\frac{1}{c^2}\frac{\partial^2\bm{A}}{\partial t^2} &=\frac{4\pi}{c}\bm{J}.
\end{align}
Para desacoplarlos se cre\'o la condici\'on de Lorenz,
\begin{equation}
\bm{\nabla}\centerdot\bm{A}+\frac{1}{c}\frac{\partial \Phi}{\partial t}=0, \label{CondicLorenz}
\end{equation}
quedando la ecuaci\'on de onda para el potencial vectorial en presencia de fuentes de campo magn\'etico,
\begin{equation}
\bm{\nabla}^2\bm{A}-\frac{1}{c^2}\frac{\partial^2\bm{A}}{\partial t^2}=-\frac{4\pi}{c}\bm{J}.
\end{equation}
Usando las transformaciones de gauge de los potenciales en la condici\'on de Lorenz,
\begin{align}
&\bm{\nabla}\centerdot\left(\bm{A}+\bm{\nabla}\Psi\right)+\frac{1}{c}\frac{\partial}{\partial t}\left(\Phi-\frac{\partial\Psi}{\partial t}\right)=0 \notag \\
&\bm{\nabla}\centerdot\bm{A}+\frac{1}{c}\frac{\partial \Phi}{\partial t}+\bm{\nabla}^2\Psi-\frac{1}{c}\frac{\partial^2\Psi}{\partial t^2}=0
\end{align}
se observa que la funci\'on escalar $\Psi$ con la que se calibra la electrodin\'amica, debe cumplir la ecuaci\'on de onda
\begin{equation}
\bm{\nabla}^2\Psi-\frac{1}{c}\frac{\partial^2\Psi}{\partial t^2}=0.
\end{equation}
Finalmente reemplazando el valor de la intensidad de campo el\'ectrico $\bm{E}=-\bm{\nabla}\Phi-\frac{1}{c}\frac{\partial\bm{A}}{\partial t}$ en la Ley de Gauss $\bm{\nabla}\centerdot\bm{E}=4\pi\rho$,
\begin{align}
&\bm{\nabla}\centerdot\left(-\bm{\nabla}\Phi-\frac{1}{c}\frac{\partial\bm{A}}{\partial t}\right)=4\pi\rho \notag \\
&\bm{\nabla}\centerdot\bm{\nabla}\Phi+\bm{\nabla}\centerdot\frac{1}{c}\frac{\partial\bm{A}}{\partial t}=-4\pi\rho \notag \\
&\bm{\nabla}^2\Phi-\frac{1}{c^2}\frac{\partial^2 \Phi}{\partial t^2}=-4\pi\rho,
\end{align}
donde se us\'o la condici\'on de Lorentz (ec. \ref{CondicLorenz}). El resultado final es que los potenciales escalar y vectorial se comportan de forma ondulatoria y que las ecuaciones de Maxwell se pueden reescribir en funci\'on de \'estos, y no de los campos el\'ectrico y magn\'etico.
\newline
Se observa adicionalmente que los campos $\bm{E}$ y $\bm{B}$ sin presencia de fuentes en el vac\'{\i}o satisfacen la ecuaci\'on de onda,
\begin{align}
\bm{\nabla}^2\bm{E} &=\frac{1}{c^2}\frac{\partial^2\bm{E}}{\partial t^2} \notag \\
\bm{\nabla}^2\bm{B} &=\frac{1}{c^2}\frac{\partial^2\bm{B}}{\partial t^2}
\end{align}
y por tanto usando la teor\'{\i}a de Fourier, el campo el\'ectrico toma la forma,
\begin{equation}
\bm{E}(\bm{r},t)=\frac{1}{(2\pi)^{\frac{3}{2}}}\int\left[\widetilde{\bm{E}}_1(\bm{k})e^{i\bm{k}\centerdot\bm{r}-i\omega t}+\widetilde{\bm{E}}_2(-\bm{k})e^{i\bm{k}\centerdot\bm{r}+i\omega t}\right]d^3k
\end{equation}
y el campo magn\'etico toma una forma similar,
\begin{equation}
\bm{B}(\bm{r},t)=\frac{1}{(2\pi)^{\frac{3}{2}}}\int\left[\widetilde{\bm{B}}_1(\bm{k})e^{i\bm{k}\centerdot\bm{r}-i\omega t}+\widetilde{\bm{B}}_2(-\bm{k})e^{i\bm{k}\centerdot\bm{r}+i\omega t}\right]d^3k
\end{equation}
donde $\frac{\omega}{c}=k\equiv\vert\bm{k}\vert$. siendo $\bm{k}$ el n\'umero de onda, y donde el caracter ondulatorio de los campos electromagn\'eticos se hace evidente.
\subsection{Simetr\'{\i}a de las Ecuaciones de Maxwell}
A pesar de que \'estas ecuaciones (ver ec. (\ref{EcMaxwell})) nos dan la idea de que la electrodin\'amica no es sim\'etrica en presencia de cargas, se puede notar que desde el punto de vista energ\'etico si existe tal simetr\'{\i}a, pues la densidad de energ\'{\i}a electromagn\'etica es $u=(8\pi)^{-1}(\bm{E}\centerdot\bm{D}+\bm{B}\centerdot\bm{H})$, donde es evidente que las densidades de energ\'{\i}a el\'ectrica y magn\'etica son sim\'etricas entre si, lo cual es a\'un m\'as notorio en la forma matem\'atica que tiene el flujo de radiaci\'on $\bm{S}=c(4\pi)^{-1}(\bm{E}\times\bm{B})$.
\par
Si se supone la existencia de fuentes magn\'eticas, las ecuaciones de Maxwell (ver ec. \ref{EcMaxwell}) se reescriben de la siguiente forma:
\begin{align}
&\bm{\nabla}\centerdot\bm{D}=4\pi\rho_e\qquad \bm{\nabla}\times\bm{H}-\frac{1}{c}\frac{\partial\bm{D}}{\partial t}=\frac{4\pi}{c}\bm{J_e} \notag \\
&\bm{\nabla}\centerdot\bm{B}=4\pi\rho_h\quad -\bm{\nabla}\times\bm{E}-\frac{1}{c}\frac{\partial\bm{B}}{\partial t}=\frac{4\pi}{c}\bm{J_h} \label{EcMaxwellPM}
\end{align}
en las cuales es evidente tal simetr\'{\i}a, donde se puede observar que aparecen las fuentes magn\'eticas (polo magn\'etico $\bm{g}$ con su densidad de carga nagn\'etica asociada $\rho_h$ y la densidad de corriente magn\'etica $\bm{J_h}$). Aunque la motivaci\'on primordial que se expone aqu\'{\i} es la simetr\'{\i}a, la existencia del monopolo es requerida en las teor\'{\i}as que estudian la estructura de la materia a escalas muy peque\~nas (del orden de $10^{-30}$ m) y para poder experimentar all\'{\i}, las energ\'{\i}as necesarias son tan grandes que no existen aceleradores para hacerlo, lo cual llev\'o a algunos cient\'{\i}ficos a trabajar en cosmolog\'{\i}a, pues en \'esta \'area se encuentran cuerpos y procesos en la naturaleza con mucha energ\'{\i}a, tal como explosiones estelares, pulsares, etc., mientras otros buscan el desarrollo te\'orico de los monopolos, desde el punto de vista de la unificaci\'on, teor\'{\i}as de cuerda, o teor\'{\i}as supersim\'etricas.
\newline
Inicialmente Dirac cre\'o el monopolo magn\'etico para cuantizar la carga el\'ectrica, pero actualmente se aplica a las interacciones fuertes y d\'ebiles, a las teor\'{\i}as de gauge abelianas y no-abelianas, tambi\'en para el estudio de los procesos de confinamiento de quarks en la cromodin\'amica cu\'antica (QCD). Actualmente el problema tiene que ver con el estudio de la singularidad intr\'{\i}nseca que tiene la teor\'{\i}a de monopolos de Dirac, o cuerda de Dirac, que al parecer debido a que nunca se ha medido y a que todos los teoremas matem\'aticos fallan pues sobre la superficie abierta se crea una l\'{\i}nea de singularidad, se concluye que hay una falla conceptual en la teor\'{\i}a de Dirac.
\chapter{Relatividad Especial} \label{TER}
Debido a que se pensaba en los a\~nos 1900 que la luz necesitaba un medio para propagarse, pues el sonido necesitaba un medio para hacerlo, las olas necesitan para propagar su perturbaci\'on un medio flu\'{\i}do, etc., los f\'{\i}sicos propusieron una substancia que llenaba el espacio, que permit\'{\i}a que la luz se propagara en el vac\'{\i}o, y lo llamaron \'eter. Las propiedades de \'este \'eter son: no tiene densidad, no tiene alguna propiedad fisico-qu\'{\i}mica u \'optica, pero produce un viento cuando la tierra se mueve a trav\'es de \'el, lo que permite determinarlo midiendo la velocidad de la luz en el marco de la tierra en posiciones tales que se puedan cumplir las transformaciones de Galileo para que al recombinar los haces de luz se observen procesos de interferencia. El dispositivo para realizar \'este experimento fue llamado interfer\'ometro de Michelson-Morley. Como se mencion\'o anteriormente se basaron en las transformaciones de Galileo que correlacionaban dos sistemas de coordenadas inerciales bajo la medici\'on de un evento en el espacio; \'estas son,
\begin{equation}
\bm{r}'=\bm{r}-\bm{v}t, \qquad t'=t. \label{TransGalileo}
\end{equation}
Debido a que los resultados del experimento de Michelson-Morley fueron negativos, a\'un con luz extraterrestre, el resultado no fue completamente entendido hasta que apareci\'o la teor\'{\i} a de la relatividad especial de Albert Einstein en la cual propuso dos postulados:
\par \textbf{Postulado de la relatividad.} En \'este postulado Einstein dijo que las leyes de la naturaleza y tambi\'en los experimentos realizados en un marco de referencia inercial espec\'{\i}fico son invariantes con respecto a cualquier otro marco de referencia inercial.
\par \textbf{Constancia de la velocidad de la luz.} La velocidad de la luz es independiente del movimiento de la fuente que lo origina.
\par
Einstein se enfrent\'o a un problema cuando al observar las transformaciones de Galileo (ver ec. \ref{TransGalileo}). se di\'o cuenta que la ecuaci\'on temporal era sospechosa, pues pensaba que el tiempo no era invariante entre los sistemas de coordenadas primado y no-primado, por lo que se deb\'{\i}an cambiar en lugar de las ecuaciones de Maxwell (ver ec. \ref{EcMaxwell}) que eran existosas experimentalmente bajo todo punto de vista.
\newline
Para aplicar \'estos postulados, se crea un experimento en el que existen dos sistemas de referencia inerciales primado (SRP) y no primado (SRNP). El SRNP est\'a est\'atico y el SRP se mueve con velocidad constante, paralelo al eje x en el sentido positivo. Cuando coinciden los dos or\'{\i}genes, en el SRNP se enciende un foco y la luz se propaga en tres dimensiones a trav\'es del espacio. Los dos sistemas verifican las ecuaciones (si se desea profundizar m\'as, consulte los textos de Jackson \cite{Jackson}, Landau \cite{Landau} y Goldstein \cite{Goldstein}),
\begin{align}
& x^2+y^2+z^2-(ct)^2=0 \notag \\
& (x')^2+(y')^2+(z')^2-(ct')^2=0
\end{align}
en donde se us\'o el segundo postulado en el cual $c=c'$. Debido a la isotrop\'{\i}a del espacio y al primer postulado se concluye que las ecuaciones de los dos sistemas deben ser iguales, y que la diferencia entre ellas incide en una constante multiplicativa que marca el cambio de escala entre los marcos de referencia. Por tanto se tiene que,
\[
\begin{pmatrix}
x' \\
y' \\
z' \\
t'
\end{pmatrix}=
\begin{pmatrix}
\lambda_{11} & \lambda_{12} & \lambda_{13} & \lambda_{14} \\
\lambda_{21} & \lambda_{22} & \lambda_{23} & \lambda_{24} \\
\lambda_{31} & \lambda_{32} & \lambda_{33} & \lambda_{34} \\
\lambda_{41} & \lambda_{42} & \lambda_{43} & \lambda_{44} \\
\end{pmatrix}
\begin{pmatrix}
x \\
y \\
z \\
t
\end{pmatrix}
\]
debido a la isoptrop\'{\i}a del espacio $\lambda_{12}=\lambda_{13}=\lambda_{21}=\lambda_{23}=\lambda_{24}=\lambda_{31}=\lambda_{32}=\lambda_{34}=\lambda_{42}=\lambda_{43}=0$, y por tanto,
\[
\begin{pmatrix}
x' \\
y' \\
z' \\
t'
\end{pmatrix}=
\begin{pmatrix}
\lambda_{11} & 0 & 0 & \lambda_{14} \\
0 & \lambda_{22} & 0 & 0 \\
0 & 0 & \lambda_{33} & 0 \\
\lambda_{41} & 0 & 0 & \lambda_{44} \\
\end{pmatrix}
\begin{pmatrix}
x \\
y \\
z \\
t
\end{pmatrix}.
\]
Las ecuaciones deben ser lineales y en $\bm{v}=\bm{0}$ deben coincidir. Por tanto $\lambda_{11}=\lambda_{22}=\lambda_{33}=\lambda_{44}=1$, y ahora si se considera el movimiento paralelo al eje x teniendo en cuenta que debe existir una similitud con las transformaciones de Galileo y por tanto $\lambda_{14}=-v\lambda_{11}$ (ver ec. \ref{TransGalileo}) se llega a
\[
\begin{pmatrix}
x' \\
y' \\
z' \\
t'
\end{pmatrix}=
\begin{pmatrix}
\lambda_{11} & 0 & 0 & -v\lambda_{11} \\
0 & 1 & 0 & 0 \\
0 & 0 & 1 & 0 \\
\lambda_{41} & 0 & 0 & \lambda_{44} \\
\end{pmatrix}
\begin{pmatrix}
x \\
y \\
z \\
t
\end{pmatrix}.
\]
Reemplazando \'estas ecuaciones en la ecuaci\'on principal $x^2+y^2+z^2-(ct)^2=(x')^2+(y')^2+(z')^2-(ct')^2$ se obtiene
\begin{equation}
x^2+y^2+z^2-(ct)^2=[\lambda_{11}(x-vt)]^2+y^2+z^2-[c(\lambda_{41}x+\lambda_{44}t)]^2. \label{EcRelat4D}
\end{equation}
Al resolver \'este sistema de ecuaciones, y con $\beta=\frac{v}{c}$, se llega a las transformaciones de Lorentz (TL),
\begin{align}
x' &= \frac{x-vt}{\sqrt{1-\beta^2}} \notag \\
y' &= y \notag \\
z' &= z \notag \\
t' &= \frac{t-\frac{\beta^2}{v}x}{\sqrt{1-\beta^2}}. \label{TransLorentz}
\end{align}
\begin{center}
\fbox{\parbox{10cm}{\textbf{Problema:} Demuestre las ecuaciones (\ref{TransLorentz}) mediante el uso de la ecuaci\'on (\ref{EcRelat4D}).}}
\end{center}
\par
Usando las notaciones $\bm{\beta}=\frac{\bm{v}}{c}$, $\gamma=(1-\beta^2)^{-1/2}$ y $x_0=ct,\, x_1=x,\, x_2=y,\, x_3=z$, se reescriben las TL como
\begin{align}
x_0' &= \gamma\left(x_0-\beta x_1\right) \notag \\
x_1' &= \gamma\left(x_1-\beta x_0\right) \notag \\
x_2' &= x_2 \notag \\
x_3' &= x_3. \label{TransLorentz2}
\end{align}
\par
Si el SRP se mueve con una velocidad en $\mathcal{R}^3$, se puede ver que las ecuaciones se escriben de forma diferente, pues la cantidad $\beta$ debe ser un vector y por tanto los productos comunes entre escalares se convierten en productos punto. Para la parte espacial se tiene que,
\begin{align}
x_1' &= \left(1+\frac{\gamma-1}{\beta^2}\beta_1^2\right)x_1+\frac{\gamma-1}{\beta^2}\left(\beta_1\beta_2 x_2+\beta_1\beta_3 x_3\right)-\gamma\beta_1x_0 \notag \\
x_2' &= \left(1+\frac{\gamma-1}{\beta^2}\beta_1^2\right)x_2+\frac{\gamma-1}{\beta^2}\left(\beta_2\beta_1 x_1+\beta_2\beta_3 x_3\right)-\gamma\beta_2x_0 \notag \\
x_3' &= \left(1+\frac{\gamma-1}{\beta^2}\beta_1^2\right)x_3+\frac{\gamma-1}{\beta^2}\left(\beta_3\beta_1 x_1+\beta_3\beta_2 x_2\right)-\gamma\beta_3x_0
\end{align}
y por tanto al reunir \'estas ecuaciones se construye la ecuaci\'on vectorial,
\begin{equation}
\quad \bm{x}'=\bm{x}+\frac{\gamma-1}{\beta^2}\left(\bm{\beta}\centerdot\bm{x}\right)\bm{\beta}-\gamma\bm{\beta}x_0. \label{EcTL1}
\end{equation}
Para la parte temporal,
\begin{equation}
x_0'= \gamma\left(x_0-\beta_1 x_1-\beta_2 x_2-\beta_3 x_3\right)
\end{equation}
que vectorialmente se reescribe, resultando
\begin{equation}
x_0' =\gamma\left(x_0-\bm{\beta}\centerdot\bm{x}\right). \label{EcTL2}
\end{equation}
\begin{center}
\fbox{\parbox{10cm}{\textbf{Problema:} Demuestre las ecuaciones (\ref{EcTL1}) y (\ref{EcTL2}) mediante el uso de la ecuaci\'on (\ref{TransLorentz2}).}}
\end{center}
\par
Teniendo en cuenta que la relatividad est\'a definida en el espacio-tiempo, se puede escribir un cuadri-vector de la forma $x'^{\mu}=x'^{\mu}(x^0,x^1,x^2,x^3),\text{ para }\mu=0,1,2,3$, y usando la m\'etrica $g^{\mu\nu}=g_{\mu\nu}$ tal que
\begin{equation}
g^{\mu\nu}=g_{\mu\nu}=
\begin{pmatrix}
1 & 0 & 0 & 0 \\
0 & -1 & 0 & 0 \\
0 & 0 & -1 & 0\\
0 & 0 & 0 & -1
\end{pmatrix}
\end{equation}
donde se verifica que $g_{\mu\nu}g^{\mu\nu}=g^2=I_{4\times4}$, y en forma general que $g_{\mu\alpha}g^{\alpha\nu}=\delta_{\mu}^{\;\nu}$. Usando las notaciones $\partial^{\mu}\equiv\frac{\partial}{\partial x_{\mu}}$, $\partial_{\mu}\equiv\frac{\partial}{\partial x^{\mu}}$, $\partial^{\mu}A_{\mu}=\partial_{\mu}A^{\mu}=\frac{\partial A^0}{\partial x^0}+\bm{\nabla}\centerdot\bm{A}$, $\square\equiv\partial_{\mu}\partial^{\mu}=\frac{\partial^2}{\partial x^{0\,2}}-\bm{\nabla}^2$, se construye una transformaci\'on relativista entre dos sistemas de coordenadas de la forma, $x'^{\mu}=\Lambda^{\mu}_{\;\;\nu}x^{\nu}+a^{\mu}$, (\'esta que es an\'aloga matem\'aticamente a la transformaci\'on de Galileo en la cual $\bm{r}'=\bm{r}-\bm{v}t$, solo que en la transformaci\'on relativista hay una rotaci\'on espacio-temporal, as\'{\i} como una traslaci\'on espacio-temporal) la que expresa propiedades que le dan el nombre de Grupo de Poincar\'e (GP). En el caso de la relatividad especial se toma el caso en el que $a^{\mu}=0$. La anterior restricci\'on convierte el GP en el Grupo de Lorentz (GL) que son dos grupos que obedecen el \'algebra de Lie, y en donde se usa la condici\'on,
\begin{equation}
\Lambda^{\mu}_{\;\;\alpha}\Lambda^{\nu}_{\;\;\beta}g_{\mu\nu}=g_{\alpha\beta}. \label{condic}
\end{equation}
\par
Adicionalmente se verifica que
\begin{equation}
x''^{\mu}=\Lambda^{\mu}_{\;\;\nu}x'^{\nu}=\Lambda^{\mu}_{\;\;\nu}\Lambda^{\nu}_{\;\;\alpha}x^{\alpha}=\Lambda^{\mu}_{\;\;\alpha}x^{\alpha},
\end{equation}
lo que significa que dos transformaciones sucesivas son tambi\'en transformaciones de Lorentz y por tanto el \'algebra es cerrada. La matriz de transformaci\'on $\Lambda$, o boost de Lorentz, se puede construir usando los postulados fundamentales de la relatividad especial y luego aplic\'andolos a un sistema tridimensional. Usando $\beta'=\frac{\gamma-1}{\beta^2}$, el boost de Lorentz queda definido por la ecuaciones ya anteriormente demostradas (ver ec. (\ref{EcTL1}) y (\ref{EcTL2})):
\begin{equation}
r'= \Lambda r
\end{equation}
donde los vectores $r'$, y $r$, son vectores columna definidos por
\begin{equation}
r'=
\begin{pmatrix}
x'_0 \\
x'_1 \\
x'_2 \\
x'_3
\end{pmatrix},
\text{ y por }
r=
\begin{pmatrix}
x_0 \\
x_1 \\
x_2 \\
x_3
\end{pmatrix},
\end{equation}
y donde el boost de Lorentz, $\Lambda$, se define como
\begin{equation}
\Lambda=
\begin{pmatrix}
\gamma & -\gamma\beta_1 & -\gamma\beta_2 & -\gamma\beta_3 \\
-\gamma\beta_1 & 1+\beta'\beta_1^2 & \beta'\beta_1\beta_2 & \beta'\beta_1\beta_3 \\
-\gamma\beta_2 & \beta'\beta_1\beta_2 & 1+\beta'\beta_2^2 & \beta'\beta_2\beta_3 \\
-\gamma\beta_3 & \beta'\beta_1\beta_3 & \beta'\beta_2\beta_3 & 1+\beta'\beta_3^2 \\
\end{pmatrix} \label{boost}
\end{equation}
\begin{center}
\fbox{\parbox{10cm}{\textbf{Problema:} Demuestre la ecuaci\'on (\ref{boost}).}}
\end{center}
\par
El tiempo propio de cada sistema, que marca su evoluci\'on espacio-temporal y que es el tiempo tal cual se ve en el marco inercial del sistema, est\'a definido por $d\tau^2\equiv c^2dt^2-dx^2=-g_{\mu\nu}dx^{\mu}dx^{\nu}$, donde se constata que en otro sistema de referencia primado, el tiempo propio definido por
\begin{align}
d\tau'^2 &= -g_{\mu\nu}dx'^{\mu}dx'^{\nu}=-g_{\mu\nu}\frac{\partial x'^{\mu}}{\partial x^{\alpha}}\frac{\partial x'^{\nu}}{\partial x^{\beta}}dx^{\alpha}dx^{\beta} \notag \\
&=-g_{\mu\nu}\Lambda^{\mu}_{\;\;\alpha}\Lambda^{\nu}_{\;\;\beta}dx^{\alpha}dx^{\beta}=-g_{\mu\nu}dx^{\mu}dx^{\nu} \notag \\
&=d\tau^2
\end{align}
es invariante bajo transformaciones de Lorentz. En la anterior ecuaci\'on se demuestra adicionalmente que $g_{\alpha\beta}=g_{\mu\nu}\frac{\partial x'^{\mu}}{\partial x^{\alpha}}\frac{\partial x'^{\nu}}{\partial x^{\beta}}$. Si se aislan las partes espaciales y temporales: $dx'^i=\Lambda^i_{\;\;0}cdt$ y $cdt'=\Lambda^0_{\;\;0}cdt$ y se dividen entre s\'{\i}, se llega a
\begin{equation}
\frac{dx'^i}{cdt'}=\frac{\Lambda^i_{\;\;0}}{\Lambda^0_{\;\;0}}=\beta_i
\end{equation}
entonces,
\begin{equation}
\Lambda^i_{\;\;0}=\beta_i\Lambda^0_{\;\;0} \label{TrRelat}
\end{equation}
lo que muestra una relaci\'on entre la parte espacio-temporal y la parte netamente temporal del GL.
\par
Si en la ecuaci\'on (\ref{condic}), se imponen los valores $\alpha=\beta=0$ entonces $\Lambda^{\mu}_{\;\;0}\Lambda^{\nu}_{\;\;0}g_{\mu\nu}=g_{00}=-1$ donde separando la parte espacial y la temporal la ecuaci\'on se transforma a $\sum\limits_{i=1}^{3}(\Lambda^i_{\;\;0})^2-(\Lambda^0_{\;\;0})^2=-1$ y usando el resultado hallado anteriormente: $\Lambda^i_{\;\;0}=\beta_i\Lambda^0_{\;\;0}$, se tiene $(\beta^2-1)(\Lambda^0_{\;\;0})^2=-1$, que al despejar se obtiene,
\[
\Lambda^{0}_{\;\;0}=\frac{1}{\sqrt{1-\beta^2}}=\gamma\quad \text{y} \quad \Lambda^{i}_{\;\;0}=\beta_i\gamma.
\]
\par
Tambi\'en se observa que $\Lambda^{\mu}_{\;\;j}\Lambda^{\nu}_{\;\;j}g_{\mu\nu}=1$, donde siguiendo un procedimiento an\'alogo al anterior resultan las ecuaciones,
\begin{align}
\Lambda^i_{\;\;j} &= \delta_{ij}+\beta_i\beta_j\frac{\gamma-1}{\beta^2} \notag \\
\Lambda^0_{\;\;j} &= \gamma \beta_j
\end{align}
que sirven para verificar el boost de Lorentz (ver ec. (\ref{boost})). Finalmente se escriben los vectores,
\begin{align}
&x'_0=\gamma(x_0-\bm{\beta}\centerdot\bm{x}) \notag \\
&\bm{x}'=\bm{x}+\frac{\gamma-1}{\beta^2}(\bm{\beta}\centerdot\bm{x})\bm{\beta}-\gamma\bm{\beta}x_0, \label{TransLorentz3}
\end{align}
que marcan la estructura general de las transformaciones de Lorentz entre dos sistemas de referencia que se mueven en cualquier direcci\'on.
\begin{center}
\fbox{\parbox{10cm}{\textbf{Problema:} Realice un an\'alisis f\'{\i}sico de las ecuaciones (\ref{TransLorentz3}) y halle las condiciones f\'{\i}sicas para las cuales \'estas son relevantes tanto para la velocidad como para la aceleraci\'on.}}
\end{center}
\par
Con \'este resumen de la relatividad especial y del grupo de Lorentz, se se sientan las bases para poder calcular la invarianza de la carga el\'ectrica y la covarianza de la electrodin\'amica en la siguiente secci\'on.
\chapter[Invarianza Carga El\'ectrica-Covarianza ED]{Invarianza de la Carga El\'ectrica y Covarianza de la Electrodin\'amica} \label{CapEDC}
Esta secci\'on es netamente educacional, y est\'a extraida primordialmente del texto de J. D. Jackson \cite{Jackson} y apoyada en los textos de L. D. Landau \cite{Landau} y de H. Goldstein \cite{Goldstein}.
\par
La importancia de hallar \'esta invarianza es que \'esta demuestra que la electrodin\'amica y las fuentes electromagn\'eticas respetan la relatividad especial.
\par
Teniendo en cuenta la fuerza de Lorentz, aplicada a una part\'{\i}cula de carga $q$,
\begin{align}
\frac{d\bm{p}}{dt} &= \frac{1}{\gamma}\frac{d\bm{p}}{d\tau}=\frac{q}{c}(c\bm{E}+\bm{v}\times\bm{B}) \notag \\
&= \frac{q}{c}(U_0\bm{E}+\bm{U}\times\bm{B})
\end{align}
donde el cuadri-momento est\'a definido por $p^{\mu}=(p_0=\frac{E}{c},\bm{p})=m(U_0=\gamma c,\bm{U}=\gamma\bm{v})$, y donde se observa que la parte temporal es proporcional a una energ\'{\i}a y transforma seg\'un la ec. (\ref{TrRelat}). 
\par
Adicionalmente,
\begin{align}
\frac{dE}{dt} &= \int_{\text vol}\bm{J}\centerdot\bm{E}d^3x=\int_{\text vol}\rho\bm{v}\centerdot\bm{E}d^3x \notag \\
&= \int_q\bm{v}\centerdot\bm{E}dq=q\bm{v}\centerdot\bm{E}
\end{align}
y por tanto, $\frac{dp_0}{dt}=\frac{q}{c}\bm{U}\centerdot\bm{E}=\frac{dE}{d\tau}$, que en notaci\'on tensorial usando  $\bm{E}\centerdot\bm{U}=F^{0\mu}U_{\mu}$, queda como $\frac{dE}{d\tau}=\frac{q}{c}F^{0\mu}U_{\mu}$. Considerando la densidad de corriente como $\bm{J}(\bm{x},t)\equiv\sum_ne_n\delta^3(\bm{x}-\bm{x}_n(t))\frac{d\bm{x}_n(t)}{dt}$, y la densidad de carga como $J^0\equiv\varepsilon_n(\delta^3(\bm{x}-\bm{x}_n))$, donde $\delta(x)$ es la funci\'on de distribuci\'on delta de Dirac (ver el ap\'endice \ref{DDirac}), construimos la cuadri-densidad de corriente $J^{\alpha}(x)\equiv\sum_ne_n\delta^3(\bm{x}-\bm{x}_n(t))\frac{dx^{\;\;\alpha}_n(t)}{dt}$, y por tanto la cuadri-divergencia estar\'a dada por
\begin{align}
\nabla\centerdot \bm{J}(\bm{x},t) &=-\sum_ne_n\frac{\partial}{\partial x_n^{\;\;i}}\delta^3(\bm{x}-\bm{x}_n(t))\frac{dx^{\;\;i}_n(t)}{dt} \notag \\
&=-\sum_ne_n\frac{\partial}{\partial t}\delta^3(\bm{x}-\bm{x}_n(t)) \notag \\
&=-\frac{\partial\varepsilon(\bm{x},t)}{\partial t}
\end{align}
que en cuatro dimensiones queda como $\frac{\partial J^{\alpha}}{\partial x^{\alpha}}\equiv\partial_{\alpha}J^{\alpha}=0$ y de \'esta manera la cuadri-corriente es invariante de Lorentz. Ahora definiendo la carga como la parte temporal de la cuadri-densidad de corriente, $Q\equiv\int d^3xJ^0(x)$ se puede analizarla en funci\'on de todo el espacio como una variaci\'on temporal de la forma $\frac{dQ}{dt}=\int d^3x\frac{\partial J^0(x)}{\partial x^0}$. Usando el teorema de la divergencia se transforma la integral volum\'etrica a una integral sobre la superficie que encierra el volumen: $\frac{dQ}{dt}=\int d^3x\bm{\nabla}\centerdot\bm{J}(x)=0$, la que se anula. En cuatro dimensiones \'esta carga queda como $Q=\int d^4xJ^{\alpha}(x)\partial_{\alpha}\theta(\eta_{\beta}x^{\beta})$ donde $\theta(x)$ es la funci\'on escal\'on o de Heaviside definida por
\[
\theta(x)=
\left\{
\begin{aligned}
&1\qquad x>0 \\
&0\qquad x<0
\end{aligned}
\right.
\]
y $\eta_{\beta}$ se define como $\eta_1\equiv\eta_2\equiv\eta_3\equiv0,\quad\eta_0\equiv+1$. Realizando una transformaci\'on de Lorentz sobre $Q$ resulta $Q'=\int d^4xJ^{\alpha}(x)\partial_{\alpha}\theta(\eta'_{\beta}x^{\beta})$ y por tanto $\eta'_{\beta}=\Lambda^{\alpha}_{\;\;\beta}\eta_{\alpha}$. Si se calcula la diferencia entre las cargas transformadas mediante la TL, se obtiene $(Q'-Q)=(Q'-Q)(\theta(\eta'_{\beta}x^{\beta})-\theta(\eta_{\beta}x^{\beta}))=0$ pues la cuadri-densidad de corriente $J^{\alpha}$ tiende a cero cuando $\vert x\vert$ tiende a infinito y $\theta(\eta'_{\beta}x^{\beta})-\theta(\eta_{\beta}x^{\beta})$ tiende a cero cuando el tiempo tiende a infinito con $x$ fijo. Por tanto $Q$ es un escalar y es entonces invariante bajo cualquier transformaci\'on.
\par
Las ecuaciones de onda para los potenciales $\bm{A}$ y $\Phi$, est\'an dadas por,
\begin{align}
&\bm{\nabla}^2\bm{A}-\frac{1}{c^2}\frac{\partial^2\bm{A}}{\partial t^2}=-\frac{4\pi\bm{J}}{c} \notag \\
&\bm{\nabla}^2\Phi-\frac{1}{c^2}\frac{\partial^2\Phi}{\partial t^2}=-4\pi\rho
\end{align}
restringidas por la condici\'on de Lorentz,
\begin{equation}
\frac{1}{c}\frac{\partial\Phi}{\partial t}+\bm{\nabla}\centerdot\bm{A}=0.
\end{equation}
y cuyas soluciones tienen la forma de Fourier,
\begin{align}
f(\bm{r},t) &=\int^{\infty}_{-\infty}d\omega f_{\omega}(\bm{r})e^{-i\omega t} \notag \\
f_{\omega}(\bm{r},t) &=\frac{1}{2\pi}\int^{\infty}_{-\infty}dt f(\bm{r},t)e^{i\omega t}
\end{align}
Si se define el cuadri-potencial $A^{\alpha}=(\Phi,\bm{A})=(A^0,A^i)$, (para $i=1,2,3$), las ecuaciones de onda se mezclan dando como resultado,
\begin{equation}
\square A^{\alpha}=-\frac{4\pi}{c}J^{\alpha}\quad \text{y} \quad \partial_{\alpha}A^{\alpha}=0.
\end{equation}
\par
En la electrodin\'amica los campos se definen mediante los potenciales de la manera usual,
\begin{align}
\bm{E} &=-\frac{1}{c}\frac{\partial \bm{A}}{\partial t}-\bm{\nabla}\Phi \notag \\
\bm{B} &=\bm{\nabla}\times\bm{A}
\end{align}
que en forma de componentes quedan,
\begin{align}
E_i &=-\frac{1}{c}\frac{\partial A_i}{\partial x_0}-\frac{\partial\Phi}{\partial x_i}=-(\partial^0A^1-\partial^1A^0) \notag \\
B_i &=\varepsilon_{ijk}\frac{\partial A_k}{\partial x_j}=-(\partial^2A^3-\partial^3A^2)
\end{align}
donde se tuvo en cuenta que $\partial^{\alpha}=(\frac{\partial}{\partial x_0},-\bm{\nabla})$. Con la anterior notaci\'on se escribe el tensor electromagn\'etico $F^{\mu\nu}=\partial^{\mu}A^{\nu}-\partial^{\nu}A^{\mu}$,
\begin{equation}
F^{\mu\nu}=
\begin{pmatrix}
0 & -E_1 & -E_2 & -E_3 \\
E_1 & 0 & -B_3 & B_2 \\
E_2 & B_3 & 0 & -B_1 \\
E_3 & -B_2 & B_1 & 0
\end{pmatrix} \label{TensorEM1}
\end{equation}
donde el tensor electromagn\'etico totalmente covariante est\'a descrito como
\begin{equation}
F_{\mu\nu}=g_{\mu\alpha}F^{\alpha\beta}g_{\beta\nu}=
\begin{pmatrix}
0 & E_1 & E_2 & E_3 \\
-E_1 & 0 & -B_3 & B_2 \\
-E_2 & B_3 & 0 & -B_1 \\
-E_3 & -B_2 & B_1 & 0
\end{pmatrix}. \label{TensorEM2}
\end{equation}
\begin{center}
\fbox{\parbox{10cm}{\textbf{Problema:} Demuestre las ecuaciones (\ref{TensorEM1}) y (\ref{TensorEM2}).}}
\end{center}
As\'{\i}, queda definida la transformaci\'on entre tensores $F^{\mu\nu}\overset{\bm{E}\to-\bm{E}}{\longrightarrow}F_{\mu\nu}$, definidos en el espacio normal. En el espacio dual tambi\'en se define el tensor electromagn\'etico dual
\[
\mathcal{F}^{\mu\nu}=\frac{1}{2}\varepsilon^{\mu\nu\gamma\delta}F_{\gamma\delta}=
\begin{pmatrix}
0 & -B_1 & -B_2 & -B_3 \\
B_1 & 0 & E_3 & -E_2 \\
B_2 & -E_3 & 0 & E_1 \\
B_3 & E_2 & -E_1 & 0
\end{pmatrix}
\]
donde el tensor $\varepsilon^{\mu\nu\gamma\delta}$ est\'a definido como,
\[
\varepsilon^{\mu\nu\gamma\delta}=
\left\{
\begin{aligned}
&+1\quad \text{\small para } (\alpha,\beta,\gamma,\delta)=(0,1,2,3) \\
&\qquad\quad\text{\small permutaci\'on par} \\
&-1\quad \text{\small para cualquier perm. impar} \\
&0\quad \text{\small cualquiera de \'{\i}ndices son iguales}
\end{aligned}
\right.
\]
con la propiedad $\varepsilon_{\mu\nu\gamma\delta}=-\varepsilon^{\mu\nu\gamma\delta}$.\newline
Dos de las Ecuaciones de Maxwell (Ec. \ref{EcMaxwell}) son
\begin{align}
&\frac{\partial E_i}{\partial x_i}=4\pi\rho=\frac{4\pi}{c}c\rho=\frac{4\pi}{c}J_0 \notag \\
&\varepsilon_{ijk}\nabla H_k-\frac{1}{c}\frac{\partial E_i}{\partial x_0}=\frac{4\pi}{c}J_i,
\end{align}
las cuales se pueden unir resultando,
\begin{align}
&\varepsilon_{ijk}\nabla H_k-\frac{1}{c}\frac{\partial E_i}{\partial x_0}-\frac{\partial E_i}{\partial x_i}=\frac{4\pi}{c}J_{\alpha} \notag \\
&\therefore\quad \partial_{\alpha}F^{\alpha\beta}=-\frac{4\pi}{c}J^{\beta}
\end{align}
Las restantes ecuaciones de Maxwell son,
\begin{align}
\frac{\partial B_i}{\partial x_i} &=0 \notag \\
\varepsilon_{ijk}\frac{\partial E_k}{\partial x_j}+\frac{1}{c}\frac{\partial B_i}{\partial x_0} &=0,
\end{align}
quedando,
\begin{align}
&\varepsilon_{ijk}\frac{\partial E_k}{\partial x_j}+\frac{1}{c}\frac{\partial B_i}{\partial x_0}-\frac{\partial B_i}{\partial x_i}=0 \notag \\
&\therefore\quad \partial_{\alpha}\mathcal{F}^{\alpha\beta}=0.
\end{align}
\par
Finalmente las cuatro ecuaciones de Maxwell se pueden escribir en funci\'on del tensor electromagn\'etico de la siguiente manera,
\begin{align}
&\partial_{\alpha}F^{\beta\gamma}+\partial_{\beta}F^{\gamma\alpha}+\partial_{\gamma}F^{\alpha\beta}=0 \notag \\
&\partial_{\alpha}F^{\alpha\beta}=-\frac{4\pi}{c}J^{\beta}
\end{align}
en donde se demuestra la ecuaci\'on vista anteriomente, $\square A^{\alpha}=-\frac{4\pi}{c}J^{\alpha}$ con la ecuaci\'on de continuidad $\partial_{\alpha}A^{\alpha}=0$ y de \'esta manera queda definida la covarianza del electromagnetismo. Con la covarianza de los campos electromagn\'eticos, \'estos transforman mediante las relaciones,
\begin{align}
\bm{E}' &= \gamma(\bm{E}+\bm{\beta}\times\bm{B})-\frac{\gamma^2}{\gamma+1}\bm{\beta}(\bm{\beta}\centerdot\bm{E}) \notag \\
\bm{B}' &= \gamma(\bm{B}-\bm{\beta}\times\bm{E})-\frac{\gamma^2}{\gamma+1}\bm{\beta}(\bm{\beta}\centerdot\bm{B}). \label{CamposEMRelat}
\end{align}
\begin{center}
\fbox{\parbox{10cm}{\textbf{Problema:} Si $v<<c$ y $v\simeq c$ entonces qu\'e pasa con las ecuaciones (\ref{CamposEMRelat})?; explique tanto matem\'atica como f\'{\i}sicamente.}}
\end{center}
\begin{center}
\fbox{\parbox{10cm}{\textbf{Problema:} Si $v=c$ en la ecuaci\'on (\ref{CamposEMRelat}), qu\'e significa para los campos electromagn\'eticos?.}}
\end{center}
\chapter{Radiaci\'on Electromagn\'etica \label{RadEM}}
El potencial escalar est\'a definido por,
\begin{equation}
\Phi(\bm{x},t)=\int d^3x'\int dt'\frac{\rho(\bm{x}',t)}{\vert\bm{x}-\bm{x}'\vert}\delta\left(t'+\frac{\vert\bm{x}-\bm{x}'\vert}{c}-t\right)
\end{equation}
donde la contribuci\'on para el monopolo el\'ectrico se obtiene reemplazando $\vert\bm{x}-\bm{x}'\vert=\vert\bm{x}\vert=r$, resultando,
\begin{align}
\Phi(\bm{x},t) &=\int d^3x'\int dt'\frac{\rho(\bm{x}',t)}{r}\delta\left(t'+\frac{r}{c}-t\right) \notag \\
&=\frac{1}{r}\int d^3x'\rho(\bm{x}',t)\int dt'\delta\left(t'+\frac{r}{c}-t\right) \notag \\
&=\frac{1}{r}\int d^3x'\rho\left(\bm{x}',t'=t-\frac{r}{c}\right) \notag \\
&=\frac{q(t'=t-\frac{r}{c})}{r}
\end{align}
pero al ser escalar, la contribuci\'on solo es \'esta. Para el potencial vectorial, bajo el gauge de Lorentz, est\'a definido por (si desea desarrollar las ecuaciones en el sistema MKS o SI, deber\'a introducir un factor $\frac{4\pi}{c}$ \'o $\frac{\mu_0c}{4\pi}$, seg\'un sea el caso),
\begin{equation}
\bm{A}(\bm{x},t)=\frac{1}{c}\int{d^3x'}\int dt'\frac{\bm{J}(\bm{x}',t')}{\vert\bm{x}-\bm{x}'\vert}\cdot\delta\left(t'+\frac{\vert\bm{x}-\bm{x}'\vert}{c}-t\right) \label{PotVectLG}
\end{equation}
donde el delta de Dirac hace que que los campos no se propagen instant\'aneamente y por tanto se conviertan en campos retardados. Pero se observa que las fuentes var\'{\i}an en el tiempo sinusoidalmente de la forma $\rho(\bm{x},t)=\rho(\bm{x})e^{-i\omega t}$ y $\bm{J}(\bm{x},t)=\bm{J}(\bm{x})e^{-i\omega t}$ y por tanto el potencial vectorial (Ec. (\ref{PotVectLG})) se modifica como,
\begin{align}
\bm{A} &=\frac{1}{c}\int{d^3x}\int dt'\frac{\bm{J}(\bm{x}')e^{-i\omega t'}}{\vert\bm{x}-\bm{x}'\vert}\delta\left(t'+\frac{\vert\bm{x}-\bm{x}'\vert}{c}-t\right) \notag \\
&=\frac{1}{c}\int{d^3x'}\frac{\bm{J}(\bm{x}')}{\vert\bm{x}-\bm{x}'\vert}\int dt'e^{-i\omega t'}\delta\left(t'+\frac{\vert\bm{x}-\bm{x}'\vert}{c}-t\right) \notag \\
&=\frac{1}{c}\int d^3x'\bm{J}(\bm{x}')\frac{e^{i\frac{\omega}{c}\vert\bm{x}-\bm{x}'\vert}}{\vert\bm{x}-\bm{x}'\vert}=\frac{1}{c}\int d^3x'\bm{J}(\bm{x}')\frac{e^{ik\vert\bm{x}-\bm{x}'\vert}}{\vert\bm{x}-\bm{x}'\vert} \label{PotencVector}
\end{align}
Si la longitud de onda es $\lambda=\frac{2\pi c}{\omega}$, y si las dimensiones de la fuente de radiaci\'on son mucho menores $d<<\lambda$, entonces se definen tres zonas espaciales que definen la \textit{zona cercana o est\'atica} para $d<<r<<\lambda$, la \textit{zona intermedia o de inducci\'on} para $d<<r\sim\lambda$, y la \textit{zona lejana o de radiaci\'on} para $d<<\lambda<<r$. Usando (ver Ap\'endice \ref{EcLaplaceCE})
\begin{align}
\frac{1}{\vert\bm{x}-\bm{x'}\vert} &=\sum^{\infty}_{l=0}\frac{r^{'l}}{r^{l+1}}P_l(\cos\gamma) \notag \\
&=4\pi\sum^{\infty}_{l=0}\sum^{l}_{m=-l}\frac{1}{2l+1}\frac{r^{'l}}{r^{l+1}}\vert Y_{lm}(\theta,\phi)\vert^2 \label{Expans}
\end{align}
donde $\frac{2l+1}{4\pi}=\sum^{l}_{m=-l}Y*_{lm}(\theta',\phi')Y_{lm}(\theta,\phi)$ (\cite{Arfken}, \cite{Jeffreys}), para la zona cercana $kr<<1$ y por tanto al analizar el l\'{\i}mite cuando $kr\to0$, la exponencial tiende al valor unitario y as\'{\i}, usando (\ref{Expans}) y los datos del Ap\'endice \ref{EcLaplaceCE}, el potencial vectorial (\ref{PotencVector}) es,
\begin{equation}
\bm{A}=\frac{1}{c}\sum^{\infty}_{l=0}\sum^{l}_{m=-l}\frac{4\pi}{2l+1}\frac{Y_{lm}(\theta,\phi)}{r^{l+1}}\cdot\int d^3x'\bm{J}(\bm{x}')r^{'l}Y*_{lm}(\theta',\phi') \label{ArmonEsfer}
\end{equation}
mostrando que en la zona cercana los campos son casi-estacionarios.
\begin{center}
\fbox{\parbox{10cm}{\textbf{Problema:} Use el ap\'endice \ref{EcLaplaceCE} y demuestre la ecuaci\'on (\ref{ArmonEsfer})}}
\end{center}
\par
En la zona de radiaci\'on donde $kr>>1$, la exponencial de la ecuaci\'on (\ref{Expans}) oscila con gran magnitud, dando las caracter\'{\i}sticas propias del potencial y de los campos de radiaci\'on. Por tanto en el l\'{\i}mite cuando $kr\to\infty$, se verifica $r-\bm{\hat{n}}\centerdot\bm{x}'=r\left(1-\frac{\bm{\hat{n}}\centerdot\bm{x}'}{r}=r\right)$, y por tanto el potencial vectorial (\ref{PotencVector}) es,
\begin{align}
\bm{A} &=\frac{1}{c}\int d^3x'\bm{J}(\bm{x}')\frac{e^{ik(r-\bm{\hat{n}}\centerdot\bm{x}')}}{r-\bm{\hat{n}}\centerdot\bm{x}'} \notag \\
&=\frac{1}{c}\frac{e^{ikr}}{r}\int d^3x'\bm{J}(\bm{x}')e^{-i\bm{\hat{n}}\centerdot\bm{x}'} \notag \\
&=\frac{e^{ikr}}{cr}\sum_{n=0}^{\infty}\frac{(-ik)^n}{n!}\int d^3x'\bm{J}(\bm{x}')\left(\bm{\hat{n}}\centerdot\bm{x}'\right)^n
\end{align}
lo anterior debido a que $d<<\lambda$ y por tanto se expande en t\'erminos de $k$. Si se tiene en cuenta solamente el primer t\'ermino, el potencial vectorial es $\bm{A}=\frac{e^{ikr}}{cr}\int d^3x'\bm{J}(\bm{x}')$, donde usando las propiedades de la derivada convectiva y de las soluciones fasoriales ($\frac{\partial}{\partial t}=i\omega$), que reescriben la ecuaci\'on de continuidad como $\bm{\nabla}\centerdot\bm{J}=i\omega\rho$ y por tanto el potencal vectorial queda como $\bm{A}(\bm{x})=-i\frac{\omega}{c}\frac{e^{ikr}}{r}\int d^3x'\bm{x}'\rho(\bm{x}')$. Usando la definici\'on del momento dipolar el\'ectrico $\bm{p}=q\bm{d}=\int d^3x'\bm{x}'\rho(\bm{x}')$ y la definici\'on del n\'umero de onda $k=\frac{\omega}{c}$, el potencial finalmente toma la forma,
\begin{equation}
\bm{A}(\bm{x})=-ik\bm{p}\frac{e^{ikr}}{r}.
\end{equation}
Finalmente los campos del dipolo el\'ectrico son,
\begin{align}
\bm{B} &=\bm{\nabla}\times\bm{A}=\bm{\nabla}\times\left(-ik\bm{p}\frac{e^{ikr}}{r}\right) \notag \\
&=k^2(\bm{\hat{n}}\times\bm{p})\frac{e^{ikr}}{r}\left(1-\frac{1}{ikr}\right)
\end{align}
y el campo el\'ectrico
\begin{align}
\bm{E} &=\frac{ic}{\omega}\bm{\nabla}\times\bm{B}=ik\bm{\nabla}\times\left(\bm{\nabla}\times\bm{A}\right) \notag \\
&=\frac{k^2e^{ikr}}{r}(\bm{\hat{n}}\times\bm{p})\times\bm{\hat{n}}+\left[3\bm{\hat{n}}(\bm{\hat{n}}\centerdot\bm{p})-\bm{p}\right] \notag \\
&\hspace{3.5cm}\cdot\left(\frac{1}{r^3}-\frac{ik}{r^2}\right)e^{ikr}
\end{align}
donde en la zona est\'atica cuando se toma el $\lim_{kr\to0}$ y as\'{\i} $\frac{e^{ikr}}{r}\to0$, $\left(1-\frac{1}{ikr}\right)\to\frac{i}{kr}$, $k^2\to0$ y $e^{ikr}\left(\frac{1}{r^3}-\frac{ik}{r^2}\right)=\frac{e^{ikr}}{r^3}(1-ikr)\to\frac{1}{r^3}$, quedando los campos como,
\begin{align}
\bm{B} &=k^2(\bm{\hat{n}}\times\bm{p})\frac{e^{ikr}}{r}\frac{i}{kr}=\frac{ik}{r^2}(\bm{\hat{n}}\times\bm{p}) \notag \\
\bm{E} &=\frac{1}{r^3}\left[3\bm{\hat{n}}(\bm{\hat{n}}\centerdot\bm{p})-\bm{p}\right],
\end{align}
mientras que en la zona de radiaci\'on cuando se toma el $\lim_{kr\to\infty}$ y por tanto $\frac{e^{ikr}}{r}$ se conserva, pues decaen m\'as o menos igual, mientras que $\frac{1}{ikr}\to0$ y $\left(\frac{1}{r^3}-\frac{ik}{r^2}\right)\to0$, toman la forma,
\begin{align}
\bm{B} &=k^2(\bm{\hat{n}}\times\bm{p})\frac{e^{ikr}}{r} \notag \\
\bm{E} &=\left[k^2(\bm{\hat{n}}\times\bm{p})\frac{e^{ikr}}{r}\right]\times\bm{\hat{n}} \notag \\
&=\bm{B}\times\bm{\hat{n}}
\end{align}
El vector de Poynting, que define la potencia radiada, est\'a dado por $\bm{S}=\frac{c}{8\pi}\bm{E}\times\bm{B}^*$, pero lo que se necesita es  calcular la potencia radiada per angulo s\'olido unitario debida a la oscilaci\'on del momento dipolar $\frac{dP}{d\Omega}$, se proyecta el vector de Poynting al vector unitario en direcci\'on de r, $\bm{\hat{n}}$ por $r^2$ que marca la caracter\'{\i}stica del cambio de espacio, resultando,
\begin{align}
\frac{dP}{d\Omega} &=\frac{r^2c}{8\pi}\bm{\hat{n}}\centerdot\bm{E}\times\bm{B}^* \notag \\
&=\frac{k^4c}{8\pi}\bm{\hat{n}}\centerdot[(\bm{\hat{n}}\times\bm{p})\times\bm{\hat{n}}]\times(\bm{\hat{n}}\times\bm{p}) \notag \\
&=\frac{k^4c}{8\pi}[\bm{\hat{n}}\times\bm{p})\times\bm{\hat{n}}]^2 \notag \\
&=\frac{k^4c}{8\pi}\vert\bm{\hat{n}}\times\bm{p}\sin\theta\vert^2=\frac{k^4c}{8\pi}\vert\bm{p}\vert^2\sin^2\theta
\end{align}
donde resolviendo para el \'angulo s\'olido se obtiene la potencia radiada total $P=\frac{k^4c}{3}\vert\bm{p}\vert^2$.
\chapter{Monopolos Magn\'eticos}
Si se analizan las ecuaciones de Maxwell (EMx) (ver ec. (\ref{EcMaxwell})), se puede observar que \'estas se dividen en macrosc\'opicas (EMM) y microsc\'opicas (EMm). Las EMm describen la electrodin\'amica que es v\'alida para variaciones espacio-temporales arbitrarias del campo electromagn\'etico acoplado, donde las fuentes describen todas las posibilidades; esto es: cargas puntuales, polarizaciones, corrientes a nivel at\'omico, etc., en otras palabras describen el marco para distribuciones de fuentes y campos arbitrarios a escalas macrosc\'opicas y microsc\'opicas. Las EMM est\'an definidas exclusivamente en cuerpos macrosc\'opicos y por tanto para una correcta descripci\'on electrodin\'amica de \'estos, se definen unos nuevos campos llamados la densidad de flujo el\'ectrico, o desplazamiento el\'ectrico $\bm{D}$ y la intensidad de campo magn\'etico o campo magnetizante $\bm{H}$, siendo ambos originados de relaciones no-lineales, por lo general dificultuosas de analizar, de los campos $\bm{E}$ y $\bm{B}$, las variables espacial, $\bm{r}$, y el tiempo, $t$. En general se ha ense\~nado en los textos promedio que las relaciones entre las EMM y las EMm son tan sencillas como $\bm{D}=\varepsilon\bm{E}$ y $\bm{H}=\mu^{-1}\bm{B}$, pues el tratamiento de dichas cantidades se realiza en el vac\'{\i}o, lo que est\'a muy por fuera de la realidad, pues en general \'estas supuestas constantes llamadas permitividad $\varepsilon$ y permeabilidad $\mu$ se analizan dentro de materiales para visualizar las susceptibilidades el\'ectrica y magn\'etica de \'estos, resultando variables tensoriales de la forma,
\begin{equation}
\varepsilon=
\begin{pmatrix}
\varepsilon_{xx} & \varepsilon_{xy} & \varepsilon_{xy} \\
\varepsilon_{yx} & \varepsilon_{yy} & \varepsilon_{yz} \\
\varepsilon_{zx} & \varepsilon_{zy} & \varepsilon_{zz} \\
\end{pmatrix}\quad
\text{y}\quad\mu=
\begin{pmatrix}
\mu_{xx} & \mu_{xy} & \mu_{xy} \\
\mu_{yx} & \mu_{yy} & \mu_{yz} \\
\mu_{zx} & \mu_{zy} & \mu_{zz} \\
\end{pmatrix}.
\end{equation}
\par
Se puede ver una falta de simetr\'{\i}a en las EMx, que se puede solucionar si se suprimen las fuentes y por tanto la teor\'{\i}a electromagn\'etica descrita se hace en ausencia de fuentes mediante las ecuaciones,
\begin{align}
&\bm{\nabla}\centerdot\bm{E}=0 \notag \\
&\bm{\nabla}\centerdot\bm{B}=0 \notag \\
&\bm{\nabla}\times\bm{E}+\frac{1}{c}\frac{\partial\bm{B}}{\partial t}=0 \notag \\
&\bm{\nabla}\times\bm{B}-\frac{1}{c}\frac{\partial\bm{E}}{\partial t}=0
\end{align}
donde la simetr\'{\i}a buscada se logra pero con la consequencia de perder las fuentes, lo que es un problema grave pues si se quiere describir correctamente la naturaleza, las fuentes deben existir. Por tanto se debe establecer la simetr\'{\i}a en presencia de fuentes, lo cual se intent\'o realizar casi desde la \'epoca de Maxwell, sin creer mucho en que las ecuaciones resultantes describir\'{\i}an algo en la naturaleza, sino que solo se ver\'{\i}an sim\'etricas matem\'aticamente. Posteriormente debido a Paul Dirac con sus famosos art\'{\i}culos acerca del polo magn\'etico \cite{Dirac1}, \cite{Dirac2} y Cap. \ref{TPMDirac}, en los que defini\'o al monopolo magn\'etico como condicionante necesario para cuantizar a la carga el\'ectrica, con el inconveniente de que resultaron divergencias diametralmente opuestas que defin\'{\i}an una cuerda de divergencia, a la que se le di\'o el nombre de cuerda de Dirac, con la que algunos autores no est\'an de acuerdo pues no se ha medido algo parecido en la naturaleza. M\'as recientemente, entre muchas investigaciones, algunos de los art\'{\i}culos que evidencian una dualidad electromagn\'etica, \cite{Goddard}, \cite{Olive}, \cite{Montonen}, \cite{Witten}, \cite{Wu}, \cite{Song},  la relacionan desde el punto de vista topol\'ogico, a\'un teniendo en cuenta la cuerda de Dirac, e incluyendo teor\'{\i}as de gran unificaci\'on como teor\'{\i}as de cuerdas o supercuerdas, y comportamientos no lineales como solitones o instantones.
\par
Para definir cl\'asicamente al monopolo magn\'etico, se realizan unas transformaciones entre los campos electromagn\'eticos (sin fuentes), cambiando $\bm{B}$ por $-\bm{E}$ y $\bm{E}$ por $\bm{B}$, de la siguiente forma,
\begin{equation}
\left.
\begin{aligned}
&\bm{\nabla}\centerdot\bm{E}=0 \\
&\bm{\nabla}\centerdot\bm{B}=0 \\
&\bm{\nabla}\times\bm{E}+\frac{1}{c}\frac{\partial\bm{B}}{\partial t}=0 \\
&\bm{\nabla}\times\bm{B}-\frac{1}{c}\frac{\partial\bm{E}}{\partial t}=0
\end{aligned}
\right\}
\left.
\begin{aligned}
&\bm{E}\to\bm{B} \\
&\bm{B}\to-\bm{E}
\end{aligned}
\right.
\left\{
\begin{aligned}
&\bm{\nabla}\centerdot\bm{B}=0 \\
&\bm{\nabla}\centerdot\bm{E}=0 \\
&\bm{\nabla}\times\bm{B}-\frac{1}{c}\frac{\partial\bm{E}}{\partial t}=0 \\
&\bm{\nabla}\times\bm{E}+\frac{1}{c}\frac{\partial\bm{B}}{\partial t}=0
\end{aligned}.
\right.
\end{equation}
Lo que no es natural, como se mencion\'o anteriormente. Si tal simetr\'{\i}a no existiera entonces por ejemplo si hubieran habido dos campos $\bm{E}_1$ y $\bm{B}_1$ y se utilizaran otros campos $\bm{E}_2=\bm{B}$ y $\bm{B}_2=-\bm{E}$, como una nueva teor\'{\i}a, al analizar la energ\'{\i}a de las dos teor\'{\i}as, se tendr\'{\i}a que
\[
\frac{1}{8\pi}\vert\bm{E}_1\vert^2+\frac{1}{8\pi}\vert\bm{B}_1\vert^2\quad\neq\quad\frac{1}{8\pi}\vert\bm{E}_2\vert^2+\frac{1}{8\pi}\vert\bm{B}_2\vert^2
\]
y por tanto se verificar\'{\i}a que las cargas magn\'eticas no existen. Para mantener la dualidad electromagn\'etica se deben verificar las transformaciones en presencia de fuentes, en donde se puede observar que $\bm{E}\to c\bm{B}$, $c\bm{B}\to-\bm{E}$, $c\bm{J}_e\to\bm{J}_h$, $\bm{J}_h\to-\bm{J}_e$, $c\rho_e\to\rho_h$ y $\rho_h\to-c\rho_e$, lo que marca transformaciones generales entre campos y fuentes electromagn\'eticas. Finalmente se tiene que dichas transformaciones llevan a las siguientes ecuaciones,
\begin{equation}
\left.
\begin{aligned}
&\bm{\nabla}\centerdot\bm{E}=4\pi\rho_e \\
&\bm{\nabla}\centerdot\bm{B}=4\pi\rho_h \\
&\bm{\nabla}\times\bm{E}+\frac{1}{c}\frac{\partial\bm{B}}{\partial t}=-\frac{4\pi}{c}\bm{J}_h \\
&\bm{\nabla}\times\bm{B}-\frac{1}{c}\frac{\partial\bm{E}}{\partial t}=\frac{4\pi}{c}\bm{J}_e
\end{aligned}
\right\}
\left.
\begin{aligned}
&\bm{E}\to\bm{B} \\
&\bm{B}\to-\bm{E} \\
&(\rho_e,\bm{J}_e)\to(\rho_h,\bm{J}_h) \\
&(\rho_h,\bm{J}_h)\to-(\rho_e,\bm{J}_e)
\end{aligned}
\right.
\left\{
\begin{aligned}
&\bm{\nabla}\centerdot\bm{B}=4\pi\rho_h \\
&\bm{\nabla}\centerdot\bm{E}=-4\pi\rho_e \\
&\bm{\nabla}\times\bm{B}-\frac{1}{c}\frac{\partial\bm{E}}{\partial t}=\frac{4\pi}{c}\bm{J}_e \\
&\bm{\nabla}\times\bm{E}+\frac{1}{c}\frac{\partial\bm{B}}{\partial t}=\frac{4\pi}{c}\bm{J}_h
\end{aligned}
\right.
\end{equation}
Se infiere que en el espacio 'normal' en el cual, en las ordenadas positivas est\'a el campo $\bm{E}$, en las ordenadas negativas el campo $-\bm{E}$, en las abscisas positivas el campo $\bm{B}$ y en las abscisas negativas el campo $-\bm{B}$, puede ser rotado una cantidad $\frac{\pi}{2}$ en contra de las manecillas del reloj, dando como resultado el espacio dual en el que en las ordenadas positivas est\'a el campo $\widetilde{\bm{E}}$, en las ordenadas negativas el campo $-\widetilde{\bm{E}}$, en las abscisas positivas el campo $\widetilde{\bm{B}}$ y en las abscisas negativas el campo $-\widetilde{\bm{B}}$. Resumiendo \'estos pasos, se verifican las transformaciones,
\begin{equation}
\widetilde{\bm{B}}=-\bm{E} \qquad \widetilde{\bm{E}}=\bm{B}
\end{equation}
Igualmente se realiza est\'a rotaci\'on a las fuentes electromagn\'eticas resultando las transformaciones,
\begin{equation}
\widetilde{\rho}_m=-\rho_e \qquad \widetilde{\rho}_e=\rho_h.
\end{equation}
\par
Tradicionalmente se define el campo magn\'etico mediante potencial vectorial usando la relaci\'on,
\[
\bm{\nabla}\centerdot\bm{B}=\bm{\nabla}\centerdot(\bm{\nabla}\times\bm{A})=0,
\]
lo cual es desafortunado, pues dicha ecuaci\'on niega al monopolo magn\'etico, pues las l\'ineas de campo ni convergen, ni divergen. Pero el potencial vectorial $\bm{A}$ es necesario para la descripci\'on mec\'anico-cu\'antica de la electrodin\'amica, por tanto no se puede prescindir de \'el. Por tanto en un volumen $\tau$ se debe verificar usando el teorema de la divergencia,
\begin{equation}
\int_{\tau}\bm{\nabla}\centerdot\bm{B}d\tau=4\pi g=\oint_S\bm{B}\centerdot d\bm{S}
\end{equation}
y por \'esta raz\'on el campo magn\'etico debe estar definido por la ecuaci\'on $\bm{B}=\bm{\nabla}\times\bm{A}+\bm{\chi}$, donde $\bm{\chi}$ es un campo que evita que no existan los polos magn\'eticos y de \'esta manera
\begin{align}
&\int_{\tau}\bm{\nabla}\centerdot\bm{\nabla}\times\bm{A}d\tau+\int_V\bm{\nabla}\centerdot\bm{\chi}d\tau=4\pi g \notag \\
&\int_{\tau}\bm{\nabla}\centerdot\bm{\chi}d\tau=4\pi g.
\end{align}
\'Esta funci\'on, debe ser infinita en un punto y cero en cualquier otro lugar, para sostener la idea de una carga puntual en el espacio. Como podemos ver \'esta es una caracter\'{\i}stica t\'{\i}pica de la funci\'on de distribuci\'on delta de Dirac (ver ap\'endice \ref{DDirac}).
\par
Si el volumen se escoge arbitrariamente, entonces $\bm{\chi}$ debe ser infinita en un punto de la superficie cerrada por el volumen formando una l\'{\i}nea infinita que conecta al monopolo. \'Esta se llama la cuerda de Dirac, y debido a sus caracter\'{\i}sticas, el campo $\bm{B}=\bm{\nabla}\times\bm{A}+\bm{\chi}$ tender\'{\i}a a una singularidad, por lo que el potencial vectorial $\bm{A}$ debe compensar la singularidad en los alrededores de la cuerda de Dirac y debido a \'esto no puede ser definido en cualquier punto del espacio.
\par
Hasta aqu\'{\i} todo parece correcto, pero el teorema de decomposici\'on de Helmholtz dice que cualquier campo vectorial que sea lo suficientemente suave y r\'apidamente decayente, se puede descomponer en un campo vectorial irrotacional (sin circulaci\'on) y en un campo vectorial solenoidal (sin divergencia) de la forma,
\begin{equation}
\bm{T}=-\bm{\nabla}\mathcal{T}(\bm{\nabla}\centerdot\bm{T})+\bm{\nabla}\times\mathcal{T}(\bm{\nabla}\times\bm{T}) \label{EcHelmholtz}
\end{equation}
donde se observa que si el campo es solenoidal, se recupera $\bm{T}=\bm{\nabla}\times\mathcal{T}(\bm{\nabla}\times\bm{T})=\bm{\nabla}\times\bm{U}$, y si el campo es irrotacional se obtiene $\bm{T}=-\bm{\nabla}\mathcal{T}(\bm{\nabla}\centerdot\bm{T})=-\bm{\nabla}\mathcal{U}$.
\par
Espec\'{\i}ficamente para la electrodin\'amica el campo magn\'etico se escribir\'a como, $\bm{B}=-\bm{\nabla}\Phi+\bm{\nabla}\times\bm{A}$, lo que es v\'alido solamente si $\bm{\nabla}\centerdot\bm{B}$ y $\bm{\nabla}\times\bm{B}$ tienden a cero m\'as r\'apido que $r^{-2}$ cuando $r\to\infty$, y si $\bm{B}$ tiende a cero cuando $r\to\infty$. De \'esta manera $\bm{A}$ y $\Phi$ son,
\begin{align}
\bm{A} &= \frac{1}{4\pi}\int\frac{\bm{\nabla}\times\bm{B}}{\vert\bm{x}-\bm{x}'\vert}d^3\bm{x}' \notag \\
\Phi &= \frac{1}{4\pi}\int\frac{\bm{\nabla}\centerdot\bm{B}}{\vert\bm{x}-\bm{x}'\vert}d^3\bm{x}' \label{PotencEM}
\end{align}
\begin{center}
\fbox{\parbox{10cm}{\textbf{Problema:} Explique por qu\'e las ecuaciones (\ref{PotencEM}) son las formas correctas?. Es simplemente una soluci\'on matem\'atica?. Existen m\'as posibilidades f\'{\i}sicamente admisibles?}}
\end{center}
Entonces se definen $\bm{\nabla}\centerdot\bm{B}=4\pi\delta^{(3)}(\bm{x})$ y $\bm{\nabla}\times\bm{B}=0$ pues si el campo magn\'etico est\'a definido mediante un polo, \'este debe tener la forma matem\'atica id\'entica a la del polo el\'ectrico, $\bm{B}=q_hr^{-2}\hat{r}$, y por tanto \'este es irrotacional, lo que nos lleva a que el campo magn\'etico est\'e definido por $\bm{B}=-\bm{\nabla}\Phi$ y por tanto para el polo magn\'etico $\bm{A}=0$, pues si existiera, violar\'{\i}a el teorema de descomposici\'on de Helmholtz, ver ec. \ref{EcHelmholtz}, y as\'{\i} aparece una l\'{\i}nea de discontinuidad. Sinembargo el uso de $\bm{\nabla}\centerdot\bm{B}=4\pi\rho_h$ indica que no es correcto utilizar el potencial $\bm{A}$. Adicionalmente la carga magn\'etica se debe considerar como una distribuci\'on y por tanto la ecuaci\'on correcta debe ser $\bm{\nabla}\centerdot\bm{B}=4\pi q_h\delta^{(3)}(\bm{x})$. Si se escoge la existencia del potencial vectorial, entonces habr\'a una singularidad que no es necesariamente recta.
\par
Usando el teorema de Stokes,
\begin{equation}
\oint_C\bm{V}\centerdot d\bm{L}=\int_S\bm{\alpha}\centerdot d\bm{S}
\end{equation}
donde $\bm{\alpha}=\bm{\nabla}\times\bm{V}$. Si se considera un flujo magn\'etico, debido a un polo de la forma usual, entonces se observa que $\psi(r,\theta)=\int_S\bm{B}\centerdot d\bm{S}=2\pi q_h(1-\cos\theta)=\oint_C\bm{A}'\centerdot d\bm{L}$. Para que la definici\'on del campo magn\'etico, $\bm{\nabla}\times\bm{A}'$, sea correcta sobre una superficie, \'esta no puede ser cualquiera debido a la l\'{\i}nea de singularidad. As\'{\i} las posibles definiciones correctas del campo deben ser,
\begin{align}
\bm{A}' &= \frac{q_h(1-\cos\theta)}{r\sin\theta}\hat{\bm{\phi}}\qquad\text{para }\theta<\pi-\varepsilon \notag \\
\bm{A}'' &= -\frac{q_h(1+\cos\theta)}{r\sin\theta}\hat{\bm{\phi}}\qquad\text{para }\theta>\varepsilon \label{CamposVect}
\end{align}
donde $\bm{A}'$ y $\bm{A}''$ est\'an definidos de forma tal que evitan la singularidad de Dirac. $\bm{A}'$ alrededor del eje negativo de las z, y $\bm{A}''$ en cualquier punto del espacio, con excepci\'on de $\theta=\varepsilon$ alrededor del eje positivo de las z. Los anteriores campos no ser\'{\i}an  potenciales vectoriales, sino campos vectoriales sin una definici\'on f\'{\i}sica clara. El origen de e\'stas ecuaciones se fundamenta  en el an\'alisis del campo de un solenoide largo y delgado. Ver Ap\'endice (\ref{Magneto}).
\newline
Con el objetivo de aclarar un poco m\'as la teor\'{\i}a del monopolo magn\'etico, se construye un campo magn\'etico Coulombiano de la forma,
\begin{equation}
\bm{B}=q_h\frac{\bm{r}}{r^3}
\end{equation}
y por tanto una part\'{\i}cula con carga el\'ectrica $q_e$ en dicho campo sufre una fuerza magn\'etica, $\bm{F}_m=q_e\bm{v}\times\bm{B}$, de la forma,
\begin{equation}
m\frac{d^2\bm{r}}{dt^2}=\frac{q_eq_h}{r^3}\left(\frac{d\bm{r}}{dt}\times\bm{r}\right) \label{FuerzaMP}
\end{equation}
Suponiendo que la energ\'{\i}a se conserva y solo es de movimiento, se puede proyectar la ec. (\ref{FuerzaMP}) en la trayectoria para analizar la \'orbita seguida por la carga el\'ectrica dentro del campo del polo magn\'etico
\begin{equation}
\bm{r}\centerdot\frac{d^2\bm{r}}{dt^2}=\frac{q_eq_h}{mr^3}\bm{r}\centerdot\left(\frac{d\bm{r}}{dt}\times\bm{r}\right)=\frac{q_eq_h}{mr^3}\bm{r}\centerdot\frac{1}{2}\frac{d}{dt}\left(\bm{r}\times\bm{r}\right)=\bm{0}
\end{equation}
pero
\begin{align}
\bm{r}\centerdot\frac{d^2\bm{r}}{dt^2} &=\frac{d}{dt}\left(\bm{r}\centerdot\frac{d\bm{r}}{dt}\right)-\left(\frac{d\bm{r}}{dt}\centerdot\frac{d\bm{r}}{dt}\right) =\frac{1}{2}\frac{d^2}{dt^2}\left(\bm{r}\centerdot\bm{r}\right)-\left(\frac{d\bm{r}}{dt}\right)^2 \notag \\
&=\frac{1}{2}\frac{d^2}{dt^2}r^2-v^2=0
\end{align}
donde se usaron los resultados anteriores para la \'ultima igualdad. Debido a que la potencia es nula se verifica la ecuaci\'on $\frac{d}{dt}\left(\frac{1}{2}mv^2\right)=0$ que al resolverla (definiendo $b^2\equiv2Ct$ con $C$ como la primera constante de integraci\'on y la segunda cero para $t=0$) resulta la ecuaci\'on $r=\sqrt{v^2t^2+b^2}$, que informa acerca del radio seguido por la part\'{\i}cula cargada, el cual no es cerrado en el sistema monopolar, sino que la cargas caen hacia el monopolo hasta una distancia m\'{\i}nima $b$, y luego es reflejado hasta el infinito.
\newline
Desde el punto de vista del momento angular $\bm{L'}=\bm{r}\times m\bm{v}=m\bm{r}\times\frac{d\bm{r}}{dt}$, se puede observar que
\begin{align}
\frac{d\bm{L'}}{dt} &=m\frac{d}{dt}\left(\bm{r}\times\frac{d\bm{r}}{dt}\right)=\bm{r}\times m\frac{d^2\bm{r}}{dt^2}=\bm{r}\times \left[\frac{q_eq_h}{r^3}\left(\frac{d\bm{r}}{dt}\times\bm{r}\right)\right] \notag \\
&= \frac{q_eq_h}{mr^3}\left[m\bm{r}\times\left(\frac{d\bm{r}}{dt}\times\bm{r}\right)\right]=\frac{q_eq_h}{mr^3}\left[\left(\bm{r}\times m\frac{d\bm{r}}{dt}\right)\times\bm{r}\right]=\frac{q_eq_h}{mr^3}\left(\bm{L'}\times\bm{r}\right) \label{VarMomentoAng}
\end{align}
donde se verifica que $\frac{d\bm{L'}}{dt}\centerdot\bm{L'}=0$ y por tanto $\frac{d\bm{L'}}{dt}=\bm{0}$ y que la magnitud del momento angular es $L'=mvb$.
\newline
Si se analiza con un problema ordinario de Coulomb se observa que en \'este problema el momento angular no es constante, como si lo es en un problema de tipo Coulomb. Modificando $\bm{L'}\times\bm{r}$ como,
\begin{align}
&\left(m\bm{r}\times\frac{d\bm{r}}{dt}\right)\times\bm{r}=m\left(\bm{r}\centerdot\bm{r}\right)\frac{d\bm{r}}{dt}-m\left(\bm{r}\centerdot\frac{d\bm{r}}{dt}\right)\bm{r} =mr\left(r\frac{d\bm{r}}{dt}-\frac{dr}{dt}\bm{r}\right) \notag \\
&=mr^3\left(\frac{r\frac{d\bm{r}}{dt}-\frac{dr}{dt}\bm{r}}{r^2}\right)=mr^3\frac{d}{dt}\left(\frac{\bm{r}}{r}\right)
\end{align}
en la ec. (\ref{VarMomentoAng}), se llega a
\begin{align}
&\frac{d\bm{L'}}{dt}-\frac{q_eq_h}{mr^3}mr^3\frac{d}{dt}\left(\frac{\bm{r}}{r}\right)=\frac{d\bm{L}}{dt}=0 \notag \\
&\frac{d}{dt}\left(\bm{L'}-q_eq_h\frac{\bm{r}}{r}\right)=\frac{d\bm{L}}{dt}=0
\end{align}
y por tanto,
\begin{equation}
\bm{L}=\bm{L'}-q_eq_h\frac{\bm{r}}{r}=\bm{L'}-q_eq_h\bm{\hat{r}}
\end{equation}
cuya magnitud es \cite{Shnir},
\begin{equation}
L^2=\bm{L'}^2-q^2_eq^2_h=(mvb)^2+(q_eq_h)^2,
\end{equation}
resultando que la aparici\'on de un t\'ermino adicional en la definici\'on del momento angular genera una contribuci\'on no-trivial de campo.
\newline
Como el vector de Poynting, $\bm{S}=\frac{1}{4\pi}\bm{E}\times\bm{B}$, tiene dimensiones de potencia por unidad de superficie, entonces el momento angular del campo el\'ectrico de una carga el\'ectrica puntual cuya posici\'on definida por un radio vector $\bm{r}$ y el campo magn\'etico de un monopolo, est\'a definido por,
\begin{align}
\bm{L'}_{q_eq_h} &=\frac{1}{4\pi}\int d^3r'\left[\bm{r'}\times\left(\bm{E}\times\bm{B}\right)\right]=\frac{q_h}{4\pi}\int d^3r'\left[\bm{r'}\times\left(\bm{E}\times\frac{\bm{r'}}{r'^3}\right)\right] \notag \\
&=\frac{q_h}{4\pi}\int d^3r'\left[\frac{\bm{E}}{r'^2}-\frac{\bm{\hat{r}'}\centerdot\bm{E}}{r'}\bm{\hat{r}'}\right].
\end{align}
Al integrar el primer t\'ermino se observa que para un volumen que incluya el comportamiento de todo el sistema, \'esta se anula. La segunda integral se integra por partes resultando,
\begin{equation}
\bm{L'}_{q_eq_h}=-\frac{q_h}{4\pi}\int d^3r'\left(\bm{\nabla'}\centerdot\bm{E}\right)\bm{\hat{r}'}.
\end{equation}
Finalmente usando $\bm{\nabla'}\centerdot\bm{E}=4\pi q_e\delta^3(\bm{r}-\bm{r'})$, se obtiene,
\begin{equation}
\bm{L'}_{q_eq_h}=-q_eq_h\bm{\hat{r}'},
\end{equation}
resultando que el valor del momento angular no es nulo a\'un en un sistema est\'atico: carga el\'ectrica-monopolo magn\'etico.
\newline
Si se definen los campos como $\bm{E}=q_{e1}\frac{\bm{r}}{r^3}+\bm{E}(q_{e2})$ y $\bm{B}=q_{h1}\frac{\bm{r}}{r^3}+\bm{B}(q_{h2})$ para un sistema de pares de cargas (el\'ectrica-magn\'etica), llamado diones (d), entonces el momento angular se dicho sistema es,
\begin{align}
\bm{L'}_{dd} &=\frac{1}{4\pi}\int d^3r'\left[\bm{r'}\times\left(\bm{E}\times\bm{B}\right)\right] \notag \\
&=\frac{1}{4\pi}\int d^3r'\left\{\bm{r'}\times\left[\left(q_{e1}\frac{\bm{r}}{r^3}+\bm{E}(q_{e2})\right)\left(q_{h1}\frac{\bm{r}}{r^3}+\bm{B}(q_{h2})\right)\right]\right\} \notag \\
&=\frac{1}{4\pi}\int d^3r'\left\{\bm{r'}\times\left[q_{e1}\frac{\bm{r}}{r^3}\times\bm{B}(q_{h2})+q_{h1}\bm{E}(q_{e2})\times\frac{\bm{r}}{r^3}\right]\right\} \notag \\
\end{align}
donde integrando de igual manera que en el caso anterior resulta un valor no nulo, $\bm{L'}_{dd}=\left(q_{e1}q_{h2}-q_{h1}q_{e2}\right)\bm{\hat{r}}$.
\newline
Si el sistema de cargas se acerca, \'este sufre una dispersi\'on del tipo Rutherford de la forma \cite{Shnir}, \cite{Jackson}, \cite{Goldstein},
\begin{equation}
\frac{d\sigma}{d\Omega}=\frac{1}{\Theta^4}\left(\frac{2q_eq_h}{mv}\right)^2
\end{equation}
\begin{center}
\fbox{\parbox{10cm}{Explique por qu\'e es una dispersi\'on de \'esta clase?. Comp\'arela con la dispersi\'on de Rutherford exacta. A cu\'ales conclusiones llega?.}}
\end{center}
Todos los resultados anteriores muestran que el comportamiento f\'{\i}sico de los monopolos es muy similar a lo que se observa en la naturaleza, y por tanto no es una incoherencia su descripci\'on.
\par
Para analizar el potencial vectorial de un campo monopolar se usan consideraciones de simetr\'{\i}a. Se supone que el campo magn\'etico se comporta de igual manera a la ley de Coulomb, y por tanto la Lagrangiana del sistema est\'a definida por $L=\frac{1}{2}m\bm{v}^2+q_e\bm{v}\centerdot\bm{A}$. Aplic\'andole las ecuaciones de Euler-Lagrange $\frac{\partial L}{\partial q_j}-\frac{d}{dt}\frac{\partial L}{\partial\dot{q}_j}=0$, se obtiene la ecuaci\'on de movimiento
\begin{equation}
m\frac{d^2\bm{r}}{dt^2}=\frac{q_eq_h}{r^3}\bm{\dot{r}}\times\bm{r},
\end{equation}
donde el campo magn\'etico se ha tomado como si fuera del tipo Coulombiano; \'esto es, $\bm{B}=q_h\frac{\bm{r}}{r^3}=\bm{\nabla}\centerdot\bm{A}$. Pero si se considera el campo magn\'etico esf\'ericamente sim\'etrico, el potencial vectorial correspondiente ser\'{\i}a $\bm{A}=A(\theta)\bm{\nabla}\phi$, donde usando la definici\'on del campo vectorial de la ec. (\ref{CamposVect}), se elige la funci\'on $A(\theta)$ como $-q_h(1+\cos\theta)$, y por tanto despu\'es de calcular el gradiente $\bm{\nabla}\varphi=(r\sin\theta)^{-1}(-\sin\varphi\hat{\bm{x}}+\cos\varphi\hat{\bm{y}})$ se obtiene el campo vectorial,
\begin{equation}
\bm{A}=\frac{q_h(1+\cos\theta)}{r\sin\theta}(\sin\varphi\hat{\bm{x}}-\cos\varphi\hat{\bm{y}}) \label{CampoVect};
\end{equation}
en \'este c\'alculo se ha usado la transformaci\'on de coordenadas esf\'ericas, $\hat{\bm{\varphi}}=-\sin\varphi\hat{\bm{x}}+\cos\varphi\hat{\bm{y}}$. Finalmente se puede escribir el campo vectorial de la forma
\begin{equation}
\bm{A}(\bm{r})=\frac{q_h}{r}\frac{\bm{r}\times\hat{\bm{n}}}{r-\bm{r}\centerdot\hat{\bm{n}}}
\end{equation}
donde el vector unitario $\hat{\bm{n}}$ est\'a dirigido hacia el eje z. Calculando el rotacional de \'este campo se obtiene,
\begin{equation}
\bm{B}=\bm{\nabla}\times\bm{A}=\bm{\nabla}\times\left[\frac{q_h}{r}\frac{\bm{r}\times\hat{\bm{n}}}{r-\bm{r}\centerdot\hat{\bm{n}}}\right]=g\frac{\bm{r}}{r^3}.
\end{equation}
lo que comprueba que el campo $\bm{A}$ obedece la forma de los campos de Coulomb o sea que es inversamente proporcional al cuadrado de la distancia ($r$).
Tambi\'en se observa que el caracter vectorial del campo vectorial desaparece. Sinembargo para $\theta=0$ o $\theta=\pi$ la ec. (\ref{CampoVect}) se anula o marca una singularidad. A \'esto se le llama una l\'{\i}nea semi-infinita de singularidad. Una posible soluci\'on a \'este problema es reescribir el potencial vectorial como
\begin{equation}
\bm{A}(\bm{r})=(1+\cos\theta)\frac{i}{q_e}U^{-1}\bm{\nabla}U
\end{equation}
donde $U=e^{-iq_eq_h\varphi}$. De \'esta forma el potencial es una transformaci\'on de gauge pura, la cual es singular y est\'a complementada por un factor con dependencia polar.
\par
Analizando la vecindad de r, bajo un intervalo $\varepsilon$ $(R^2=r^2+\varepsilon^2)$, el potencial vectorial se puede escribir como,
\begin{equation}
\bm{A}(\bm{r},\varepsilon)=\frac{q_h}{R}\frac{\bm{r}\times\hat{\bm{n}}}{R-\bm{r}\centerdot\hat{\bm{n}}},
\end{equation}
que en el l\'{\i}mite cuando $\varepsilon^2\to0$, se obtiene el campo magn\'etico
\begin{equation}
\bm{B}(\bm{r})=q_h\frac{\bm{r}}{r^3}-4q_h\pi\hat{\bm{n}}\theta(z)\delta(x)\delta(y)
\end{equation}
que resuelve todos los problemas con las ecuaciones de Maxwell (ec. (\ref{EcMaxwell})).
\subsection{Teor\'{\i}a del Monopolo sin Cuerdas de Dirac \label{MonopSinCD}}
Se esperar\'{\i}a que la teor\'{\i}a de monopolos y campos deber\'{\i}a tomar una forma que es completamente dual a la teor\'{\i}a de cargas y campos y adicionalmente los campos electromagn\'eticos de un sistema de monopolos y los de un sistema de cargas deber\'{\i}an tener propiedades din\'amicas id\'enticas, pero \'estas propiedades deseables no se sostienen simultaneamente \cite{Comay3}, \cite{Comay4}.
\newline
Las ecuaciones de Maxwell son $\frac{\partial F^{\mu\nu}_{(e)}}{\partial x^{\nu}}=-4\pi J^{\mu}_{(e)}$ y $\frac{\partial F^{*\mu\nu}_{(e)}}{\partial x^{\nu}}=0$, con la fuerza de Lorentz dada por $ma^{\mu}_{(e)}=q_e F^{\mu\nu}_{(e)}v_{\nu}$, donde $m$ es la masa en reposo de la part\'{\i}cula y $v_{\nu}$ es su cuadri-velocidad. Todos los campos son derivados desde un cuadri-potencial regular mediante la ecuaci\'on, $F_{(e)\mu\nu}=\frac{\partial A_{(e)\nu}}{\partial x^{\mu}}-\frac{\partial A_{(e)\mu}}{\partial x^{\nu}}$, donde el cuadri-potencial $A_{(e)\alpha}$ como generador de los campos cumple una importancia fundamental en la din\'amica del sistema tal que forma la Lagrangiana de interacci\'on, $\mathcal{L}_{\text{int}}=-J^{\mu}_{(e)}A_{(e)\mu}$. Intentando construir una teor\'{\i}a dual de la forma, $\frac{\partial F^{*\mu\nu}_{(m)}}{\partial x^{\nu}}=-4\pi J^{\mu}_{(m)}$ y $\frac{\partial -F^{\mu\nu}_{(m)}}{\partial x^{\nu}}=0$, con la fuerza de Lorentz $ma^{\mu}_{(m)}=q_m F^{*\mu\nu}_{(m)}v_{\nu}$. Sus campos se derivan de cuadri-potenciales mediante $F^*_{(m)\mu\nu}=\frac{\partial A_{(m)\nu}}{\partial x^{\mu}}-\frac{\partial A_{(m)\mu}}{\partial x^{\nu}}$. Teniendo en cuenta un potencial vectorial de la forma, (ver Cap. \ref{RadEM}), $\bm{A}_{(e)}=-iAe^{i\omega\left(\frac{z}{v}-t\right)}\bm{\hat{i}}$, los campos est\'an definidos por \begin{equation}
\bm{E}=-\frac{\partial \bm{A}_{(e)}}{\partial t}=\omega Ae^{i\omega\left(\frac{z}{v}-t\right)}\bm{\hat{i}}
\end{equation}
y
\begin{align}
\bm{B} &=\bm{\nabla}\times\bm{A}_{(e)}=
\begin{pmatrix}
\bm{\hat{i}} &\bm{\hat{j}} & \bm{\hat{k}} \\
\frac{\partial}{\partial x} &\frac{\partial}{\partial x} &\frac{\partial}{\partial x} \\
-iAe^{i\omega\left(\frac{z}{v}-t\right)} & 0 & 0 \\
\end{pmatrix} \notag \\
&=\omega Ae^{i\omega\left(\frac{z}{v}-t\right)}\bm{\hat{j}}.
\end{align}
Con el monopolo magn\'etico el potencial dual se define de igual forma, como $\bm{A}_{(m)}=-iAe^{i\omega\left(\frac{z}{v}-t\right)}\bm{\hat{j}}$, y aplicando las transformaciones duales $\bm{E}\to\bm{B}$ y $\bm{B}\to-\bm{E}$, se hallan los campos,
\begin{align}
\bm{E} &=-\bm{\nabla}\times\bm{A}_{(m)}=
\begin{pmatrix}
\bm{\hat{i}} &\bm{\hat{j}} & \bm{\hat{k}} \\
\frac{\partial}{\partial x} &\frac{\partial}{\partial x} &\frac{\partial}{\partial x} \\
0 & -iAe^{i\omega\left(\frac{z}{v}-t\right)} & 0 \\
\end{pmatrix} \notag \\
&=\omega Ae^{i\omega\left(\frac{z}{v}-t\right)}\bm{\hat{j}}.
\end{align}
y
\begin{equation}
\bm{B}=-\frac{\partial \bm{A}_{(m)}}{\partial t}=\omega Ae^{i\omega\left(\frac{z}{v}-t\right)}\bm{\hat{i}}.
\end{equation}
Por tanto se oberva que
\begin{equation}
\bm{E}=-\frac{\partial \bm{A}_{(e)}}{\partial t}=-\bm{\nabla}\times\bm{A}_{(m)}=\bm{E}
\end{equation}
y que
\begin{equation}
\bm{B}=\bm{\nabla}\times\bm{A}_{(e)}=-\frac{\partial \bm{A}_{(m)}}{\partial t}\bm{B},
\end{equation}
concluyendo que se pueden identificar los campos de radiaci\'on de cargas con los campos de radiaci\'on de monopolos, lo que hace innecesaria la definici\'on de los monopolos con cuerdas de Dirac. Se puede observar adicionalmente que debido a que la acci\'on es un escalar de Lorentz, todos los t\'erminos de la densidad Lagrangiana lo deben ser tambi\'en; como la electrodin\'amica es lineal, el t\'ermino de interacci\'on carga-monopolo debe ser una suma de cantidades bilineales que contienen dos factores: uno relacionado con cargas y el otro con monopolos y que un sistema de una carga sin movimiento - monopolo sin movimiento, no cambia en el tiempo.
\newline
Las cargas no interact\'uan con campos ligados de monopolos y los monopolos no interact\'uan con campos ligados de cargas. Las cargas interact\'uan con todos los campos de radiaci\'on de las cargas y con los campos de radiaci\'on de los monopolos. Los monopolos interactuan con todos los campos de radiaci\'on de los monopolos y con los campos de radiaci\'on de las cargas. Los campos de radiaci\'on de las cargas y de los monopolos son considerados como una sola entidad y se denotan con el sub\'{\i}ndice $w$ para formar el tensor ligado de campos de radiaci\'on de cargas, $F^{\mu\nu}_{(e,w)}$ y el tensor ligado de campos de radiaci\'on de monopolos, $F^{\mu\nu}_{(m,w)}$. Finalmente la forma de la fuerza de Lorentz ejercida sobre las cargas es, $ma^{\mu}_{(e)}=q_m F^{\mu\nu}_{(e,w)}v_{(e)\nu}$ y la ejercida correspondientemente sobre los monopolos es, $ma^{\mu}_{(m)}=q_m F^{*\mu\nu}_{(m,w)}v_{(m)\nu}$, donde a los campos monopolares se les denomina campos magnetoel\'ectricos \cite{Comay1}, \cite{Comay2}.
\newline
Para desarrollar una teor\'{\i}a de monopolos de \'esta clase, se debe tener en cuenta que \'esta debe reducir a la teor\'{\i}a conocida de cargas y ondas para un sistema sin monopolos y a una teor\'{\i}a dual para sistemas donde las cargas no existen. La teor\'{\i}a se debe originar de una densidad Lagrangiana cuyos t\'erminos son funciones regulares de los potenciales, de sus derivadas, de las corrientes de carga y de monopolos.
\newline
Los t\'erminos de dicha teor\'{\i}a deben ser escalares de Lorentz y deber\'{\i}an ser\'{\i}an invariantes bajo traslaciones espacio-temporales asociadas al grupo de Poincar\'e; adicionalmente las ecuaciones de movimiento de sus campos deben ser lineales y deben sostener el principio de superposici\'on.
\newline
 Por \'ultimo se debe asumir que un sistema de una carga y un monopolo, donde las dos part\'{\i}culas no se mueven, no cambia con el tiempo.
\chapter{La Teor\'{\i}a de los Polos Magn\'eticos \label{TPMDirac}}
En \'este cap\'{\i}tulo se pretende explicar y desarrollar detalladamente las ecuaciones y los procesos matem\'aticos del art\'{\i}culo original de P. A. M. Dirac \cite{Dirac2}. Lo que se busca aqu\'{\i} es presentar la teor\'{\i}a original del monopolo y su cuantizaci\'on con el objetivo que el lector entienda los conceptos originales y la matem\'atica involucrada que llevaron a Dirac a la regla de cuantizaci\'on de la carga el\'ectrica, usando la teor\'{\i}a de polos magn\'eticos.
\newline
El art\'{\i}culo de Dirac \cite{Dirac2}: 'La teor\'{\i}a de los polos magn\'eticos', el autor describe la forma general de las ecuaciones cl\'asicas de movimiento enmarcada en el tensor electromagn\'etico y construye la din\'amica electromagn\'etica mediante la ecuaci\'on de Lorentz, as\'{\i} como la teor\'{\i}a de los polos magn\'eticos comparando continuamente con el caso el\'ectrico de forma tal que conceptualmente adiciona t\'erminos a las ecuaciones para solucionar problemas relacionados principalmente con las cuerdas de discontinuidad o cuerdas de Dirac, sin que \'estos afecten las ecuaciones de movimiento. El m\'etodo usado es el de cambiar el punto de referencia al de las l\'{\i}neas de mundo de las part\'{\i}culas y nuevas variables sin car\'acter f\'{\i}sico que orientan dichas cuerdas contenidas en l\'aminas asociadas a cada polo. Los potenciales retardados se definen en \'estas l\'aminas pues son contribuciones de los polos y son del tipo Lienard-Wiechert. El autor modifica tales potenciales para que sean consistentes con la teor\'{\i}a b\'asica. El paso siguiente es la definici\'on del principio de acci\'on de la electrodin\'amica, que debe ser modificado para que no hallar inconsistencias con las cuerdas de Dirac usando la condici\'on que una cuerda nunca debe pasar a trav\'es de una part\'{\i}cula cargada. \'Esta modificaci\'on da como resultado un t\'ermino relacionado con la variaci\'on del potencial que no afecta las ecuaciones de movimiento. Con todo lo anterior el autor define la funci\'on Hamiltoniana para poder cuantizar la teor\'{\i}a y por tanto se escriben los momentos tanto de las part\'{\i}culas cargadas como de los polos usando la t\'ecnica de Fourier. Las ecuaciones de movimiento est\'an descritas entre dos superficies tridimensionales: de los campos y de los polos, de forma tal que las part\'{\i}culas no existen sino hasta cuando llegan a la superfice de los campos y lo mismo para los polos suponiendo que \'estos est\'an unidos a las cuerdas y que su superficie tridimensional vas m\'a alla. Sinembargo a pesar que algunos campos dejan de evolucionar din\'amicamente, el resto del sistema continua hasta que todo \'este evoluciona totalmente. \'Este comportamiento para las acciones de ciertas part\'{\i}culas o campos y otros no es \'util para el manejo de la teor\'{\i}a pues, el sistema no tiene que detenerse din\'amicamente cuando solamente un campo lo hace. Con las anteriores consideraciones se construye las ecuaciones que relacionan la din\'amica de las coordenadas y los momentos (tanto para las part\'{\i}culas cargadas como para los polos) y con \'estas se hallan las ecuaciones de Hamilton-Jacobi con las que se cuantiza la teor\'{\i}a, obteniendo la condici\'on de cuantizaci\'on de Dirac que define la necesidad de la existencia del polo magn\'etico para poder cuantizar la carga el\'ectrica.
\newline
\newline
\textbf{I-II. Introducci\'on y ecuaciones cl\'asicas de movimiento}
\newline
\newline
El autor describe la elecrodin\'amica desde el punto de vista de la simetr\'{\i}a de los campos el\'ectricos y magn\'eticos concluyendo que para que \'esta sea coherente el polo magn\'etico debe exisitir y el soporte de dicha afirmaci\'on es que hay teor\'{\i}as exitosas que se han basado en part\'{\i}culas predichas te\'oricamente y que despu\'es de cierto tiempo se han descubierto. Un ejemplo para las que se descubrieron son los neutrinos y para las que no, es el bos\'on de Higgs. Tambi\'en se apoya en el hecho de que el polo magn\'etico (monopolo) es necesario para la cuantizaci\'on de la carga el\'ectrica, dada por,
\begin{equation}
eg=\frac{1}{2}n\hbar c
\end{equation}
donde $n$ es un entero y $g$ es la carga magn\'etica; por tanto la carga el\'ectrica para su cuantizaci\'on requiere que sea definida mediante $\frac{1}{2}\frac{n\hbar c}{g}$. \newline
En la secci\'on \textit{Las ecuaciones cl\'asicas de movimiento} se hace una breve descripci\'on de la teor\'{\i}a de los campos electromagn\'eticos usando tanto el tensor electromagn\'etico $F_{\mu\nu}=-F_{\nu\mu}$ (ver Cap\'{\i}tulo \ref{CapEDC}), como su dual $F^{\dagger}_{\mu\nu}$ (En el Cap\'{\i}tulo \ref{CapEDC} el tensor dual est\'a definido por $\mathcal{F}^{\mu\nu}$), con las propiedades $F^{\dagger}_{01}=F_{23}$, $F^{\dagger}_{23}=-F_{01}$ y $F^{\dagger\dagger}_{\mu\nu}=-F_{\mu\nu}$. Adicionalmente se usa un nuevo tensor electromagn\'etico $G_{\mu\nu}$ que cumple la condici\'on $F^{\dagger}_{\mu\nu}G^{\mu\nu}=F_{\mu\nu}G^{\dagger\mu\nu}$. Las ecuaciones de Maxwell est\'an dadas por $\frac{\partial F_{\mu\nu}}{\partial x_{\nu}}=-4\pi j_{\mu}$, donde $j_{\mu}$ es la cuadri-densidad de corriente el\'ectrica y para el espacio dual se cumple que la cuadri-divergencia se anula; esto es $\frac{\partial F^{\dagger}_{\mu\nu}}{\partial x_{\nu}}=0$, resultando una falta de simetr\'{\i}a en las ecuaciones de Maxwell que es corregida si la ecuaci\'on en el espacio dual se define como $\frac{\partial F^{\dagger}_{\mu\nu}}{\partial x_{\nu}}=-4\pi k_{\mu}$, donde $k_{\mu}$ es la cuadri-densidad de corriente magn\'etica. Usando la l\'{\i}nea de mundo de la part\'{\i}cula cargada que determina su evoluci\'on espacio-temporal y est\'a definida por $z_{\mu}=z_{\mu}(s)$, donde $s$ es el tiempo propio de la part\'{\i}cula, una carga puntual $e$ que se desplaza en una l\'{\i}nea de mundo infinita genera una cuadri-densidad de corriente el\'ectrica dada por
\begin{equation}
j_{\mu}(x)=\sum_e e\int{\frac{dz_{\mu}}{ds}\delta^4(x-z)ds}.
\end{equation}
Para el caso magn\'etico, la cuadri-corriente magn\'etica est\'a definida de forma similar:
\begin{equation}
k_{\mu}(x)=\sum_g g\int{\frac{dz_{\mu}}{ds}\delta^4(x-z)ds}.
\end{equation}
Las funciones delta $\delta^4(x)=\delta(x_0)\delta(x_1)\delta(x_2)\delta(x_3)$ (ver ap\'endice \ref{DDirac}) son usadas para describir un punto de discontinuidad en el espacio de las cargas y $\sum_{e,g}$ describe la suma sobre las part\'{\i}culas cargadas ($e$) o sobre los polos ($g$).\newline
Con el fin de describir la din\'amica del sistema se escribe la fuerza de Lorentz $\vec{F}=e(\vec{E}+\vec{v}\times\vec{B})$ en funci\'on del tensor electromagn\'etico quedando,
\begin{equation}
m\frac{d^2z_ {\mu}}{ds^2}=e\frac{dz^{\nu}}{ds}F_{\mu\nu}(z).
\end{equation}
Dicha ecuaci\'on describe el comportamiento din\'amico de una part\'{\i}cula el\'ectrica cargada a lo largo de su l\'{\i}nea de mundo y la  descripci\'on de los campos se realiza en el punto $z$ donde est\'a situada la part\'{\i}cula. Para una part\'{\i}cula cargada magn\'eticamente, usando consideraciones de simetr\'{\i}a, se asume una cuadri-fuerza de Lorentz, 
\begin{equation}
m\frac{d^2z_ {\mu}}{ds^2}=g\frac{dz^{\nu}}{ds}F^{\dagger}_{\mu\nu}(z).
\end{equation}
Debido al caracter de los campos si la medici\'on de \'estos se hace exactamente en el punto $z$, dicha magnitud originar\'a una divergencia; por \'este motivo las variaciones se deben realizar en magnitudes bajas, pero debido a que la funci\'on delta var\'{\i}a br\'uscamente, es conveniente reemplazarla por una funci\'on m\'as suave que cumpla las propiedades de la funci\'on delta o imponer condiciones l\'{\i}mite tales que la teor\'{\i}a solo es invariante de Lorentz en el l\'{\i}mite mismo y despu\'es de \'este, pero no antes (\'esta es la elecci\'on hecha por Dirac). Dirac modific\'o el tensor electromagn\'etico $F_{\mu\nu}$ al tensor $F^*_{\mu\nu}$ que cumple las condiciones requeridas, resultando las ecuaciones de movimiento para las cargas el\'ectricas y para los polos,
\begin{align}
\frac{\partial F^*_{\mu\nu}}{\partial x_{\nu}} &=-4\pi\sum_e e\int{\frac{dz_{\mu}}{ds}\delta^4(x-z)ds} \label{DivElect} \\
\frac{\partial F^{\dagger}_{\mu\nu}}{\partial x_{\nu}} &=-4\pi\sum_g g\int{\frac{dz_{\mu}}{ds}\delta^4(x-z)ds} \label{DivPolo}.
\end{align}
Las cuadri-fuerzas de Lorentz tambi\'en se modifican, para las part\'{\i}culas cargadas y para los polos a,
\begin{align}
m\frac{d^2z_ {\mu}}{ds^2} &=e\frac{dz^{\nu}}{ds}F_{\mu\nu}(z) \\
m\frac{d^2z_ {\mu}}{ds^2} &=g\frac{dz^{\nu}}{ds}F^{\dagger *}_{\mu\nu}(z)
\end{align}
donde solo falta estructurar al nuevo tensor electromagn\'etico.
\newline
\newline
\textbf{III. Potenciales electromagn\'eticos}
\newline
\newline
Para crear una teor\'{\i}a que pueda llevarse a la mec\'anica cu\'antica, se necesita llegar al principio de acci\'on (ver ap\'endice \ref{PpioAcc}) mediante las ecuaciones de movimiento y para \'esto se necesitan los potenciales electromagn\'eticos.
\newline
Mediante la definici\'on $F_{\mu\nu}=\frac{\partial A_{\nu}}{\partial x^{\mu}}-\frac{\partial A_{\mu}}{\partial x^{\nu}}$, en funci\'on del cuadri-potencial, se observa que se cumplen las ecuaciones $\frac{\partial F_{\mu\nu}}{\partial x^{\nu}}=-4\pi j_{\mu}$ y $\frac{\partial F^{\dagger}_{\mu\nu}}{\partial x^{\nu}}=0$, pero \'estas no cumplen con la condici\'on del polo magn\'etico pues su cuadri-divergencia es nula. Por dicho motivo la definici\'on del tensor electromagn\'etico debe variar de alguna manera debido a que \'este falla en algunos puntos de la superficie que encierra al polo, formando una 'especie' de cuerda que se extender\'{\i}a desde el polo hacia el inifinito uniendo todos los puntos sobre cada una de las superficies que encierran al polo; el inconveniente es que dicha cuerda no tendr\'{\i}a significado f\'{\i}sico y debido a que es infinita no existir\'{\i}an variables para describirla.
\newline
Las variables necesarias para fijar las posiciones de las cuerdas son coordenadas din\'amicas y momentos conjugados a ellas. Dichas variables son necesarias para una teor\'{\i}a din\'amica pero no corresponden con observables y no afectan el fen\'omeno f\'{\i}sico.
\newline
Cada cuerda esquemaliza una l\'amina bidimensional en el espacio-tiempo; \'estas l\'aminas ser\'an las regiones donde el tensor electomagn\'etico $F_{\mu\nu}$ falla y cada l\'amina podr\'{\i}a ser descrita expresando un punto general $y_{\mu}$, sobre ella como funci\'on de dos par\'ametros $\tau_0$ y $\tau_1$ ($y_{\mu}=y_{\mu}(\tau_0,\tau_1)$). Se supone que cada l\'amina se extiende al infinito y por tanto los par\'ametros $\tau_0$ y $\tau_1$ pueden ser asignados con valores de forma tal que $\tau_1=0$ sobre la l\'{\i}nea de mundo del polo y se extiende al infinito siguiendo la cuerda y $-\infty<\tau_0<+\infty$ como valores que describen desde el pasado infinito hasta el futuro infinito.
\par
La caracter\'{\i}stica de \'esta cuerda es que en sus extremos deben estar localizados los polos. As\'{\i} el nuevo tensor se define como,
\begin{equation}
F_{\mu\nu}=\frac{\partial A_{\nu}}{\partial x^{\mu}}-\frac{\partial A_{\mu}}{\partial x^{\nu}}+4\pi\sum_gG^{\dagger}_{\mu\nu}, \label{DivPolCorr}
\end{equation}
donde el campo $G^{\dagger}_{\mu\nu}$ es un campo que desaparece en todo sitio excepto sobre una de las l\'aminas y su suma se realiza sobre todas las l\'aminas cada una de las cuales est\'a asociada a un polo. Como se mencion\'o anteriormente, cada una de estas l\'aminas que encierran al polo puede ser descrita por un conjunto de coordenadas ($\tau_0, \tau_1$), formando una funci\'on $y_{\mu}=y_{\mu}(\tau_0,\tau_1)$ con el objetivo de orientar los puntos en los cuales existe la divergencia debido a la cuerda. Reemplazando la ecuaci\'on (\ref{DivPolCorr}) en la ecuaci\'on (\ref{DivPolo}), se obtiene el siguiente desarrollo: primero calculando  la cuadri-divergencia del tensor electromagn\'etico del polo
\begin{align}
F^{\dagger}_{\mu\nu} &=\frac{\partial A^{\dagger}_{\nu}}{\partial x^{\mu}}-\frac{\partial A^{\dagger}_{\mu}}{\partial x^{\nu}}+4\pi\sum_gG^{\dagger\dagger}_{\mu\nu} \notag \\
&\equiv\partial_{\mu} A^{\dagger}_{\nu}-\partial_{\nu} A^{\dagger}_{\mu}+4\pi\sum_gG^{\dagger\dagger}_{\mu\nu}
\end{align}
y por tanto,
\begin{equation}
\partial^{\nu}F^{\dagger}_{\mu\nu}=\partial^{\nu}\partial_{\mu} A^{\dagger}_{\nu}-\partial^{\nu}\partial_{\nu} A^{\dagger}_{\mu}+4\pi\sum_g\partial^{\nu}G^{\dagger\dagger}_{\mu\nu}
\end{equation}
pero debido a que $G^{\dagger\dagger}_{\mu\nu}=-G_{\mu\nu}$ y a que $\partial^{\nu}F^{\dagger}_{\mu\nu}=-4\pi\sum_g g\int{\frac{dz_{\mu}}{ds}\delta^4(x-z)ds}$ entonces, $-4\pi\sum_g \partial^{\nu}G_{\mu\nu}=-4\pi\sum_g g\int{\frac{dz_{\mu}}{ds}\delta^4(x-z)ds}$ y por tanto es obvio que
\begin{equation}
\partial^{\nu}G_{\mu\nu}=g\int{\frac{dz_{\mu}}{ds}\delta^4(x-z)ds}; \label{Ec Gz}
\end{equation}
su soluci\'on se logra usando el teorema de Stokes
\begin{equation}
\iint\left(\frac{\partial U}{\partial x}\frac{\partial V}{\partial y}-\frac{\partial U}{\partial y}\frac{\partial V}{\partial x}\right)dxdy=\int U\left(\frac{\partial V}{\partial x}dx+\frac{\partial V}{\partial y}dy\right),
\end{equation}
para $x=\tau_0$, $y=\tau_1$, $U=\delta^4(x-y)$ y $V=y_{\mu}$,
\begin{multline}
\underset{S}{\int\int}\left(\frac{\partial \delta^4(x-y)}{\partial\tau_0}\frac{\partial y_{\mu}}{\partial\tau_1}-\frac{\partial y_{\mu}}{\partial\tau_1}\frac{\partial \delta^4(x-y)}{\partial\tau_0}\right)d\tau_0\tau_1 \\
=\int_C \delta^4(x-y)\left(\frac{\partial y_{\mu}}{\partial\tau_0}d\tau_0+\frac{\partial y_{\mu}}{\partial\tau_1}d\tau_1\right)
\end{multline}
reorganizando el primer miembro se llega a
\begin{equation}
-\underset{S}{\int\int}\left(\frac{\partial y_{\mu}}{\partial\tau_1}\frac{\partial \delta^4(x-y)}{\partial\tau_0}-\frac{\partial \delta^4(x-y)}{\partial\tau_0}\frac{\partial y_{\mu}}{\partial\tau_1}\right)d\tau_0\tau_1
\end{equation}
usando la regla de la cadena para $y_{\nu}$
\begin{equation}
-\underset{S}{\int\int}\left(\frac{\partial y_{\mu}}{\partial\tau_1}\frac{\partial y_{\nu}}{\partial\tau_0}-\frac{\partial y_{\nu}}{\partial\tau_0}\frac{\partial y_{\mu}}{\partial\tau_1}\right)\frac{\partial \delta^4(x-y)}{\partial y_{\nu}}d\tau_0\tau_1
\end{equation}
ahora se intercambia la variable de integraci\'on para la funci\'on delta quedando
\begin{equation}
\underset{S}{\int\int}\left(\frac{\partial y_{\mu}}{\partial\tau_1}\frac{\partial y_{\nu}}{\partial\tau_0}-\frac{\partial y_{\nu}}{\partial\tau_0}\frac{\partial y_{\mu}}{\partial\tau_1}\right)\frac{\partial \delta^4(x-y)}{\partial x_{\nu}}d\tau_0\tau_1
\end{equation}
y por tanto queda
\begin{equation}
\frac{\partial}{\partial x_{\nu}}\underset{S}{\int\int}\left(\frac{\partial y_{\mu}}{\partial\tau_1}\frac{\partial y_{\nu}}{\partial\tau_0}-\frac{\partial y_{\nu}}{\partial\tau_0}\frac{\partial y_{\mu}}{\partial\tau_1}\right)\delta^4(x-y)d\tau_0\tau_1.
\end{equation}
Finalmente al incluir el segundo miembro del teorema de Stokes y teniendo en cuenta la carga magn\'etica se obtiene,
\begin{equation}
\frac{\partial G_{\mu\nu}}{\partial x_{\nu}}=g\int \delta^4(x-y)\left(\frac{\partial y_{\mu}}{\partial\tau_0}d\tau_0+\frac{\partial y_{\mu}}{\partial\tau_1}d\tau_1\right)
\end{equation}
recordando que $y_{\mu}=y_{\mu}(\tau_0,\tau_1)$ y tomando $\tau_1=0$, debido a que se integra solo sobre una sola l\'amina del anillo y que no tienda al infinito, entonces
\begin{equation}
\frac{\partial G_{\mu\nu}}{\partial x_{\nu}}=\int \delta^4(x-y)\left(\frac{\partial y_{\mu}(\tau_0,0)}{\partial\tau_0}\right)d\tau_0=\partial^{\nu}G_{\mu\nu},
\end{equation}
haciendo las correspondencias $\tau_0=s$ y $\frac{\partial y_{\mu}(\tau_0,0)}{\partial\tau_0}=\frac{dz_{\mu}}{ds}$, $y_{\mu}(\tau_0,0)$ es la l\'{\i}nea de mundo de la part\'{\i}cula, se observa que corresponde con la ecuaci\'on (\ref{Ec Gz}). Tambi\'en aqu\'{\i} se puede observar que el campo que est\'a definido solamente sobre las l\'aminas y que le da consistencia a la electrodin\'amica conel polo magn\'etico incluido tiene la forma:
\begin{equation}
G_{\mu\nu}=g\iint\left(\frac{\partial y_{\mu}}{\partial\tau_0}d\tau_0\frac{\partial y_{\nu}}{\partial\tau_1}+\frac{\partial y_{\mu}}{\partial\tau_1}\frac{\partial y_{\nu}}{\partial\tau_1}\right)d\tau_0d\tau_1 \label{Gmunu}.
\end{equation}
Las ecuaciones (\ref{DivElect}) y (\ref{DivPolo}) están directamente relacionadas mediante la ecuaci\'on $F_{\mu\nu}=\frac{\partial A_{\nu}}{\partial x^{\mu}}-\frac{\partial A_{\mu}}{\partial x^{\nu}}+4\pi\sum_gG^{\dagger}_{\mu\nu}$ para definir los potenciales retardados (PR) en las l\'{\i}neas de mundo de las part\'{\i}culas previamente definidas y su contribuci\'on a cada part\'{\i}cula depende solamente de la l\'{\i}nea de mundo de ella y de la l\'amina que est\'a unida a \'esta en el caso que halla polo. La contribuci\'on de los PR se define mediante una funci\'on invariante de Lorentz definida por
\begin{equation}
J(x)=
\begin{cases}
2\delta(x^{\mu}_{\mu}) & \text{para } x_0>0 \\
0 & \text{para } x_0<0
\end{cases}
\end{equation}
o por
\begin{align}
J(x) &= \frac{1}{r}\delta(x_0-r), \\
r &=(x_1^2+x_2^2+x_3^2)^{\frac{1}{2}}.
\end{align}
La funci\'on $\Delta(x)$ de Jordan y Pauli (ver ap\'endice \ref{PJFunction}) se relaciona con $J(x)$ mediante
\begin{equation}
\Delta(x)=J(x)-J(-x) \label{FunctPauli-Jordan}.
\end{equation}
Se verifica directamente alrededor del origen usando la integral de $\square J(x)$ sobre un peque\~no volumen cuadridimensional alrededor del origen como una integral superficial tridimensional sobre la frontera de ese volumen que,
\begin{equation}
\square J(x)=4\pi\delta_4(x).
\end{equation}
La contribuci\'on de una part\'{\i}cula cargada a los PR est\'a definida por,
\begin{equation}
A_{\nu,\text{ret}}^*(x)=e\int^{\infty}_{-\infty}J(x-z)\frac{dz_{\nu}}{ds}ds \label{ContribPR}
\end{equation}
basada en  los campos y potenciales dados por Li\'enard-Wiechert,
\begin{equation}
A^{\alpha}(x)=\frac{4\pi}{c}\int d^4x'D_{\text{ret}}(x-x')J^{\alpha}(x') \label{Lien-Wiech},
\end{equation}
donde $D_{\text{ret}}(x-x')$ es el propagador retardado (ret) o avanzado (av),
\begin{equation}
D_{i}(x-x')=
\begin{cases}
i=\text{ret} & \frac{1}{2\pi}\theta(x_0-x'_0)\delta[(x-x')^2] \\
i=\text{av} & \frac{1}{2\pi}\theta(x'_0-x_0)\delta[(x-x')^2],
\end{cases}
\end{equation}
y donde $\theta(x_0-x'_0)$ es la funci\'on de Heaviside o funci\'on escal\'on. La contribuci\'on correspondiente de un polo est\'a dada por la integraci\'on sobre toda la l\'amina, y debido a que su origen tiene un caracter antisim\'etrico, debe estar presente el tensor totalmente antisim\'etrico $\varepsilon_{\alpha\beta\gamma\delta}$:
\begin{equation}
A_{\nu,\text{ret}}(x)=g\varepsilon_{\nu\lambda\rho\sigma}\iint\frac{\partial y^{\lambda}}{\partial\tau_0}\frac{\partial y^{\rho}}{\partial\tau_1}\frac{\partial J(x-y)}{\partial x_{\sigma}}d\tau_0d\tau_1 \label{ContribPole}.
\end{equation}
Para evidenciar como es la variaci\'on de ese PR en la l\'amina, se realiza una variaci\'on cuadridimensional quedando,
\begin{equation}
\varepsilon^{\mu\nu\alpha\beta}\partial_{\mu}A_{\nu,\text{ret}}(x)=g\varepsilon^{\mu\nu\alpha\beta}\varepsilon_{\nu\lambda\rho\sigma}.\iint\frac{\partial y^{\lambda}}{\partial\tau_0}\frac{\partial y^{\rho}}{\partial\tau_1}\frac{\partial^2 J(x-y)}{\partial y_{\sigma}\partial y^{\mu}}d\tau_0d\tau_1.
\end{equation}
Usando la relaci\'on
\begin{equation}
\varepsilon^{\mu\nu\alpha\beta}\varepsilon_{\nu\lambda\rho\sigma}=\delta_{\lambda}^{\mu}\delta_{\rho}^{\alpha}\delta_{\sigma}^{\beta}+\delta_{\rho}^{\mu}\delta_{\sigma}^{\alpha}\delta_{\lambda}^{\beta}+\delta_{\sigma}^{\mu}\delta_{\rho}^{\alpha}\delta_{\lambda}^{\beta}-\delta_{\lambda}^{\mu}\delta_{\rho}^{\beta}\delta_{\sigma}^{\alpha}-\delta_{\rho}^{\mu}\delta_{\sigma}^{\beta}\delta_{\lambda}^{\alpha}-\delta_{\sigma}^{\mu}\delta_{\rho}^{\beta}\delta_{\lambda}^{\alpha}
\end{equation}
la ecuaci\'on anterior se modifica a,
\begin{multline}
\varepsilon^{\mu\nu\alpha\beta}\partial_{\mu}A_{\nu,\text{ret}}(x)=g\iint\left\{\frac{\partial y^{\mu}}{\partial\tau_0}\frac{\partial y^{\alpha}}{\partial\tau_1}\frac{\partial^2 J(x-y)}{\partial y^{\mu}\partial y_{\beta}}+\frac{\partial y^{\beta}}{\partial\tau_0}\frac{\partial y^{\mu}}{\partial\tau_1}\frac{\partial^2 J(x-y)}{\partial y^{\mu}\partial y_{\alpha}}\right. \\
+\frac{\partial y^{\beta}}{\partial\tau_0}\frac{\partial y^{\alpha}}{\partial\tau_1}\frac{\partial^2 J(x-y)}{\partial y^{\mu}\partial y_{\rho}}-\frac{\partial y^{\mu}}{\partial\tau_0}\frac{\partial y^{\beta}}{\partial\tau_1}\frac{\partial^2 J(x-y)}{\partial y^{\mu}\partial y_{\alpha}} \\
\left.-\frac{\partial y^{\alpha}}{\partial\tau_0}\frac{\partial y^{\mu}}{\partial\tau_1}\frac{\partial^2 J(x-y)}{\partial y^{\mu}\partial y_{\beta}}.-\frac{\partial y^{\alpha}}{\partial\tau_0}\frac{\partial y^{\beta}}{\partial\tau_1}\frac{\partial^2 J(x-y)}{\partial y^{\mu}\partial y_{\rho}}\right\}d\tau_0d\tau_1
\end{multline}
\begin{multline}
=g\iint\left\{\frac{\partial y^{\alpha}}{\partial\tau_1}\frac{\partial^2 J(x-y)}{\partial \tau_0\partial y_{\beta}}+\frac{\partial y^{\beta}}{\partial\tau_0}\frac{\partial^2 J(x-y)}{\partial \tau_1\partial y_{\alpha}}+\frac{\partial y^{\beta}}{\partial\tau_0}\frac{\partial y^{\alpha}}{\partial\tau_1}\square J(x-y)\right. \\
\left.-\frac{\partial y^{\beta}}{\partial\tau_1}\frac{\partial^2 J(x-y)}{\partial \tau_0\partial y_{\alpha}}-\frac{\partial y^{\alpha}}{\partial\tau_0}\frac{\partial^2 J(x-y)}{\partial \tau_1\partial y_{\beta}}-\frac{\partial y^{\alpha}}{\partial\tau_0}\frac{\partial y^{\beta}}{\partial\tau_1}\square J(x-y)\right\}d\tau_0d\tau_1
\end{multline}
\begin{multline}
=g\iint\left\{\frac{\partial y^{\alpha}}{\partial\tau_1}\frac{\partial^2 J(x-y)}{\partial \tau_0\partial y_{\beta}}+\frac{\partial y^{\beta}}{\partial\tau_0}\frac{\partial^2 J(x-y)}{\partial \tau_1\partial y_{\alpha}}-\frac{\partial y^{\beta}}{\partial\tau_1}\frac{\partial^2 J(x-y)}{\partial \tau_0\partial y_{\alpha}}\right. \\
\left.-\frac{\partial y^{\alpha}}{\partial\tau_0}\frac{\partial^2 J(x-y)}{\partial \tau_1\partial y_{\beta}}\right\}d\tau_0d\tau_1+4\pi g\iint\left\{\frac{\partial y^{\beta}}{\partial\tau_0}\frac{\partial y^{\alpha}}{\partial\tau_1}\delta_4(x-y)\right.  \\
\left.-\frac{\partial y^{\alpha}}{\partial\tau_0}\frac{\partial y^{\beta}}{\partial\tau_1}\delta_4(x-y)\right\}d\tau_0d\tau_1,
\end{multline}
donde $\square J(x-y)=4\pi\delta_4(x-y)$. Por tanto usando, $\iint\frac{\partial y^{\alpha}}{\partial\tau_1}\frac{\partial^2 J(x-y)}{\partial \tau_0\partial y_{\beta}}d\tau_0d\tau_1=\int dy^{\alpha}\frac{\partial J(x-y)}{\partial y_{\beta}}$ y $dy^{\alpha}\frac{\partial J(x-y)}{\partial y_{\beta}}=\left[\frac{\partial J(x-y)}{\partial y_{\beta}}\right]_{y=z}\frac{dz^{\alpha}}{ds}ds,$ se obtiene
\begin{multline}
\varepsilon^{\mu\nu\alpha\beta}\partial_{\mu}A_{\nu,\text{ret}}(x)=g\iint\left\{\frac{\partial J(x-y)}{\partial y_{\beta}}dy^{\alpha}+\frac{\partial J(x-y)}{\partial y_{\alpha}}dy^{\beta}\right.  \\
\left.-\frac{\partial J(x-y)}{\partial y_{\alpha}}dy^{\beta}-\frac{\partial J(x-y)}{y_{\beta}}dy^{\alpha}\right\}+4\pi g\iint\left\{\frac{\partial y^{\beta}}{\partial\tau_0}\frac{\partial y^{\alpha}}{\partial\tau_1}\delta_4(x-y) \right. \\
\left.-\frac{\partial y^{\alpha}}{\partial\tau_0}\frac{\partial y^{\beta}}{\partial\tau_1}\delta_4(x-y)\right\}d\tau_0d\tau_1.
\end{multline}
Integrando mediante el uso de la funci\'on delta y debido a que,
\begin{equation}
g\int\left\{\frac{\partial J(x-y)}{\partial y_{\alpha}}dy^{\beta}+\frac{\partial J(x-y)}{y_{\beta}}dy^{\alpha}\right\}
\end{equation}
es un intercambio entre $\alpha\text{ y }\beta$ y teniendo en cuenta que,
\begin{equation}
g^{-1}G^{\alpha\beta}= \int\left\{\frac{\partial y^{\beta}}{\partial\tau_0}\frac{\partial y^{\alpha}}{\partial\tau_1}\delta_4(x-y)-\frac{\partial y^{\alpha}}{\partial\tau_0}\frac{\partial y^{\beta}}{\partial\tau_1}\delta_4(x-y)\right\}d\tau_0d\tau_1
\end{equation}
la anterior ecuaci\'on se escribe abreviadamente de la forma,
\begin{equation}
\varepsilon^{\mu\nu\alpha\beta}\partial_{\mu}A_{\nu,\text{ret}}(x)=g\int\left\{\frac{\partial J(x-y)}{\partial y_{\beta}}dy^{\alpha}+\frac{\partial J(x-y)}{\partial y_{\alpha}}dy^{\beta} \right\}-(\alpha\beta)
+4\pi G^{\alpha\beta}
\end{equation}
cambiando la referencia a la l\'{\i}nea de mundo de la part\'{\i}cula descrita con las cuadri-coordenadas $z_{\mu}(s)$, donde $s$ es el tiempo propio de \'esta y $\varepsilon^{\mu\nu\alpha\beta}\partial_{\mu}A_{\nu,\text{ret}}(x)-4\pi G^{\alpha\beta}=F^{\dagger\alpha\beta}_{\text{ret}}$, la ecuaci\'on se transforma a
\begin{align}
F^{\dagger\alpha\beta}_{\text{ret}} &=g\int\left[\frac{\partial J(x-y)}{\partial z_{\beta}}\right]_{y=z}\frac{dz^{\alpha}}{ds}ds-(\alpha\beta) \notag \\
&=g\int\frac{\partial J(x-z)}{\partial z_{\beta}}\frac{dz^{\alpha}}{ds}ds-(\alpha\beta).
\end{align}
Usando el hecho que $B^{\alpha}_{\text{ret}}=g\int\frac{\partial J(x-z)}{\partial z_{\beta}}\frac{dz^{\alpha}}{ds}ds$, o sea que se mantiene la correspondencia del potencial de Lienard-Wiechert con los potenciales magn\'eticos para un campo retardado producido por un polo magn\'etico (campos magn\'eticos retardados $B^{\sigma}_{\text{ret}}$), la ecuaci\'on finalmente toma la forma
\begin{align}
F^{\dagger\alpha\beta}_{\text{ret}}=-\frac{\partial B^{\alpha}_{\text{ret}}}{\partial x_{\beta}}+\frac{\partial B^{\beta}_{\text{ret}}}{\partial x_{\alpha}}.
\end{align}
\newline
\newline
\textbf{IV. Principio de acci\'on}
\newline
\newline
El principio de acci\'on (ver ap\'endice \ref{PpioAcc}) se enuncia sobre tres t\'erminos: el de la part\'{\i}cula ($I_1$), el del campo ($I_2$) y el de las interacciones de las cargas el\'ectrica y magn\'etica con el campo ($I_3$). Entonces la acci\'on se define $I=I_1+I_2+I_3$, donde
\begin{align}
I_1 &=\sum_{e+g} m\int{ds} \notag \\
I_2 &=\frac{1}{16\pi}\int{F^*_{\mu\nu}(x)F^{\mu\nu}(x)d^4x} \notag \\
I_3 &=\sum_ee\int{A^{\nu}(z)\frac{dz_{\nu}}{ds}ds} \label{Acc123}
\end{align}
donde $F^*_{\mu\nu}=\int{F_{\mu\nu}(x')\gamma(x-x')d^4x'}$ con $\gamma(x-x')$ definida como una funci\'on par que se aproxima a la funci\'on delta $\delta^4(x)$ y que tiene como objetivo evitar infinitos en la ecuaci\'on de movimiento debido a los campos infinitos producidos por las cargas puntuales y por los monopolos magn\'eticos; por \'esta raz\'on el delta $I_2$ se redefine como $I'_2=\frac{1}{16\pi}\iint{F_{\mu\nu}(x')\gamma(x-x')F^{\mu\nu}(x')d^4xd^4x'}$.
\newline
La variaci\'on de la acci\'on $I_1$ est\'a dada por,
\begin{equation}
\delta I_1 =-\sum_{e+g} m\int{\frac{d^2z_{\mu}}{ds^2}\delta z^{\mu}\delta s}.
\end{equation}
Para $I_3$,
\begin{equation}
\delta I_3=\sum_ee\int\left[\left(\frac{\partial A_{\nu}}{\partial x^{\mu}}-\frac{\partial A_{\mu}}{\partial x^{\nu}}\right)+\delta A_{\nu}\right]_{x=z}\frac{dz^{\nu}}{ds}ds
\end{equation}
Finalmente la variaci\'on en la acci\'on del campo es
\begin{align}
\delta I'_2 &=\frac{1}{16\pi}\iint\left[F_{\mu\nu}(x)\delta F^{\mu\nu}(x)\gamma(x-x')+\delta F_{\mu\nu}(x)F^{\mu\nu}(x)\gamma(x-x')\right]d^4x d^4x' \notag \\
&=\frac{1}{8\pi}\int\left[F^*_{\mu\nu}(x)\delta F^{\mu\nu}(x)\right]d^4x
\end{align}
pero $F^{\mu\nu}(x)=\frac{\partial A^{\nu}}{\partial x_{\mu}}-\frac{\partial A^{\mu}}{\partial x_{\nu}}+4\pi\sum_g G^{\dagger\mu\nu}$ y $\delta F^{\mu\nu}=\frac{\partial \delta A^{\nu}}{\partial x_{\mu}}-\frac{\partial \delta A^{\mu}}{\partial x_{\nu}}+4\pi\sum_g \delta G^{\dagger\mu\nu}$, por tanto reemplazando de obtiene
\begin{align}
\delta I_2 &=-\frac{1}{4\pi}\int F^*_{\mu\nu}(x)\frac{\partial \delta A^{\mu}}{\partial x_{\nu}}d^4x+\frac{1}{2}\sum_g \int F^*_{\mu\nu}(x)\delta G^{\dagger}_{\mu\nu}d^4x \notag \\
&=\frac{1}{4\pi}\int \frac{\partial F^*_{\mu\nu}(x)}{\partial x_{\nu}} \delta A^{\mu}d^4x+\frac{1}{2}\sum_g \int F^{\dagger *}_{\mu\nu}(x)\delta G_{\mu\nu}d^4x \label{deltaI2}
\end{align}
Usando la definici\'on $G_{\mu\nu}(x)=2g\iint\frac{\partial y_{\mu}}{\partial \tau_0}\frac{\partial y_{\mu}}{\partial \tau_1}\delta_4(x-y)d\tau_0d\tau_1$ su variaci\'on est\'a dada por,
\begin{align}
\delta G_{\mu\nu}(x) &=2g\iint\delta\left(\frac{\partial y_{\mu}}{\partial \tau_0}\frac{\partial y_{\mu}}{\partial \tau_1}\right)\delta_4(x-y)d\tau_0d\tau_1 \notag \\
&\hspace{5.2cm}+2g\iint\frac{\partial y_{\mu}}{\partial \tau_0}\frac{\partial y_{\mu}}{\partial \tau_1}\frac{\partial\delta_4(x-y)}{\partial y^{\sigma}}dy^{\sigma}d\tau_0d\tau_1 \notag \\
&=2g\iint\left(\frac{\partial \delta y_{\mu}}{\partial \tau_0}\frac{\partial y_{\mu}}{\partial \tau_1}+\frac{\partial y_{\mu}}{\partial \tau_0}\frac{\partial \delta y_{\mu}}{\partial \tau_1}\right)\delta_4(x-y)d\tau_0d\tau_1 \notag \\
&\hspace{5.2cm}+2g\iint\frac{\partial y_{\mu}}{\partial \tau_0}\frac{\partial y_{\mu}}{\partial \tau_1}\frac{\partial\delta_4(x-y)}{\partial y^{\sigma}}\delta y^{\sigma}d\tau_0d\tau_1 \notag \\
\end{align}
por tanto $\delta I'_2$ es
\begin{align}
\delta I'_2 &=\frac{1}{4\pi}\int F^*_{\mu\nu}(x)\frac{\partial \delta A^{\mu}}{\partial x_{\nu}}d^4x \notag \\
&\hspace{2cm}+\sum_g g\int F^*_{\mu\nu}(x)\left\{\iint\left[\left(\frac{\partial \delta y_{\mu}}{\partial \tau_0}\frac{\partial y_{\mu}}{\partial \tau_1}+\frac{\partial y_{\mu}}{\partial \tau_0}\frac{\partial \delta y_{\mu}}{\partial \tau_1}\right)\delta_4(x-y)\right.\right. \notag \\
&\hspace{4.5cm}\left.\left.+\frac{\partial y_{\mu}}{\partial \tau_0}\frac{\partial y_{\mu}}{\partial \tau_1}\frac{\partial\delta_4(x-y)}{\partial y^{\sigma}}\delta y^{\sigma}\right]d\tau_0d\tau_1\right\}d^4x.
\end{align}
Integrando sobre $x$,
\begin{align}
\delta I'_2 &=\frac{1}{4\pi}\int F^*_{\mu\nu}(x)\frac{\partial \delta A^{\mu}}{\partial x_{\nu}}d^4x+\sum_g g
\iint \left[F^{\dagger *}_{\mu\nu}(y)\left(\frac{\partial \delta y_{\mu}}{\partial \tau_0}\frac{\partial y_{\mu}}{\partial \tau_1}+\frac{\partial y_{\mu}}{\partial \tau_0}\frac{\partial \delta y_{\mu}}{\partial \tau_1}\right)\right. \notag \\
&\hspace{6cm}\left.+\frac{\partial y_{\mu}}{\partial \tau_0}\frac{\partial y_{\mu}}{\partial \tau_1}\frac{\partial F^{\dagger *}_{\mu\nu}(y)}{\partial y^{\sigma}}\delta y^{\sigma}\right]d\tau_0d\tau_1,
\end{align}
pero $\frac{\partial}{\partial \tau_0}[F^{\dagger *}_{\mu\nu}(y)\delta y^{\mu}]=\frac{\partial F^{\dagger *}_{\mu\nu}(y)}{\partial \tau_0}\delta y^{\mu}+F^{\dagger *}_{\mu\nu}(y)\frac{\partial \delta y^{\mu}}{\partial \tau_0}$, entonces,
\begin{align}
\delta I_2 &=\frac{1}{4\pi}\int F^*_{\mu\nu}(x)\frac{\partial \delta A^{\mu}}{\partial x_{\nu}}d^4x+\sum_g g
\iint \left[F^{\dagger *}_{\mu\nu}(y)\left(\frac{\partial \delta y_{\mu}}{\partial \tau_0}\frac{\partial y_{\mu}}{\partial \tau_1}+\frac{\partial y_{\mu}}{\partial \tau_0}\frac{\partial \delta y_{\mu}}{\partial \tau_1}\right)\right. \notag \\
&\hspace{3.5cm}\left.+\frac{\partial y_{\mu}}{\partial \tau_0}\frac{\partial y_{\mu}}{\partial \tau_1}\left(\frac{\partial}{\partial \tau_0}[F^{\dagger *}_{\mu\nu}(y)\delta y^{\mu}]-F^{\dagger *}_{\mu\nu}(y)\frac{\partial \delta y^{\mu}}{\partial \tau_0}\right)\right]d\tau_0d\tau_1 \notag \\
&=\frac{1}{4\pi}\int F^*_{\mu\nu}(x)\frac{\partial \delta A^{\mu}}{\partial x_{\nu}}d^4x+\sum_g g
\iint \left[\frac{\partial}{\partial \tau_1}[F^{\dagger *}_{\mu\nu}(y)\delta y^{\mu}]\frac{\partial \delta y_{\mu}}{\partial \tau_0}\right. \notag \\
&\hspace{1.5cm}\left.-\frac{\partial F^{\dagger *}_{\mu\nu}(y)}{\partial \tau_1}\delta y^{\mu}\frac{\partial \delta y_{\mu}}{\partial \tau_0}+\frac{\partial}{\partial \tau_0}[F^{\dagger *}_{\mu\nu}(y)\delta y^{\mu}]\frac{\partial \delta y_{\mu}}{\partial \tau_1}-\frac{\partial F^{\dagger *}_{\mu\nu}(y)}{\partial \tau_0}\delta y^{\mu}\frac{\partial \delta y_{\mu}}{\partial \tau_1}\right. \notag \\
&\hspace{1.5cm}\left.+\frac{\partial y_{\mu}}{\partial \tau_0}\frac{\partial y_{\mu}}{\partial \tau_1}\frac{\partial}{\partial \tau_0}[F^{\dagger *}_{\mu\nu}(y)\delta y^{\mu}]-\frac{\partial y_{\mu}}{\partial \tau_0}\frac{\partial y_{\mu}}{\partial \tau_1}F^{\dagger *}_{\mu\nu}(y)\frac{\partial \delta y^{\mu}}{\partial \tau_0}\right]d\tau_0d\tau_1
\end{align}
Siguiendo el desarrollo algebraico, se realiza el cambio a la l\'{\i}nea de mundo tal como se hizo anteriormente y usando el teorema de Stokes,
\begin{equation}
\iint\left(\frac{\partial U}{\partial d\tau_0}\frac{\partial V}{\partial d\tau_1}-\frac{\partial U}{\partial d\tau_1}\frac{\partial V}{\partial d\tau_0}\right)d\tau_0d\tau_1=\int U\left(\frac{\partial V}{\partial \tau_0}d\tau_0+\frac{\partial V}{\partial \tau_1}d\tau_1\right),
\end{equation}
se obtiene la acci\'on total,
\begin{equation}
\delta I=\delta I'_2-\frac{1}{4\pi}\int F^*_{\mu\nu}\frac{\partial \delta A^{\mu}}{\partial x_{\nu}}d^4x+\delta I_1+\delta I_2=\delta I'_2+\frac{1}{4\pi}\int\frac{\partial F^*_{\mu\nu}}{\partial x_{\nu}}\delta A^{\mu}d^4x+\delta I_1+\delta I_2
\end{equation}
donde,
\begin{align}
\delta I_2 &=\sum_e e\int\left[\left(\frac{\partial A_{\nu}}{\partial x^{\mu}}-\frac{\partial A_{\mu}}{\partial x^{\nu}}\right)_{x=z}\delta z^{\mu}+(\delta A_{\nu})_{x=z}\right]\left(\frac{dz^{\nu}}{ds}\right)ds \notag \\
\delta I'_2 &=\sum_g g\int F^{\dagger *}_{\mu\nu}(z)\delta z^{\mu}\frac{dz^{\nu}}{ds}ds-\sum_g g\iint\left(\frac{\partial F^{\dagger *}_{\mu\nu}}{dy^{\rho}}+\frac{\partial F^{\dagger *}_{\nu\rho}}{dy^{\mu}}\right. \notag \\ &\hspace{6cm}\left.+\frac{\partial F^{\dagger *}_{\rho\mu}}{dy^{\nu}}\right)\frac{\partial y^{\rho}}{\partial \tau_0}\frac{\partial y^{\nu}}{\partial \tau_1}\delta y^{\mu}d\tau_0d\tau_1 \label{deltaI2TS}.
\end{align}
Si $\sum_g g\iint\left(\frac{\partial F^{\dagger *}_{\mu\nu}}{dy^{\rho}}+\frac{\partial F^{\dagger *}_{\nu\rho}}{dy^{\mu}}+\frac{\partial F^{\dagger *}_{\rho\mu}}{dy^{\nu}}\right)\frac{\partial y^{\rho}}{\partial \tau_0}\frac{\partial y^{\nu}}{\partial \tau_1}\delta y^{\mu}d\tau_0d\tau_1=0$, entonces $\frac{\partial F^{\dagger *}_{\mu\nu}}{dy^{\rho}}+\frac{\partial F^{\dagger *}_{\nu\rho}}{dy^{\mu}}+\frac{\partial F^{\dagger *}_{\rho\mu}}{dy^{\nu}}=0$, \'esto significa que todos los puntos se mantendr\'an en la l\'amina. Igualando a cero los otros coeficientes se puede observar que se reproducen las ecuaciones de movimiento iniciales, lo cual le da consistencia a la teor\'{\i}a y se concluye que las cuerdas de Dirac nunca deben pasar por las part\'{\i}culas cargadas.
\newline
De $\frac{\partial F^{\dagger *}_{\mu\nu}}{dy^{\nu}}=0$ se observa que al aplicar el principio de acci\'on no se llega a una ecuaci\'on de movimiento lo que significa que las cuerdas de Dirac no poseen ecuaci\'on de movimiento y por tanto no tienen car\'acter f\'{\i}sico.
\newline
Con la acci\'on anterior existe el problema de la anulaci\'on del momento conjugado a $A_0$ en la formulaci\'on Hamiltoniana. La soluci\'on de \'esto se logra adicionando un t\'ermino a la acci\'on de la forma,
$I_4=\frac{1}{8\pi}\int{\frac{\partial A^*_{\nu}}{\partial x_{\mu}}\frac{\partial A_{\mu}}{\partial x_{\nu}}}d^4x$, cuya variaci\'on es
\begin{align}
\delta I_4 &=\frac{1}{8\pi}\int{\frac{\partial \delta A^*_{\nu}}{\partial x_{\mu}}\frac{\partial A_{\mu}}{\partial x_{\nu}}+\frac{\partial A^*_{\nu}}{\partial x_{\mu}}\frac{\partial \delta A_{\mu}}{\partial x_{\nu}}}d^4x \notag \\
&=\frac{1}{4\pi}\int{\frac{\partial A^*_{\nu}}{\partial x_{\mu}}\frac{\partial \delta A_{\mu}}{\partial x_{\nu}}}d^4x \label{deltaI4},
\end{align}
integrando por partes con $U=\frac{\partial A^*_{\nu}}{\partial x_{\mu}}$, $dV=\frac{\partial \delta A_{\mu}}{\partial x_{\nu}}dx^4$, $dU=\frac{\partial^2 A^*_{\nu}}{\partial x^{\mu}\partial x^{\nu}}dx^4$ y $V=\delta A^{\mu}$ y usando la condici\'on de integral extremal, se obtiene,
\begin{equation}
\delta I_4=-\frac{1}{4\pi}\int{\frac{\partial^2 A^*_{\nu}}{\partial x^{\mu}\partial x_{\nu}}\delta A^{\mu}}d^4x \label{I4},
\end{equation}
pero debido a la condici\'on suplementaria que $\frac{\partial A_{\nu}}{\partial x_{\mu}}=0$ o $\frac{\partial A^*_{\nu}}{\partial x_{\mu}}=0$, entonces el t\'ermino $\frac{\partial^2 A^*_{\nu}}{\partial x^{\mu}\partial x_{\nu}}$ no afecta las ecuaciones de movimiento. En la formulaci\'on Hamiltoniana es necesario distiguir entre las ecuaciones que se mantienen en virtud de las condiciones suplementarias y las que son independientes de \'estas; entonces,
\begin{equation}
\frac{\partial F^*_{\mu\nu}}{\partial x_{\nu}}=-4\pi\sum_e e\int{\frac{dz_{\mu}}{ds}\delta_4(x-z)}ds
\end{equation}
con $F_{\mu\nu}=\frac{\partial A_{\nu}}{\partial x^{\mu}}-\frac{\partial A_{\mu}}{\partial x^{\nu}}+4\pi\sum_gG^{\dagger}_{\mu\nu}$ y $G_{\mu\nu}=g\iint\left(\frac{\partial y_{\mu}}{\partial \tau_0}\frac{\partial y_{\nu}}{\partial \tau_1}-\frac{\partial y_{\mu}}{\partial \tau_1}\frac{\partial y_{\nu}}{\partial \tau_0}\right)\delta_4(x-z)d\tau_0d\tau_1$ entonces reemplazando,
\begin{equation}
-\frac{\partial^2 \partial A^*_{\mu}}{\partial x_{\nu}\partial x^{\nu}}+4\pi\sum_g\frac{\partial (G^{\dagger}_{\mu\nu})^*}{\partial x_{\nu}}=-4\pi\sum_e e\int{\frac{dz_{\mu}}{ds}\delta_4(x-z)}ds,
\end{equation}
debido a que $\frac{\partial^2 A^*_{\nu}}{\partial x_{\nu}\partial x^{\mu}}$ se anula por la condici\'on suplementaria. Finalmente se obtiene el D'Alembertiano,
\begin{equation}
\square A^*_{\mu}=4\pi\sum_e e\int{\frac{dz_{\mu}}{ds}\delta_4(x-z)}ds+4\pi\sum_g\frac{\partial (G^{\dagger}_{\mu\nu})^*}{\partial x_{\nu}}
\end{equation}
\newline
\newline
\textbf{V. El m\'etodo para pasar a la formulaci\'on Hamiltoniana}
\newline
\newline
Para cuantizar una teor\'{\i}a se usa convertir las ecuaciones de movimiento de un sistema din\'amico en forma de principio de acci\'on en forma Halmiltoniana. El m\'etodo es tomar la integral de acci\'on previa a un tiempo $t$ y hallar su variaci\'on permitiendo la variaci\'on de $t$. Entonces la variaci\'on de la acci\'on $\delta I$ es funci\'on lineal de $\delta t$ y $\delta q$ en las coordenadas din\'amicas del tiempo. Los t\'erminos adicionales en $\delta I$ se cancelan cuando se usan las ecuaciones de movimiento. Se introduce la variaci\'on total en los $q$ finales, $\Delta q=\delta q+\dot{q}\delta t$ y usando la transformaci\'on de Legendre $\delta I=\sum p_r(q_r,\dot{q}_r) \Delta q_r-W(q_r,\dot{q}_r)\delta t$, se definen los momentos $p_r$ y la energ\'{\i}a $W$ que son funci\'on de las coordenadas generalizadas $q_r$ y de las velocidades generalizadas $\dot{q}_r$. \'Estas variables son constantes en el tiempo si la Hamiltoniana transformada es nula y que satisface la relaci\'on
\begin{equation}
W-H(p,q)=0
\end{equation}
que es la ecuaci\'on de Hamilton-Jacobi cuyo argumento son derivadas parciales de la acci\'on respecto a $q$ y a $t$. Debido a que existen varias ecuaciones que relacionan a $q$ y a $p$, entonces deben existir varias ecuaciones de Hamilton-Jacobi que originen varias ecuaciones de onda.
\newline
Relativistamente es el mismo procedimiento pero se evalua la integral de la acci\'on sobre el espacio-tiempo relacionado con una superficie tridimensional que se extiende al infinito. Lo anterior se puede modificar debido a que en lugar de detener la integral de acci\'on en un tiempo determinado o en una superficie $S$ espacial-oide tridimensional definida, se podr\'{\i}an detener diferentes t\'erminos en la acci\'on en diferentes tiempos. Lo anteriormente expuesto es posible si se describe la din\'amica del sistema enmarcada en difentes partes a diferentes tiempos y despu\'es de que algunas partes del sistema han dejado de evolucionar voluntaria o involuntariamente desde el punto de vista din\'amico, el resto del sistema lo sigue haciendo hasta que eventualmente, termine la din\'amica del sistema. Cada vez que una parte del sistema para, se puede definir para \'esta una ecuaci\'on de Hamilton-Jacobi. \'Este m\'etodo no siempre es conveniente para soluci\'on de muchos problemas por tanto se generalizar\'a pasando a un espacio en el que una superficie espacial-oide tridimensional $S$ en la que se puede detener diferentes t\'erminos de la integral y no toda su acci\'on. El hecho es que si se eval\'ua cuidadosamente la din\'amica del sistema, se puede observar si parte de \'este se comporta de manera no natural o no f\'{\i}sica; una vez hecha la evaluaci\'on se detiene esa parte de la integral de acci\'on y puede hacerse dicho proceso a diferentes tiempos de evoluci\'on del sistema y el resto de \'este seguir\'a su desarollo regido por la ecuaci\'on de movimiento hasta cuando \'esta se detenga, colapse, etc.
\newline
Una forma de detener la integral de acci\'on cuando existe interacci\'on entre las part\'{\i}culas y los campos es suponer que las part\'{\i}culas no existen en el espacio-tiempo en donde se traslapan los conos de luz de diferentes part\'{\i}culas y posteriormente detener el campo donde el intervalo de tiempo para hacer \'esto debe ser de gran magnitud para dejar que \'este se estabilize. Entonces se var\'{\i}a la integral de acci\'on detenida mediante la variaci\'on sobre los puntos de la l\'{\i}nea de mundo de la part\'{\i}cula, $z_{\mu}$, en el espacio-tiempo donde \'esta desaparece y tambi\'en en la superficie tridimensional de los campos $S_F$, donde el campo deja de existir. Intentando emular el principio de Hamilton o el teorema fundamental del c\'alculo variacional, se iguala a cero la parte de la variaci\'on de la integral de acci\'on detenida y que no est\'a conectada con las variaciones de frontera, se obtiene la ecuaci\'on de movimiento de las part\'{\i}culas antes que desaparezcan y de \'esta manera se originan las ecuaciones de campo que rigen a la part\'{\i}cula a\'un despu\'es de haber dejado de existir. Debido a las variaciones realizadas en las l\'{\i}neas de mundo, $\Delta z_{\mu}$, localizadas dentro del campo, se obtienen ecuaciones m\'as apropiadas para manipular que las del m\'etodo tradicional en donde las part\'{\i}culas y el campo dejan de existir simult\'aneamente.
\newline
Con la nueva electrodin\'amica se supone que todas las part\'{\i}culas y tambi\'en las cuerdas unidas a los polos dejan de existir en la superficie tridimensional de los polos $S_P$, y que los campos electromagn\'eticos dejan de existir mucho m\'as tarde sobre la superficie de los campos $S_F$, resultando que las integrales de acci\'on dadas por las ecuaciones $I_1$ e $I_3$ de la ec. (\ref{Acc123}), tienen que detenerse cuando las l\'{\i}neas de mundo alcancen a $S_P$, mientras que las acciones dadas por las ecuaciones $I_2$ de la ec. (\ref{Acc123}) y (\ref{I4}), dejan de existir en la frontera $S_F$. El hecho de que se detengan \'estas cantidades no significa que el sistema colapse, pues el resto de la din\'amica de \'este seguir\'a evolucionando hasta $S_F$ y a\'un despu\'es cuando llegue a $S_F$.
\newline
Si la conexi\'on entre $U$ y $U^*$ dada por $U^*(x)=\int U(x')\gamma(x-x')d^3x'$, y es tal que el valor de cualquiera de ellos en un punto $x$ es determinado por los valores del otro sobre los puntos espacio-temporales cercanos a $x$ formando una especie de recursi\'on. As\'{\i} si uno se anula en cierta regi\'on espacio-temporal, el otro tambi\'en lo har\'a en esa regi\'on excepto en las cercan\'{\i}as de \'esta.
\newline
Debido a que $G_{\mu\nu}$ se anula en cualquier lugar excepto sobre las l\'aminas, $G^*_{\mu\nu}$ debe hacerlo tambi\'en en la regi\'on entre $S_P$ y $S_F$, excepto en los puntos cerca de donde las cuerdas dejan de existir. En \'esta regi\'on se tiene que deja de existir la primera suma sobre $\square A^*_{\mu}=4\pi\sum_e e\int{\frac{dz_{\mu}}{ds}\delta_4(x-z)}ds+4\pi\sum_g\frac{\partial (G^{\dagger}_{\mu\nu})^*}{\partial x_{\nu}}$, debido a que la integral se detiene en $S_P$; por tanto $\square A^*_{\mu}(x)=0$, y se seg\'un lo analizado tambi\'en se tiene que $\square A_{\mu}(x)=0$ en la regi\'on entre $S_P$ y $S_F$ con excepci\'on de los puntos cercanos a donde las part\'{\i}culas cargadas dejan de existir. En las regiones donde los D'Alembertianos de $A_{\mu}(x)$ y $A^*_{\mu}(x)$ se mantienen se pueden definir \'estas mediante transformaciones de Fourier de la forma:
\begin{equation}
A_{\mu}(x)=\sum_{k_0}\int A_{k_{\mu}}e^{ikx}\frac{d^3k}{k_0} \label{TransfFourierA1}
\end{equation}
\begin{equation}
A^*_{\mu}(x)=\sum_{k_0}\int A^*_{k_{\mu}}e^{ikx}\frac{d^3k}{k_0} \label{TransfFourierA2}
\end{equation}
donde $kx=k_0x_0-k_1x_1-k_2x_2-k_3x_3$, $d^3k=dk_1dk_2dk_3$, $k_0=\pm (k^2_1+k^2_2+k^2_3)^{\frac{1}{2}}$ y la suma se realiza sobre los valores positivos y negativos de $k_0$. El factor $\frac{d^3k}{k_0}$ se introduce debido a que \'este es invariante de Lorentz. La condici\'on que $A_{\mu}$ y $A^*_{\mu}$ sean reales da las condiciones,
\begin{equation}
A_{-k_{\mu}}=-\bar{A}_{k_{\mu}},\qquad A^*_{-k_{\mu}}=\bar{A}^*_{k_{\mu}}.
\end{equation}
Sea la resoluci\'on de Fourier de la funci\'on $\gamma(x)$ dada por,
\begin{equation}
\gamma(x)=\frac{1}{2\pi}\int \gamma_le^{ilx}d^4l\qquad (\gamma_{-l}=\bar{\gamma}_l),
\end{equation}
entonces la condici\'on de funci\'on par $\gamma(-x)=\gamma(x)$ da $\gamma_{-l}=\gamma_l$ debido a que $\gamma_l\; \epsilon\; \mathbb{R}$. Se tiene entonces que
\begin{align}
\gamma_kA_{k_{\mu}} &=\frac{1}{2\pi}\int \gamma_ke^{ikx}d^4k\left(\sum_{k_0}\int A_{k_{\mu}}e^{ikx}\frac{d^3k}{k_0}\right) \notag \\
&=\frac{1}{2\pi}\sum_{k_0}\iint \gamma_kA_{k_{\mu}}e^{i(k+k')x}\frac{d^3kd^4k'}{k_0}
\end{align}
donde usando el delta de Dirac se obtiene,
\begin{align}
\gamma_kA_{k_{\mu}}(x) &=\sum_{k_0}\int \gamma_kA_{k_{\mu}}e^{ikx}\frac{d^3k}{k_0} \notag \\
&=\sum_{k_0}\int A^*_{k_{\mu}}e^{ikx}\frac{d^3k}{k_0}=A^*_{\mu}(x).
\end{align}
Es necesario conservar la transformaci\'on de Fourier de $A_{\mu}(x)$ en cada punto $z$ donde una part\'{\i}cula cargada desaparece y mantener la transformaci\'on de Fourier de $A^*_{\mu}(x)$ en $y$ donde la cuerda deja de existir. Es probable que lo anterior se pueda manejar de tal forma que al elegir una funci\'on $\gamma$ que origine una condici\'on para que el punto $y$ nunca est\'e demasiado cerca a $z$, debido a que \'esto significar\'{\i}a que la totalidad del sistema deja de existir. Si se asume un campo $U(x)$, est\'e definido por $U^*(x')$ en los puntos $x'$ muy cercanos a los puntos $x$ y por fuera del cono de luz de $x$. As\'{\i} la transformaci\'on de Fourier de $A_{\mu}(z)$ es definida por la transformaci\'on de Fourier de $A^*_{\mu}(x')$ sobre los puntos $x'$ para los cuales la transformaci\'on de Fourier de $A^*_{\mu}(z)$ es v\'alida y est\'a definida. De la misma forma la transformaci\'on de Fourier de $A^*_{\mu}(y)$ ser\'a v\'alida solamente si $U^*(x)$ est\'a definida por $U(x')$ en los puntos $x'$ muy cercanos a a $x$ fuera del cono de luz de de $x$.
\newline
La condici\'on suplementaria $\frac{\partial A_{\nu}}{\partial x_{\nu}}=0,\;\frac{\partial A^*_{\nu}}{\partial x_{\nu}}=0$, se modifica en la regi\'on entre $S_P$ y $S_F$. Con las integrales dadas en las ecuaciones (\ref{ContribPR}) y (\ref{ContribPole}) detenidas en $S_P$ y usando $z'$ para $z(s')$:
\begin{align}
\frac{\partial A_{\nu,\text{ret}}^*}{\partial x_{\nu}} &=\sum_e e\int^{z}_{-\infty}\frac{\partial J(x-z')}{\partial x_{\nu}}\frac{dz'_{\nu}}{ds'}ds' \notag \\
&=-\sum_e e\int^{z}_{-\infty}\frac{\partial J(x-z')}{\partial z'_{\nu}}\frac{dz'_{\nu}}{ds'}ds' \notag \\
&=-\sum_e eJ(x-z) \label{CondicSuplem}
\end{align}
donde $-\sum_e eJ(x-z)$ es diferente de cero debido cuando $x$ est\'e sobre el cono de luz futura de cualquier punto $z$ donde una part\'{\i}cula cargada no exista.
\newline
\newline
\textbf{VI. La formulaci\'on Hamiltoniana}
\newline
\newline
Dejando variar a $S_P$ y no a $S_F$ con el objetivo de formar la variaci\'on de la integral de acci\'on acotada justamente como en la secci\'on anterior y evaluando los t\'erminos en $\delta I$ conectados con los l\'{\i}mites, resulta que los t\'erminos originados en la electrodin\'amica,
\begin{equation}
\sum_{e+g}m\left(\frac{dz_{\mu}}{ds}\right)\Delta z^{\mu}+\sum_e eA_{\mu}\Delta z^{\mu}
\end{equation}
donde $\Delta z^{\mu}$ son los cambios totales de las coordenadas del punto donde la part\'{\i}cula deja de existir y que vienen de $\delta J_1$ y $\delta J_2$. Ya no es posible usar la acci\'on (\ref{deltaI2}) sino la acci\'on
\begin{align}
\delta I_2 &=\frac{1}{8\pi}\int_{-\infty}^{S_F}\left[F_{\mu\nu}^*\left(\frac{\partial\delta A^{\mu}}{\partial x_{\nu}}\right)+F_{\mu\nu}\left(\frac{\partial\delta A^{\mu *}}{\partial x_{\nu}}\right)\right]d^4x \notag \\
&\hspace{5.5cm}+\frac{1}{4}\sum_{g}\int_{-\infty}^{S_F}\left[F_{\mu\nu}^* \delta G^{\mu\nu\dagger}+F_{\mu\nu} \delta G^{\mu\nu\dagger *}\right]d^4x \notag \\
&=\frac{1}{8\pi}\int_{-\infty}^{S_F}\left[F_{\mu\nu}^*\left(\frac{\partial\delta A^{\mu}}{\partial x_{\nu}}\right)+F_{\mu\nu}\left(\frac{\partial\delta A^{\mu *}}{\partial x_{\nu}}\right)\right]d^4x+\frac{1}{4}\sum_{g}\int_{-\infty}^{\infty}F_{\mu\nu}^* \delta G^{\mu\nu\dagger}d^4x
\end{align}
La raz\'on del cambio en el segundo t\'ermino es que si $S_F$ est\'a lo suficientemente lejos de $S_P$, entonces $\gamma(x-x')=0$ para los $x$ anteriores a $S_P$ y los $x'$ posteriores a $S_F$. Se usar\'an los m\'etodos que se usaron para obtener (\ref{deltaI2TS}) sobre las l\'aminas que se extienden solamente sobre las partes de las l\'aminas anteriores a $S_P$, y entonces se tendr\'an extra-t\'erminos, en forma de integrales de l\'{\i}nea a lo largo de las l\'{\i}neas donde se encuentran las l\'aminas, originados debido a la aplicaci\'on del teorema de Stokes. Organizando la parametrizaci\'on de las l\'aminas de forma tal que la l\'{\i}nea donde la l\'amina encuentre a $S_P$ es dada para $\tau_0=$ constante y la l\'{\i}nea donde la l\'amina variada halle la $S_P$ variada es dada para $\tau_0=$ la misma constante, \'estas integrales de l\'{\i}nea toman la forma
\begin{equation}
\sum_g g \int_0^{\infty}F^{\dagger *}_{\mu\nu}\delta y^{\mu}\left(\frac{dy^{\nu}}{d\tau_1}\right)d\tau_1 \label{TermGFmunuReparam}.
\end{equation}
Las l\'{\i}neas de integraci\'on son las posiciones de las cuerdas cuando dejan de existir. Para la definici\'on de $\delta I_4$, ya no es posible usar (\ref{deltaI4}). En su lugar se usa
\begin{equation}
\delta I_4=\frac{1}{8\pi}\int_{-\infty}^{S_F}\left(\frac{\partial A_{\nu}^*}{\partial x^{\mu}}\frac{\partial\delta A^{\mu}}{\partial x_{\nu}}+\frac{\partial A_{\nu}}{\partial x^{\mu}}\frac{\partial\delta A^{\mu *}}{\partial x_{\nu}}\right)d^4x \label{DeltaI4},
\end{equation}
que es parecido al primer t\'ermino de
\begin{multline}
\delta I'_2=-\frac{1}{8\pi}\int_{-\infty}^{S_F}\left(F_{\mu\nu}^*\frac{\partial\delta A^{\mu}}{\partial x_{\nu}}+F_{\mu\nu}\frac{\partial\delta A^{\mu *}}{\partial x_{\nu}}\right)d^4x \\
+\frac{1}{4}\sum_g\int\left(F_{\mu\nu}^*\delta G^{\dagger\mu\nu}+F_{\mu\nu}\delta G^{\dagger\mu\nu*}\right)d^4x; \label{deltaI2prima}
\end{multline}
sum\'andolos da
\begin{equation}
\frac{1}{8\pi}\int_{-\infty}^{S_F}\left[\left(\frac{\partial A_{\nu}^*}{\partial x^{\mu}}-F_{\mu\nu}^*\right)\frac{\partial\delta A^{\mu}}{\partial x_{\nu}}+\left(\frac{\partial A_{\nu}}{\partial x^{\mu}}-F_{\mu\nu}\right)\frac{\partial\delta A^{\mu *}}{\partial x_{\nu}}\right]d^4x+\cdots
\end{equation}
Usando $F_{\mu\nu}=\frac{\partial A_{\nu}}{\partial x^{\mu}}-\frac{\partial A_{\mu}}{\partial x^{\nu}}+4\pi\sum_g G_{\mu\nu}^{\dagger}$, y su conjugada, se reescriben las cantidades anteriores como
\begin{equation}
\frac{\partial A_{\nu}^*}{\partial x^{\mu}}-F_{\mu\nu}^*\frac{\partial\delta A^{\mu}}{\partial x_{\nu}}=\frac{\partial A_{\nu}^*}{\partial x^{\mu}}-\frac{\partial A^*_{\nu}}{\partial x^{\mu}}+\frac{\partial A^*_{\mu}}{\partial x^{\nu}}-4\pi\sum_g G_{\mu\nu}^{\dagger *}=\frac{\partial A^*_{\mu}}{\partial x^{\nu}}-4\pi\sum_g G_{\mu\nu}^{\dagger *}
\end{equation}
y tambi\'en
\begin{equation}
\frac{\partial A_{\nu}}{\partial x^{\mu}}-F_{\mu\nu}\frac{\partial\delta A^{\mu}}{\partial x_{\nu}}=\frac{\partial A_{\nu}}{\partial x^{\mu}}-\frac{\partial A_{\nu}}{\partial x^{\mu}}+\frac{\partial A_{\mu}}{\partial x^{\nu}}-4\pi\sum_g G_{\mu\nu}^{\dagger}=\frac{\partial A_{\mu}}{\partial x^{\nu}}-4\pi\sum_g G_{\mu\nu}^{\dagger}.
\end{equation}
Por tanto la ecuaci\'on queda
\begin{align}
&\frac{1}{8\pi}\int_{-\infty}^{S_F}\left[\left(\frac{\partial A^*_{\mu}}{\partial x^{\nu}}-4\pi\sum_g G_{\mu\nu}^{\dagger *}\right)\frac{\partial\delta A^{\mu}}{\partial x_{\nu}}+\left(\frac{\partial A_{\mu}}{\partial x^{\nu}}-4\pi\sum_g G_{\mu\nu}^{\dagger}\right)\frac{\partial\delta A^{\mu *}}{\partial x_{\nu}}\right]d^4x+\cdots \notag \\
&=\frac{1}{8\pi}\int_{-\infty}^{S_F}\left[\left(\frac{\partial A^*_{\mu}}{\partial x^{\nu}}\frac{\partial\delta A^{\mu}}{\partial x_{\nu}}-4\pi\sum_g G_{\mu\nu}^{\dagger *}\frac{\partial\delta A^{\mu}}{\partial x_{\nu}}\right)\right. \notag \\
&\hspace{5.5cm}\left.+\left(\frac{\partial A_{\mu}}{\partial x^{\nu}}\frac{\partial\delta A^{\mu *}}{\partial x_{\nu}}-4\pi\sum_g G_{\mu\nu}^{\dagger}\frac{\partial\delta A^{\mu *}}{\partial x_{\nu}}\right)\right]d^4x+\cdots
\end{align}
teniendo en cuenta solamente los potenciales, entonces
\begin{equation}
\frac{1}{8\pi}\left[\int_{-\infty}^{S_F}\left(\frac{\partial A^*_{\mu}}{\partial x^{\nu}}\frac{\partial\delta A^{\mu}}{\partial x_{\nu}}\right)d^4x+\int_{-\infty}^{S_F}\left(\frac{\partial A_{\mu}}{\partial x^{\nu}}\frac{\partial\delta A^{\mu *}}{\partial x_{\nu}}\right)d^4x\right]+\cdots
\end{equation}
integrando por partes el primer t\'ermino usando $U=\frac{\partial A_{\mu}^*}{\partial x^{\nu}}$, $dV=\frac{\partial \delta A^{\mu}}{\partial x_{\nu}}d^4x$, entonces $dU=\frac{\partial^2 A_{\mu}^*}{\partial x^{\alpha}\partial x^{\nu}}d^4x$, $V=\delta A^{\mu}$ y por tanto,
\begin{align}
\int_{-\infty}^{S_F}\left(\frac{\partial A^*_{\mu}}{\partial x^{\nu}}\frac{\partial\delta A^{\mu}}{\partial x_{\nu}}\right)d^4x &=\left[\frac{\partial A_{\mu}^*}{\partial x^{\nu}}\delta A^{\mu}\right]_{-\infty}^{S_F}-\int_{-\infty}^{S_F}\delta A^{\mu}\frac{\partial^2 A_{\mu}^*}{\partial x^{\alpha}\partial x^{\nu}}d^4x \notag \\
&=\left[\frac{\partial A_{\mu}^*}{\partial x^{\nu}}\delta A^{\mu}\right]_{-\infty}^{S_F}-\int_{-\infty}^{S_F}\frac{\partial^2 A_{\mu}^*}{\partial x^{\alpha}\partial x^{\nu}}\delta A^{\mu}dS^{\nu}dx^{\alpha} \notag \\
&=-\int\frac{\partial A_{\mu}^*}{\partial x^{\nu}}\delta A^{\mu}dS^{\nu}
\end{align}
donde se ha integrado sobre la superficie tridimensional $S_F$ usando $d^4x=dS^{\nu}dx^{\alpha}$ y se ha usado el hecho que los potenciales se anulan en el infinito y en $S_F$ (condici\'on de integral extremal). El resultado es
\begin{align}
&\frac{1}{8\pi}\int\left(\frac{\partial A_{\mu}^*}{\partial x^{\nu}}\delta A^{\mu}+\frac{\partial A_{\mu}}{\partial x^{\nu}}\delta A^{\mu *}\right)dS^{\nu} \notag \\
&\hspace{4cm}-\frac{1}{8\pi}\int4\pi\sum_g \left(G_{\mu\nu}^{\dagger *}\frac{\partial\delta A^{\mu}}{\partial x_{\nu}}+ G_{\mu\nu}^{\dagger}\frac{\partial\delta A^{\mu *}}{\partial x_{\nu}}\right)d^4x+\cdots \notag \\
&=\frac{1}{8\pi}\int\left(\frac{\partial A_{\mu}^*}{\partial x^{\nu}}\delta A^{\mu}+\frac{\partial A_{\mu}}{\partial x^{\nu}}\delta A^{\mu *}\right)dS^{\nu} \notag \\
&\hspace{3.6cm}-\frac{1}{2}\int\sum_g \left(G_{\mu\nu}^{\dagger *}\frac{\partial\delta A^{\mu}}{\partial x_{\nu}}+ G_{\mu\nu}^{\dagger}\frac{\partial\delta A^{\mu *}}{\partial x_{\nu}}\right)d^4x+\cdots \label{VariacionA}
\end{align}
Los restantes t\'erminos en $\delta I$, se cancelan debido cuando se usa la ecuaci\'on de movimiento, sin que $S_F$  est\'e muy cerca a $S_P$. Por tanto se llega a,
\begin{multline}
\delta I=\sum_{e+g}m\frac{dz_{\mu}}{ds}\Delta z^{\mu}+\sum_e eA_{\mu}(z)\Delta z^{\mu}+\sum_g g\int F^{\dagger *}_{\mu\nu}\delta y^{\mu}\frac{dy^{\nu}}{d\tau_1}d\tau_1 \\
+\frac{1}{8\pi}\int\left(\frac{\partial A_{\mu}^*}{\partial x^{\nu}}\delta A^{\mu}+\frac{\partial A_{\mu}}{\partial x^{\nu}}\delta A^{\mu *}\right)dS^{\nu} \label{deltaI}.
\end{multline}
Pero con la expresi\'on anterior no es posible encontrar los momentos con la transformaci\'on original $\delta I=\sum p_r\Delta q_r-W\delta t$, pues las variaciones de $A^{\mu}$ y $A^{\mu *}$ dadas en (\ref{VariacionA}) no son independientes entre si y debido a que la superficie de los campos $S_F$ no ha variado.
\par
Para hallar los momentos se pasa a los componentes de Fourier de los potenciales para los que se pudiera utilizar las transformaciones de Fourier dadas por las ecuaciones (\ref{TransfFourierA1}) y (\ref{TransfFourierA2}), debido a que existe inter\'es en la expresi\'on (\ref{VariacionA}) con los potenciales sobre la superficie $S_F$. Se considera un movimiento tal que satisface las ecuaciones de movimiento de forma tal que las transformaciones de Fourier (\ref{TransfFourierA1}) y (\ref{TransfFourierA2}) son v\'alidas sobre $S_P$ tambi\'en para dicho movimiento variado. As\'{\i} la expresi\'on (\ref{VariacionA}) se convierte, hallando las variaciones de los potenciales,
\begin{equation}
\frac{\partial A^*_{\mu}}{\partial x^{\nu}}=\sum_{k_0}\int A^*_{k_{\mu}}(ik_{\nu})e^{ikx}\frac{d^3k}{k_0}
\end{equation}
pero debido a que $\delta A^{\mu}=\sum_{k'_0}\int \delta A_{k'_{\mu}}e^{ik'x}\frac{d^3k'}{k'_0}$ entonces,
\begin{align}
\frac{\partial A^*_{\mu}}{\partial x^{\nu}}\delta A^{\mu} &=i\sum_{k_0,k'_0}\iint k_{\nu}A^*_{k_{\mu}}\delta A^{k_{'\mu}}e^{i(k+k')x}\frac{d^3k}{k_0}\frac{d^3k'}{k'_0} \notag \\
\frac{\partial A_{\mu}}{\partial x^{\nu}}\delta A^{\mu *} &=i\sum_{k_0,k'_0}\iint k^*_{\nu}A_{k_{\mu}}\delta A^{*k_{'\mu}}e^{i(k+k')x}\frac{d^3k}{k_0}\frac{d^3k'}{k'_0}
\end{align}
usando $A^*_{k_{\mu}}=\gamma_k A_{k_{\mu}}$, entonces
\begin{align}
\frac{\partial A^*_{\mu}}{\partial x^{\nu}}\delta A^{\mu} &=i\sum_{k_0,k'_0}\iint k_{\nu}\gamma_k A_{k_{\mu}}\delta A^{k_{'\mu}}e^{i(k+k')x}\frac{d^3k}{k_0}\frac{d^3k'}{k'_0} \notag \\
\frac{\partial A_{\mu}}{\partial x^{\nu}}\delta A^{\mu *} &=i\sum_{k_0,k'_0}\iint k_{\nu}A_{k_{\mu}}\gamma_{k'} \delta A^{k_{'\mu}}e^{i(k+k')x}\frac{d^3k}{k_0}\frac{d^3k'}{k'_0}
\end{align}
y su suma es
\begin{equation}
\frac{\partial A^*_{\mu}}{\partial x^{\nu}}\delta A^{\mu}+\frac{\partial A_{\mu}}{\partial x^{\nu}}\delta A^{\mu *}=i\sum_{k_0,k'_0}\iint k_{\nu}(\gamma_k+\gamma_{k'})A_{k_{\mu}}\delta A^{k_{'\mu}}e^{i(k+k')x}\frac{d^3k}{k_0}\frac{d^3k'}{k'_0}.
\end{equation}
Al llevar la integral a la superficie tridimensional para $S_F$ tal que $x_0=$ constante, se tiene que el \'ultimo t\'ermino de la ecuaci\'on (\ref{deltaI}) se modifica a
\begin{equation}
\frac{i}{8\pi}\sum_{k_0,k'_0}\iiint k_{\nu}(\gamma_k+\gamma_{k'})A_{k_{\mu}}\delta A^{k_{'\mu}}e^{i(k+k')x}\frac{d^3k}{k_0}\frac{d^3k'}{k'_0}dS^{\nu}
\end{equation}
y por tanto solucionado parcialmente \'esta ecuaci\'on se llega a
\begin{equation}
i\pi^2\sum_{k_0,k'_0}\iint (\gamma_k+\gamma_{k'})A_{k_{\mu}}\delta A^{k_{'\mu}}e^{i(k_0+k'_0)x_0}\delta^3(k+k')d^3k\frac{d^3k'}{k'_0}.
\end{equation}
\'Este resultado se logra con $x_0=$ constante y $k_{\nu}\frac{dS^{\nu}}{k_0}=k_ik_0\frac{dS^{\nu}}{k_0}=k_idS^{\nu}dx_i$, para $x_1$ se integra resultando,
\begin{equation}
\int e^{i(k_{x_1}+k'_{x_1})x_1}k_{\nu}dx_1=2\pi\delta(k_{x_1}+k'_{x_1})e^{(k_0+k'_0)x_0}.
\end{equation}
El resto de las integrales para $x_2$ y $x_3$ tienen resultados id\'enticos matem\'aticamente, resultando finalmente
\begin{equation}
8\pi^3\delta(k+k')e^{(k_0+k'_0)x_0}
\end{equation}
pues $\delta^3(k)=\delta(k_1)\delta(k_2)\delta(k_3)$.
\par
Se observa claramente que el delta de Dirac hace que el integrando se anule para todos los puntos excepto para $k'_i=-k_i$, lo cual implica que $k'_0=\pm k_0$. Entonces usando $\gamma_{-l}=\gamma_l$ la integral se reduce a
\begin{equation}
i\pi^2\sum_{k_0,k'_0}\iint \left(\frac{\gamma_k+\gamma_{k'}}{k'_0}\right)A_{k_{\mu}}\delta A^{k_{'\mu}}e^{i(k_0+k'_0)x_0}\delta^3(k+k')d^3kd^3k'
\end{equation}
con $\gamma_{-l}=\gamma_l$, si $k'_i=-k_i$ y $k'_0=\pm k_0$ entonces $\gamma_{k'}=\gamma_{-k}=\gamma_k$ y por tanto $\delta^3(k+k')=1$. As\'{\i} se integra dando
\begin{multline}
i\pi^2\sum_{k_0,k'_0}\iint \left(\frac{\gamma_k+\gamma_{k'}}{k'_0}\right)A_{k_{\mu}}\delta A^{k_{'\mu}}e^{i(k_0+k'_0)x_0}\delta^3(k+k')d^3kd^3k' \\
=i\pi^2\sum_{k_0}\int \left(\frac{2\gamma_k}{-k'_0}\right)A_{k_{\mu}}\delta A^{k_{'\mu}}d^3k
\end{multline}
y si si $k'_0=+k_i$ y $k'_i=- k_i$ entonces se integra dando
\begin{multline}
i\pi^2\sum_{k_0,k'_0}\iint \left(\frac{\gamma_k+\gamma_{k'}}{k'_0}\right)A_{k_{\mu}}\delta A^{k_{'\mu}}e^{i(k_0+k'_0)x_0}\delta^3(k+k')d^3kd^3k' \\
=i\pi^2\sum_{k_0}\int \left(\frac{\gamma_k+\gamma_{k_0,k_i}}{k_0}\right)A_{k_{\mu}}\delta A_{k_0,-k^{\mu}_i}e^{2ik_0x_0}d^3k,
\end{multline}
finalmente la soluci\'on es
\begin{multline}
i\pi^2\sum_{k_0,k'_0}\iint \left(\frac{\gamma_k+\gamma_{k'}}{k'_0}\right)A_{k_{\mu}}\delta A_{k^{'\mu}}e^{i(k_0+k'_0)x_0}\delta^3(k+k')d^3kd^3k' \\
=-2i\pi^2\sum_{k_0}\int \left(\frac{2\gamma_k}{-k_0}\right)A_{k_{\mu}}\delta A_{k^{'\mu}}d^3k \\
+i\pi^2\sum_{k_0}\int \left(\frac{\gamma_k+\gamma_{k_0,k_i}}{k_0}\right)A_{k_{\mu}}\delta A_{k_0,-k^{\mu}_i}e^{2ik_0x_0}d^3k.
\end{multline}
Pero $\left(\gamma_k+\gamma_{k_0,k_i}\right)A_{k_{\mu}}\delta A_{k_0,-k^{\mu}_i}=\delta\left(\gamma_kA_{k_{\mu}}A_{k_0,-k^{\mu}_i}\right)$, pues $\gamma_k+\gamma_{k_0,k_i}=\delta\gamma_k$ y $ A_{k_{\mu}}\delta A_{k_0,-k^{\mu}_i}=\delta A_{k_{\mu}}$, lo que forma un diferencial exacto quedando
\begin{multline}
i\pi^2\sum_{k_0,k'_0}\iint \left(\frac{\gamma_k+\gamma_{k'}}{k'_0}\right)A_{k_{\mu}}\delta A^{k_{'\mu}}e^{i(k_0+k'_0)x_0}\delta^3(k+k')d^3kd^3k' \\
=-2i\pi^2 \sum_{k_0}\int \left(\frac{2\gamma_k}{-k_0}\right)A_{k_{\mu}}\delta A^{k_{'\mu}}d^3k+i\pi^2 \delta\sum_{k_0}\int \gamma_kA_{k_{\mu}}A_{k_0}e^{2ik_0x_0}\frac{d^3k}{k_0}.
\end{multline}
El segundo t\'ermino se descarta porque es un diferencial exacto y no afecta las ecuaciones de movimiento. Finalmente,
\begin{multline}
i\pi^2\sum_{k_0,k'_0}\iint \left(\frac{\gamma_k+\gamma_{k'}}{k'_0}\right)A_{k_{\mu}}\delta A^{k_{'\mu}}e^{i(k_0+k'_0)x_0}\delta^3(k+k')d^3kd^3k' \\
=-2i\pi^2 \sum_{k_0}\int \left(\frac{2\gamma_k}{-k_0}\right)A_{k_{\mu}}\delta A^{k_{\mu}}d^3k.
\end{multline}
Si $k_0 > 0$, usando 
$
\begin{cases}
A_{-k_{\mu}}=-\overline{A}_{k_{\mu}} \\
A^*_{-k_{\mu}}=-\overline{A}^*_{k_{\mu}} \\
\gamma_{-l}=\gamma_l
\end{cases}
$
entonces la ecuaci\'on se modifica a,
\begin{multline}
i\pi^2\sum_{k_0,k'_0}\iint \left(\frac{\gamma_k+\gamma_{k'}}{k'_0}\right)A_{k_{\mu}}\delta A^{k_{'\mu}}e^{i(k_0+k'_0)x_0}\delta^3(k+k')d^3kd^3k' \\
=-2i\pi^2 \sum_{k_0}\int \gamma_k\left(\overline{A}_{k_{\mu}}\delta A^{k_{\mu}}-A_{k_{\mu}}\delta \overline{A}^{k_{\mu}}\right)\frac{d^3k}{k_0}.
\end{multline}
Debido a que $A_{-k_{\mu}}\left(\gamma_{-k}\delta A^{k_{\mu}}\right)=-A_{+k_{\mu}}\gamma _{+k}\delta\overline{A}^{k_{\mu}}$, $\gamma_{-k}A_{-k_{\mu}}=\gamma_k\overline{A}_{k_{\mu}}\delta A^{k_{\mu}}$ y usando $\delta\left(\overline{A}_{k_{\mu}}A^{k_{\mu}}\right)=\delta\overline{A}_{k_{\mu}}A^{k_{\mu}}+\overline{A}_{k_{\mu}}\delta A^{k_{\mu}}$ se llega a $\delta\left(\overline{A}_{k_{\mu}}A^{k_{\mu}}\right)-2\overline{A}_{k_{\mu}}\delta A^{k_{\mu}}=\delta\overline{A}_{k_{\mu}}A^{k_{\mu}}-\overline{A}_{k_{\mu}}\delta A^{k_{\mu}}$ y as\'{\i},
\begin{multline}
-2i\pi^2\int \gamma_k\left(\overline{A}_{k_{\mu}}\delta A^{k_{\mu}}-A_{k_{\mu}}\delta \overline{A}^{k_{\mu}}\right)\frac{d^3k}{k_0} \\
=-2i\pi^2\delta\int \gamma_k\overline{A}_{k_{\mu}}A^{k_{\mu}}\frac{d^3k}{k_0} -4i\pi^2\int \gamma_k\overline{A}_{k_{\mu}}\delta A^{k_{\mu}}\frac{d^3k}{k_0}
\end{multline}
donde el primer t\'ermino de la integral no se tiene en cuenta debido a las razones anteriormente expuestas. Finalmente la variaci\'on de la acci\'on queda,
\begin{multline}
\delta I=\sum_{e+g}m\frac{dz_{\mu}}{ds}\Delta z^{\mu}+\sum_e eA_{\mu}(z)\Delta z^{\mu}+\sum_g g\int_0^{\infty}F^{\dagger *}_{\mu\nu}\delta y^{\mu}\left(\frac{dy^{\mu}}{d\tau_1}\right)d\tau_1 \\
-4i\pi^2\int\gamma_k\overline{A}_{k_{\mu}}\delta^{k_{\mu}}\frac{d^3k}{k_0} \label{deltaIb},
\end{multline}
donde $z_{\mu}$ son las coordenadas din\'amicas de las part\'{\i}culas cuando dejan de existir; $y_{\mu}(\tau_1)$ son las coordenadas de los puntos sobre las cuerdas cuando \'estas dejan de existir y que construyen un continuo unidimensional de coordenadas para cada polo; para cada valor de $\mu$ y $A_{k_{\mu}}$ son las componentes de Fourier (con $k_0>0$) de los potenciales despu\'es que las part\'{\i}culas y las cuerdas han dejado de existir. Los coeficientes en la ecuaci\'on (\ref{deltaIb}) ser\'an los momentos conjugados. As\'{\i} los momentos de una part\'{\i}cula cargada son $p_{\mu}=m\frac{dz_{\mu}}{ds}+eA_{\mu}(z)$ y los de una part\'{\i}cula con polo ser\'an $p_{\mu}=m\frac{dz_{\mu}}{ds}$. El momento conjugado a las variables de la cuerda $\beta^{\mu}(\tau_1)\equiv y^{\mu}(\tau_1)$ son $\beta^{\mu}(\tau_1)=gF^{\dagger *}_{\mu\nu}\frac{dy^{\mu}}{d\tau_1}$ y el momento conjugado a $A_{k^{\mu}}$ est\'a dado por $-\frac{4i\pi^2}{k_0}\gamma_k\overline{A}_{k_{\mu}}$. El momento $\beta^{\mu}(\tau_1)$ forma un cont\'{\i}nuo unidimensional de variables que corresponde al continuo unidimensional de las coordenadas $y^{\mu}(\tau_1)$ y el momento de campo conjugado a $A_{k^{\mu}}$ forma un continuo tridimensional que corresponde al continuo tridimensional de las coordenadas de campo.
\par
Para las coordenadas y los momentos de cada part\'{\i}cula se definen los brackets de Poisson de la forma
\begin{align}
\left\{p_{\mu},z_{\nu}\right\}=\left\{m\frac{dz_{\mu}}{ds}+eA_{\mu}(z),z_{\nu}\right\} &=\left\{m\frac{dz_{\mu}}{ds},z_{\nu}\right\}+\left\{eA_{\mu}(z),z_{\nu}\right\} \notag \\
&=\left\{m\frac{dz_{\mu}}{ds},z_{\nu}\right\}=g_{\mu\nu}, \label{PBPart}
\end{align}
para las coordenadas y los momentos de la cuerda
\begin{equation}
\left\{\beta^{\mu}(\tau_1),y_{\nu}(\tau'_1)\right\}=g_{\mu\nu}\delta(\tau_1-\tau'_1), \label{BPCuerda}
\end{equation}
para las variables de campo
\begin{align}
\left\{\overline{A}_{k_{\mu}},A_{k'_{\nu}}\right\} &=\left\{-\frac{4i\pi^2}{k_0}\gamma_k\overline{A}_{k_{mu}},A_{k'_{\nu}}\right\}=\frac{ik_0}{4\pi^2\gamma_k}\left\{A^*_{k_{mu}},A_{k'_{\nu}}\right\} \notag \\
&=\frac{ik_0}{4\pi^2\gamma_k}g_{\mu\nu}\delta^3(k-k') \label{PBVarCampo}
\end{align}
y todos los otros brackets de Poisson son cero.
\par
Usando $p_{\mu}-eA_{\mu}(z)=m\frac{dz_{\mu}}{ds}$ y $p_{\mu}=m\frac{dz_{\mu}}{ds}$, con el fin de eliminar la velocidad para cada part\'{\i}cula cargada, se obtiene
\begin{multline}
\left[p_{\mu}-eA_{\mu}(z)\right]^2=\left[p_{\mu}-eA_{\mu}(z)\right]\left[p^{\mu}-eA^{\mu}(z)\right]=\left(m\frac{dz_{\mu}}{ds}\right)^2=m^2 \\
\;\Rightarrow\; \left[p_{\mu}-eA_{\mu}(z)\right]\left[p^{\mu}-eA^{\mu}(z)\right]-m^2=0 \label{CoordDin1}
\end{multline}
y para cada polo la ecuaci\'on es
\begin{equation}
p_{\mu}p^{\mu}=m^2\;\Rightarrow\; p_{\mu}p^{\mu}-m^2=0 \label{CoordDin2},
\end{equation}
que deben estar unidas con
\begin{equation}
\beta^{\mu}(\tau_1)-gF^{\dagger *}_{\mu\nu}(y)\frac{dy^{\mu}}{d\tau_1}=0 \label{CoordDin3},
\end{equation}
donde los potenciales $A_{\mu}(z)$ en las anteriores ecuaciones est\'an dados por las componentes de Fourier, $A_{k_{\mu}}(z)$ y $\overline{A}_{k_{\mu}}(z)$. Las ecuaciones (\ref{CoordDin1}), (\ref{CoordDin2}) y (\ref{CoordDin3}) solo relacionan coordenadas y momentos din\'amicos. Son ecuaciones diferenciales satisfechas por la integral de acci\'on $I$ cuando los momentos se ven como derivadas de $I$ y son ecuaciones de Hamilton-Jacobi de la teor\'{\i}a de los polos magn\'eticos. Las condiciones suplementarias (\ref{CondicSuplem}) deben ser tratadas como ecuaciones de Hamilton-Jacobi y \'estas ecuaciones obtenidas tomando diferentes puntos del campo $x$ no son independientes de las ecuaciones de movimiento o entre si y de esta manera se obtiene una serie completa e independiente de ellas haciendo una transformaci\'on de Fourier en la regi\'on entre $S_P$ y $S_F$. En \'esta regi\'on, usando la funci\'on de Pauli-Jordan (\ref{FunctPauli-Jordan}), se pede reemplazar $J(x-z)$ por $\Delta(x-z)$ cuya trasformada de Fourier est\'a dada por
\begin{equation}
\Delta(x-z)=-\frac{i}{4\pi^2}\sum_{k_0}\int e^{ik(x-z)}\frac{d^3k}{k_0}
\end{equation}
entonces se obtienen,
\begin{equation}
k^{\nu}\gamma_k A_{k_{\nu}}-\frac{1}{4\pi^2}\sum_e ee^{-ikz}=0, \qquad k^{\nu}\gamma_k \overline{A}_{k_{\nu}}-\frac{1}{4\pi^2}\sum_e ee^{+ikz}=0
\end{equation}
que son las ecuaciones que envuelven solamente variables din\'amicas de coordenadas y momentos y que son conformadas de la manera correcta para construir las ecuaciones de Hamilton-Jacobi.
\newline
\newline
\textbf{VII. Cuantizaci\'on}
\newline
\newline
Para cuantizar se reemplazan las coordenadas din\'amicas y el momento de la teor\'{\i}a cl\'asica por operadores que satisfagan las reglas de conmutaci\'on
\begin{align}
& \left\{p_{\mu},z_{\nu}\right\}=g_{\mu\nu} \notag \\
& \left\{\beta^{\mu}(\tau_1),y_{\nu}(\tau'_1)\right\}=g_{\mu\nu}\delta(\tau_1-\tau'_1) \notag \\
& \left\{\overline{A}_{k_{\mu}},A_{k'_{\nu}}\right\}=\frac{ik_0}{4\pi^2\gamma_k}g_{\mu\nu}\delta^3(k-k')
\end{align}
y entonces se reemplaza las ecuaciones de Hamilton-Jacobi por ecuaciones de onda obtenidas igualando a cero las miembros izquierdos de las ecuaciones de Hamilton-Jacobi, que est\'an escritas en funci\'on de operadores , y que que est\'an aplicadas a la funci\'on de onda $\psi$. Las ecuaciones de onda obtenidas de \'esta manera son consistentes entre si debido a que los operadores sobre $\psi$ en sus miembros izquierdos conmutan y esto es debido a la anulaci\'on de los miembros izquierdos de los brackets de Poisson de las ecuaciones de Hamilton-Jacobi. La soluci\'on da como resultado ecuaciones de onda para part\'{\i}culas sin esp\'{\i}n; para los electrones se reemplazan por las ecuaciones de onda de esp\'{\i}n $\frac{1}{2}h$. Debido a la falta de informaci\'on sobre los polos, se supone con esp\'{\i}n $\frac{1}{2}h$.
\newline
Con las matrices de esp\'{\i}n usuales $\alpha_1,\;\alpha_2,\;\alpha_3,\;\alpha_m$, para cada part\'{\i}cula cargada existe una ecuaci\'on de la forma,
\begin{equation}
\left\{p_0-eA_0(z)-\alpha_r\left[p_r-eA_r(z)\right]-\alpha_m m\right\}\psi=0 \label{PartCarg},
\end{equation}
para cada part\'{\i}cula con un polo,
\begin{equation}
\left[p_0-\alpha_rp_r-\alpha_m m\right]\psi=0 \label{PartPolo},
\end{equation}
para cada cuerda,
\begin{equation}
\left[\beta_{\mu}(\tau_1)-gF^{\dagger *}_{\mu\nu}(y)\frac{y^{\mu}}{d\tau_1}\right]\psi=0 \label{Cuerda},
\end{equation}
y para las variables de campo,
\begin{equation}
\left(k^{\nu}\gamma_k A_{k_{\nu}}-\frac{1}{4\pi^2}\sum_e ee^{-ikz}\right)\psi=0, \qquad \left(k^{\nu}\gamma_k \overline{A}_{k_{\nu}}-\frac{1}{4\pi^2}\sum_e ee^{+ikz}\right)\psi=0. \label{VarCampo}
\end{equation}
La funci\'on de onda $\psi$ se podr\'{\i}a tomar como como una funci\'on de las variables de la part\'{\i}cula $z_{\mu}$, de las variables de esp\'{\i}n disponibles para cada part\'{\i}cula, de las variables de la cuerda $y_{\mu}(\tau_1)$ con rango $0<\tau_1<\infty$ y de las variables de campo $A_{k_{\mu}}$. Dicha funci\'on estar\'a definida solamente cuando los puntos $z_{\mu}$, $y_{\mu}(\tau_1)$ est\'en fuera de los conos de luz de las otras part\'{\i}culas.
\newline
La ecuaci\'on (\ref{PartPolo}) sugiere que el campo electromagn\'etico no act\'ua sobre los polos pero si sobre las cuerdas (ec. (\ref{Cuerda})) y debido a que los polos est\'an restringidos a los extremos de las cuerdas entonces el campo no afecta el movimiento de los polos.
\newline
\newline
\textbf{VIII. La carga y el polo unitarios}
\newline
\newline
La integral de acci\'on de la teor\'{\i}a cl\'asica podr\'{\i}a ser considerada como una funci\'on de puntos espacio-temporales $z_{\mu}$ donde las part\'{\i}culas dejan de existir, de las l\'{\i}neas $y_{\mu}(\tau_1)$ en el espacio-tiempo donde las cuerdas dejan de existir y de las variables de campo apropiadas. \'Esta acci\'on est\'a definida siempre que las cuerdas no pasen a trav\'es de cualquiera de los puntos $z_{\mu}$ donde las part\'{\i}culas cargadas dejan de existir.
\newline
Se mantendr\'an todos los puntos de las part\'{\i}culas ($z_{\mu}$) fijos y tambi\'en fijas todas las cuerdas excepto una que variar\'a continuamente manteni\'endola siempre  en la superficie tridimensional $S_P$ situada alrededor de uno de los puntos $z_{\mu}$ donde una part\'{\i}cula cargada est\'a situada justamente antes que deje de existir y luego regrese a su posici\'on original. Al mismo tiempo los potenciales $A_{\mu}(x)$ son variados continuamente con el objetivo de preservar las ecuaciones (\ref{DivPolCorr}) y (\ref{Gmunu}) con valores fijos para el campo $F_{\mu\nu}(x)$ y que retoman los valores originales junto con la cuerda. Debido a que la deformaci\'on no puede ser restituida continuamente en la acci\'on entonces una cuerda no puede pasar a trav\'es de una part\'{\i}cula cargada. La cuerda se extiende sobre una superficie cerrada bidimensional $\sigma$, permaneciendo en $S_P$ y encerrando el punto $z_{\mu}$ donde se sit\'ua la carga y dicha superficie no puede ser encogida continuamente a cero debido a que no debe pasar a trav\'es de una carga. Sinembargo se espera variar la acci\'on $I$ bajo \'este proceso de deformaci\'on.
\newline
Una peque\~na variaci\'on de una cuerda y de los potenciales con los $z_{\mu}$ fijos lleva a una variaci\'on de $I$ ($DI$), dada por la suma de los miembros derechos de las ecuaciones (\ref{DeltaI4}) y (\ref{deltaI2prima}) dada por
\begin{align}
DI &=\frac{1}{8\pi}\int_{-\infty}^{S_F}\left[\left(\frac{\partial A_{\nu}^*}{\partial x^{\mu}}-F_{\mu\nu}^*\right)\frac{\partial\delta A^{\mu}}{\partial x_{\nu}}+\left(\frac{\partial A_{\nu}}{\partial x^{\mu}}-F_{\mu\nu}\right)\frac{\partial\delta A^{\mu *}}{\partial x_{\nu}}\right]d^4x \notag \\
& +\frac{1}{4}\sum_g\int_{-\infty}^{S_F}\left(F_{\mu\nu}^*\delta G^{\dagger\mu\nu}+F_{\mu\nu}\delta G^{\dagger\mu\nu*}\right)d^4x;
\end{align}
debido a que los tensores electromagn\'eticos normal y dual se mantienen fijos y que los cuadri-potenciales normal y dual son regresados a sus valores originales, entonces la variaci\'on de la acci\'on se modifica a
\begin{equation}
DI=\frac{1}{4}\sum_g\int_{-\infty}^{S_F}\left(F_{\mu\nu}^*\delta G^{\dagger\mu\nu}+F_{\mu\nu}\delta G^{\dagger\mu\nu*}\right)d^4x,
\end{equation}
que como ya anteriormente se explic\'o se puede reparametrizar como (ver la ec. (\ref{TermGFmunuReparam}))
\begin{equation}
DI=\sum_g g \int_0^{\infty}F^{\dagger *}_{\mu\nu}\delta y^{\mu}\left(\frac{dy^{\nu}}{d\tau_1}\right)d\tau_1.
\end{equation}
Se observa que para un proceso de deformaci\'on cerrado los par\'ametros de los que depende el punto sobre la l\'amina $y^{\mu}$ ya no son importantes debido a que no se desea la descripci\'on de la l\'{\i}nea de mundo del polo y por tanto
\begin{equation}
DI=g \int F^{\dagger *}_{\mu\nu}\delta y^{\nu}dy^{\mu},
\end{equation}
donde $\delta y^{\nu}dy^{\mu}$ es un elemento diferencial de la superficie bidimensional extendida por la cuerda y estar\'a definido por $d\sigma^{\mu\nu}$. De esta manera la variaci\'on de la acci\'on ser\'a,
\begin{equation}
DI=g \int F^{\dagger *}_{\mu\nu}\sigma^{\nu\mu}.
\end{equation}
La integral resultante tiene la forma matem\'atica de una divergencia y por tanto definir\'a el flujo el\'ectrico total que pasa a trav\'es de la superficie $\sigma$ dando como resultado: $DI=4\pi ge$.
\newline
Se podr\'{\i}a rodear cualquier carga con una cuerda por un n\'umero indefinido de ciclos y por tanto la incertidumbre total en $I$ est\'a dada por la suma (suma sobre todas las cargas $e$ y sobre todos los polos $g$) $4\pi\sum_{ge}m_{ge}ge$, donde $m_{ge}$ es un coeficiente arbitrario que define un an\'alogo al peso estad\'{\i}stico de cada uno de los t\'erminos de la suma.
\newline
Asociando el comportamiento no univaluado de la integral de acci\'on $I$ de la teor\'{\i}a de los polos con el de la acci\'on de algunas de la cantidades f\'{\i}sicas, se puede observar que es posible usar la regla de quantizaci\'on de Bohr para cada carga y polo, resultando (para $c=1$)
\begin{equation}
4\pi ge=nh, \label{Cuantizge}
\end{equation}
donde $n$ es un entero. Usando el momento angular de Planck $\hbar=\frac{h}{2\pi}$ la ecuaci\'on queda como
\begin{equation}
ge=\frac{n}{2}\hbar c. \label{CuantizCargaEl}
\end{equation}
\'Este resultado puede ser obtenido por la teor\'{\i}a desarrollada en la cap\'{\i}tulo \textbf{VII. Cuantizaci\'on} sin la regla de cuantizaci\'on de Bohr que comprende cuatro postulados que determinan: la forma de la \'orbita del electr\'on alrededor del n\'ucleo (sin radiar energ\'{\i}a electromagn\'etica) y su relaci\'on con la Ley de Coulomb, su limitaci\'on entre la infinita cantidad de \'orbitas (mec\'anica cl\'asica) y solo en las que el momento angular es un m\'ultiplo entero de la constante de Planck y el cambio de \'orbita mediante la radiaci\'on de energ\'{\i}a electromagn\'etica.
\newline
Aprovechando el segundo postulado se puede usar la condici\'on que la funci\'on de onda debe ser univaluada. Para las coordenadas y  momentos de la cuerda, cuyo conmutador est\'a dado por la ec. (\ref{BPCuerda}), $\left\{\beta^{\mu}(\tau_1),y_{\nu}(\tau'_1)\right\}=g_{\mu\nu}\delta(\tau_1-\tau'_1)$, (usando el hecho que $\beta^{\mu}(\tau_1)=i\hbar\frac{\partial}{\partial y_{\mu}(\tau_1)}$, debido a que \'este es un operador momento) y donde se observa que el momento de la cuerda puede ser definido mediante la ec. (\ref{Cuerda}), $\left[\beta_{\mu}(\tau_1)-gF^{\dagger *}_{\mu\nu}(y)\frac{y^{\mu}}{d\tau_1}\right]\psi=0$ como
\begin{equation}
i\hbar\frac{\partial \psi}{\partial y_{\mu}(\tau_1)}=gF^{\dagger *}_{\mu\nu}(y)\frac{dy^{\mu}}{d\tau_1}\psi, \label{EcOndaCoordCuerda}
\end{equation}
mostrando la manera como la funci\'on de onda var\'{\i}a cuando las coordenadas de la cuerda se modifican. Si una cuerda es desplazada y se curva sobre una superficie bidimensional $\sigma$, usando la ec. (\ref{EcOndaCoordCuerda}) se puede observar que
\begin{equation}
\frac{d\psi}{\psi}=\frac{g}{i\hbar}F^{\dagger *}_{\mu\nu}(y)\frac{dy^{\mu}}{d\tau_1}dy_{\nu}(\tau_1)\quad \therefore \quad \int\frac{d\psi}{\psi}=-\frac{ig}{\hbar}\int F^{\dagger *}_{\mu\nu}(y)\frac{dy^{\mu}}{d\tau_1}dy_{\nu}.
\end{equation}
Integrando se obtiene
\begin{equation}
\psi=e^{-\frac{ig}{\hbar}\int F^{\dagger *}_{\mu\nu}(y)\frac{dy^{\mu}}{d\tau_1}dy_{\nu}},
\end{equation}
por tanto la funci\'on de onda $\psi$ queda multiplicada por
\begin{equation}
e^{-\frac{ig}{\hbar}\int F^{\dagger *}_{\mu\nu}d\sigma^{\mu\nu}}, \label{fase}
\end{equation}
donde se puede observar que $F^{\dagger *}_{\mu\nu}$ ocurriendo en diferentes puntos del integrando conmutan entre si, lo que podr\'{\i}a ser una analog\'{\i}a a una transformaci\'on local de campos. Debido a que $\psi$ es univaluada, \'este debe retomar su valor inicial y por tanto la fase (\ref{fase}) debe ser unitaria. Para que esto sea posible se debe cumplir la igualdad $\frac{g}{\hbar}\int F^{\dagger *}_{\mu\nu}d\sigma^{\mu\nu}=2\pi n$ siendo $n$ un entero. Se puede observar que \'esta condici\'on vuelve a definir la ecuaci\'on (\ref{Cuantizge}), $4\pi ge=n\hbar$. Finalmente la conclusi\'on a la que se llega es que \cite{Dirac2}:
\newline
\textbf{ \emph{...la cuantizaci\'on de las ecuaciones de movimiento de las part\'{\i}culas cargadas y de las part\'{\i}culas con polo, es posible solamente condicionando a las cargas y a los polos a que deben ser m\'ultiplos integrales de una unidad de carga $e_0$ y de una unidad de polo $g_0$ satisfaciendo, $e_0g_0=\frac{1}{2}\hbar c$}...}.
\chapter[Incoherencias Teor\'{\i}a Monopolo Magn\'etico]{Incoherencias en la Teor\'{\i}a del Monopolo Magn\'etico}
Durante muchos a\~nos los f\'{\i}sicos han intentado hallar el monopolo infructuosamente. El intento de explicar las discontinuidades inherentes en la teor\'{\i}a de Dirac o cuerdas de Dirac y su resultados negativos en la medici\'on experimental, han llevado a artificios f\'{\i}sicos y matem\'aticos que han hecho que la teor\'{\i}a sea poco natural. Actualmente existen sistemas de monopolos y campos que son duales a la electrodin\'amica cl\'asica (EDC) con los cuales se pueden explicar la supuesta falta de simetr\'{\i}a de \'esta (ver cap\'{\i}tulo (\ref{MonopSinCD})). A pesar de \'esto muchos cient\'{\i}ficos han seguido investigando y usando las cuerdas de Dirac en la teor\'{\i}a del monopolo. 
\par
Las ecuaciones de Maxwell, que gobiernan el movimiento de los campos electromagn\'eticos son,
\begin{align}
\partial_{\nu}F^{\mu\nu}_e &=-4\pi J^{\mu}_e \notag \\
\partial_{\nu}F^{*\mu\nu}_e &=0
\end{align}
y en ellas est\'an contenidas las ecuaciones de Maxwell y la relatividad especial de Einstein (ver Cap\'{\i}tulos \ref{TER} y \ref{CapEDC}). La fuerza ejercida sobre toda la materia que posee carga est\'a definida por la fuerza de Lorentz en cuatro dimensiones o la cuadri-fuerza de Lorentz,
\begin{equation}
q_eF^{\mu\nu}_ev_{\nu}=ma^{\mu}_e
\end{equation}
donde el tensor electromagn\'etico $F^{\mu\nu}$, est\'a definido seg\'un lo dicho en el Cap\'{\i}tulo \ref{CapEDC}. El tensor electromagn\'etico dual (ver Cap. \ref{CapEDC}) se define como
\begin{equation}
F^{*\mu\nu}=\frac{1}{2}\varepsilon^{\mu\nu\alpha\beta}F_{\alpha\beta}
\end{equation}
el cual es totalmente antisim\'etrico, justamente como $F^{\mu\nu}$ tambi\'en lo es. La relaci\'on intr\'{\i}nseca entre estos dos tensores y la definici\'on de la densidad Lagrangiana
\begin{equation}
L=-\frac{1}{16\pi}F^{\mu\nu}_eF_{e,\mu\nu}-J^{\mu}_eA_{e,\mu}
\end{equation}
lleva a realizar transformaciones duales que conducen a las ecuaciones de Maxwell duales, las cuales son usadas para definir la teor\'{\i}a de un sistema de monopolos con sus campos electromagn\'eticos de la forma,
\begin{align}
\partial_{\nu}F^{*\mu\nu}_m &=-4\pi J^{\mu}_m \notag \\
-\partial_{\nu}F^{\mu\nu}_m &=0,
\end{align}
en las cuales est\'an contenidas las ecuaciones de movimiento de Maxwell para los campos monopolares. La cuadri-fuerza sobre la materia monopolar es de forma an\'aloga a la ejercida sobre la materia cargada
\begin{equation}
q_hF^{*\mu\nu}_mv_{\nu}=ma^{\mu}_m.
\end{equation}
La densidad Lagrangiana del monopolo est\'a dada por
\begin{equation}
L=-\frac{1}{16\pi}F^{*\mu\nu}_mF*_{m,\mu\nu}-J^{\mu}_mA_{m,\mu}.
\end{equation}
\par
Con \'estas dos teor\'{\i}as se puede observar que \cite{Comay} ninguna de \'estas relaciona las dos cargas (el\'ectrica y magn\'etica), pues cada teor\'{\i}a contiene las ecuaciones de movimiento as\'{\i} como la ecuaci\'on de la din\'amica para una carga y no la relaciona con la otra. Dichas teor\'{\i}as deber\'{\i}an conformar una nueva teor\'{\i}a carga el\'ectrica-monopolo magn\'etico en la cual, en los l\'{\i}mites en los que una de las cargas no exista sobreviva la otra. Pero en la teor\'{\i}a de los monopolos de Dirac si se calcula el l\'{\i}mite en el que la carga el\'ectrica no exista, no se recuperan las ecuaciones de Maxwell en el espacio normal. A pesar que la teor\'{\i}a de Dirac se construya bajo la suposici\'on (deseable) que tanto la carga el\'ectrica como el monopolo tengan propiedades din\'amicas iguales, \'esto lleva a inconsistencias f\'{\i}sicas tales como las cuerdas de Dirac.
\newline
Por ejemplo los campos electromagn\'eticos radian con las condiciones $E^2=B^2$ y $\bm{E}\centerdot\bm{B}=0$ que son bi\'en conocidas en las ondas electromagn\'eticas y cuyo significado es que los campos se mueven oscilatoriamente de manera perpendicularmente a la velocidad de radiaci\'on o de propagaci\'on de la onda electromagn\'etica lo que muestra de nuevo un grado de simetr\'{\i}a que est\'a en total desacuerdo con una singularidad de cuerda.
\newline
Adicionalmente aunque con la teor\'{\i}a de Dirac al modificar a 'mano' algunas de las ecuaciones de la din\'amica electromagn\'etica y/o de la acci\'on (ver el cap\'{\i}tulo \ref{TPMDirac}) para reproducir la electrodin\'amica llam\'andola nueva electrodin\'amica y que no es posible derivarla de una funci\'on Lagrangiana electromagn\'etica corriente por lo que las variables can\'onicas no estar\'an bi\'en definidas y por tanto la funci\'on Hamiltoniana ser\'a muy dif\'{\i}cil de obtener; adicionalmente la carga magn\'etica transforma como un seudoescalar y no como un escalar lo que dificulta el entendimiento f\'{\i}sico.
\par
Existen muchos art\'{\i}culos de investigaci\'on donde se describen los inconvenientes de los polos de Dirac para casos espec\'{\i}ficos como el de Daniel Zwazinger \cite{Zwanziger} donde explica que \'estos son prohibidos en la teor\'{\i}a de la matriz de dispersi\'on debido a que dicha teor\'{\i}a es invariante de Lorentz y los polos de Dirac no respetan dicha invarianza lo cual est\'a muy bi\'en determinado en \cite{Hagen} donde analiza la no-covarianza del monopolo de Dirac resultando una incompatibilidad con la teor\'{\i}a cu\'antica de campos que para solucionarla habr\'{\i}a que redefinir muchos conceptos b\'asicos de \'esta. En fin hay mucha literatura al respecto pero el problema fundamental es la definici\'on f\'{\i}sica de las cuerdas de Dirac y el entendimiento matem\'atico de la aplicaci\'on del teorema de Stokes debido a que \'este no se define en todo el espacio por la divergencia generada por la cuerda de Dirac.
\chapter{Ap\'endices}
\section{Delta de Dirac \label{DDirac}}
La funci\'on delta de Dirac es una funci\'on de distribuci\'on que permite definir puntos matem\'aticos de discontinuidades en el espacio. Un ejemplo muy claro es su uso para definir una carga puntual. El delta de Dirac est\'a definido por las relaciones $\delta(x)=0$ para $x=0$ y $\int\delta(x)dx=1$ si $x\neq0$ y donde la regi\'on de integraci\'on incluye $x=0$. Para una funci\'on arbitraria $\int f(x)\delta(x)dx=f(0)$, donde la regi\'on de integraci\'on incluye $x=0$. La representaci\'on del delta de Dirac est\'a dado por $\delta(y)=\lim_{h\to\infty}\frac{\sin (hy)}{\pi y}$, \cite{Cohen}, [\cite{Schiff}.
\newline
La integraci\'on se logra usando
\begin{multline}
\int^{\infty}_{-\infty}e^{(\omega_z-\sigma_z)z}dz=\lim_{h\to\infty}\int^{h}_{-h}e^{(\omega_z-\sigma_z)z}dz=2\lim_{h\to\infty}\frac{\sin[ h(\omega_z-\sigma_z)]}{(\omega_z-\sigma_z)} \\
=2\pi\delta((\omega_z-\sigma_z)).
\end{multline}
La relaci\'on de ortonormalidad se construye
\begin{equation}
\int{u^*_i(\mathbf{r})u_j(\mathbf{r})d^3r}=\delta(\omega_x-\sigma_x)\delta(\omega_y-\sigma_y)\delta(\omega_z-\sigma_z)\equiv\delta^3(\mathbf{\omega}-\mathbf{\sigma}).
\end{equation}
En un intervalo $(-x,x)$ siendo $\alpha$ una cantidad positiva se impone la condici\'on,
\begin{equation}
\delta^{\alpha}(x)=\left\{
\begin{aligned}
&\frac{1}{\alpha} \quad \text{para}\quad -\frac{\alpha}{2}<x<\frac{\alpha}{2} \\
&0 \quad \text{para}\quad \vert x\vert>\frac{\alpha}{2}
\end{aligned}\right.
\end{equation}
de forma tal que en dicho intervalo la funci\'on delta tomar\'a el valor $\frac{1}{\alpha}$ forma una especie de pico. Si se integra sobre las $x$ en ese mismo intervalo y sobre una funci\'on arbitraria $f(x)$ (bi\'en definida en $x=0$), entonces se tiene que $\int_{+x}^{-x}\delta^{\alpha}(x)f(x)dx$ y teniendo en cuenta que $f(x)\backsimeq f(0)$ entonces $\int_{+x}^{-x}\delta^{\alpha}(x)f(x)dx\backsimeq f(0)\int_{+x}^{-x}\delta^{\alpha}(x)dx=f(0)$. Si $\alpha$ es muy peque\~na entonces la aproximaci\'on ser\'a mejor y por tanto al evaluar el l\'{\i}mite cuando tiende a cero se verifica
\begin{equation}
\int_{+x}^{-x}\delta(x)f(x)dx=f(0),
\end{equation}
y de forma m\'as general
\begin{equation}
\int_{+\infty}^{-\infty}\delta(x)f(x-x_0)dx=f(x_0) \label{deltaDiracGen},
\end{equation}
Algunas de las propiedades de la funci\'on delta de Dirac son:
\begin{align}
&\delta(x)=\delta(-x) \notag \\
&\frac{d\delta(x)}{dx}=-\frac{d\delta(-x)}{dx} \notag \\
&x\delta(x)=0 \notag \\
&x\frac{d\delta(x)}{dx}=-\delta(x) \notag \\
&\delta(ax)=\frac{1}{a}\delta(x) \notag \\
&\delta(x^2-a^2)=\frac{1}{2a}[\delta(x-a)+\delta(x+a)] \notag \\
&\int\delta(a-x)\delta(x-b)dx=\delta(a-b)\notag \\
&f(x)\delta(x-a)=f(a)\delta(x-a)
\end{align}
Usando la ecuaci\'on (\ref{deltaDiracGen}) y
\begin{align}
\widetilde{F}(k) &=(2\pi)^{-\frac{1}{2}}\int^{+\infty}_{-\infty}e^{-ikx}F(x)dx \notag \\
F(x) &=(2\pi)^{-\frac{1}{2}}\int^{+\infty}_{-\infty}e^{+ikx}\widetilde{F}(k)dk,
\end{align}
la relaci\'on entre el delta de Dirac y Fourier est\'a dada por
\begin{equation}
\widetilde{\delta}(k)=(2\pi)^{-\frac{1}{2}}\int^{+\infty}_{-\infty}e^{-ikx}\delta(x-x_0)dx.
\end{equation}
En los textos \cite{Arfken}, \cite{Cohen}, \cite{Schiff}, \cite{Jeffreys} se puede profundizar m\'as sobre \'este tema.
\section{Funci\'on de Pauli-Jordan\label{PJFunction}}
La funci\'on de conmutaci\'on de Pauli-Jordan (FPJ) est\'a dada por,
\begin{equation}
\Delta(x-y)=i[\varphi(x),\varphi(y)]
\end{equation}
donde
\begin{multline}
\Delta(x)=\frac{i}{(2\pi)^3}\int{e^{-ikx}\varepsilon(k^0)\delta\left(k^2-m^2\right)}dk \\
=\frac{i}{(2\pi)^3}\int\frac{dk}{(k^2+m^2)^{\frac{1}{2}}e^{ikx}\sin\left[x^0(k^2-m^2)\right]^{\frac{1}{2}}}
\end{multline}
donde se verifica que su variaci\'on temporal est\'a relacionada con la funci\'on delta de Dirac de la forma:
\begin{equation}
\left.\frac{\partial \Delta(x^0,\mathbf{x})}{\partial x^0}\right\vert_{x^0=0}=\delta(\mathbf{x}).
\end{equation}
\section[Ecuaci\'on Laplace en CE]{Ecuaci\'on de Laplace en Coordenadas Esf\'ericas (CE) \label{EcLaplaceCE}}
La ecuaci\'on de Laplace en CE se escribe como,
\begin{equation}
\frac{1}{r}\frac{\partial^2}{\partial r^2}\left(r\mathcal{F}\right)+\frac{1}{r^2\sin\theta}\frac{\partial}{\partial\theta}\left(\sin\theta\frac{\partial\mathcal{F}}{\partial\theta}\right)+\frac{1}{r^2\sin^2\theta}\frac{\partial^2\mathcal{F}}{\partial\phi^2}=0
\end{equation}
Si se asume una soluci\'on de variables separables de la forma $\mathcal{F}=\frac{R(r)}{r}\Theta(\theta)\Phi(\phi)$, al reemplazarla en la ecuaci\'on, y despu\'es de reordenar los t\'erminos \'esta queda,
\begin{equation}
r^2\sin^2\theta\left[\frac{1}{R}\frac{d^2R}{dr^2}+\frac{1}{r^2\sin\theta\Theta}\frac{d}{d\theta}\left(\sin\theta\frac{d\Theta}{d\theta}\right)\right]+\frac{1}{\Phi}\frac{d^2\Phi}{d\phi^2}=0.
\end{equation}
Si se elige una constante de separaci\'on $-m^2$, entonces se verifica la ecuaci\'on diferencial,
\begin{equation}
\frac{d^2\Phi}{d\phi^2}=-m^2\Phi,
\end{equation}
cuya soluci\'on es,
\begin{equation}
\Phi=e^{\pm im\phi}
\end{equation}
donde $m$ debe ser un entero, pues si no lo es, la funci\'on $\Phi(\phi)$ puede ser multivaluada. El resto de la ecuaci\'on es entonces,
\begin{equation}
r^2\sin^2\theta\left[\frac{1}{R}\frac{d^2R}{dr^2}+\frac{1}{r^2\sin\theta\Theta}\frac{d}{d\theta}\left(\sin\theta\frac{d\Theta}{d\theta}\right)\right]=-m^2
\end{equation}
y por tanto
\begin{equation}
\frac{r^2}{R}\frac{d^2R}{dr^2}+\frac{1}{\sin\theta\Theta}\frac{d}{d\theta}\left(\sin\theta\frac{d\Theta}{d\theta}\right)=-\frac{m^2}{\sin^2\theta}.
\end{equation}
Separando variables de nuevo con una nueva constante de separaci\'on $l(l+1)$, se llega a
\begin{equation}
\frac{r^2}{R}\frac{d^2R}{dr^2}=l(l+1)
\end{equation}
despejando se obtiene,
\begin{equation}
\frac{d^2R}{dr^2}=\frac{l(l+1)}{r^2}R
\end{equation}
donde $l(l+1)$ es otra constante. La soluci\'on de \'esta ecuaci\'on est\'a dada por funciones hiperb\'olicas,
\begin{equation}
R=\mathcal{R}_1e^{l+1}+\mathcal{R}_2e^{-l}.
\end{equation}
finalmente se tiene la \'ultima ecuaci\'on diferencial,
\begin{equation}
\frac{1}{\sin\theta\Theta}\frac{d}{d\theta}\left(\sin\theta\frac{d\Theta}{d\theta}\right)-\frac{m^2}{\sin^2\theta}=-l(l+1)
\end{equation}
la cual se reorganiza a
\begin{equation}
\frac{1}{\sin\theta}\frac{d}{d\theta}\left(\sin\theta\frac{d\Theta}{d\theta}\right)=\left[-l(l+1)+\frac{m^2}{\sin^2\theta}\right]\Theta.
\end{equation}
Si se realiza el cambio de variable $x=\cos\theta$, entonces $dx=-\sin\theta d\theta=-(1-x^2)^{\frac{1}{2}}d\theta$, y $\sin^2\theta=1-x^2$. Por tanto la ecuaci\'on queda,
\begin{equation}
-\frac{d}{dx}\left[(1-x^2)\frac{d\Theta}{dx}\right]=\left[-l(l+1)+\frac{m^2}{1-x^2}\right]\Theta.
\end{equation}
Finalmente haciendo $P(x)=\Theta(x)$, se tiene la ecuaci\'on,
\begin{equation}
-\frac{d}{dx}\left[(1-x^2)\frac{dP}{dx}\right]=\left[-l(l+1)+\frac{m^2}{1-x^2}\right]P,
\end{equation}
a la que se le llama ecuaci\'on generalizada de Legendre. La soluci\'on de \'esta ecuaci\'on se escribe como un serie de potencias,
\begin{equation}
P(x)=x^{\alpha}\sum_{j=0}^{\infty}a_jx^j
\end{equation}
donde el par\'ametro $\alpha$ no se conoce. Al derivarla da,
\begin{equation}
\frac{dP(x)}{dx}=\alpha x^{\alpha-1}\sum_{j=0}^{\infty}a_jx^j+x^{\alpha}\sum_{j=0}^{\infty}ja_jx^{j-1}
\end{equation}
y por tanto
\begin{align}
(1-x^2)\frac{dP(x)}{dx}=& \alpha x^{\alpha-1}\sum_{j=0}^{\infty}a_jx^j+x^{\alpha}\sum_{j=0}^{\infty}ja_jx^{j-1} \notag \\
&\hspace{-1.2cm}-\alpha x^{\alpha+1}\sum_{j=0}^{\infty}a_jx^j-x^{\alpha+2}\sum_{j=0}^{\infty}ja_jx^{j-1}.
\end{align}
Derivando nuevamente y reemplazando en la ecuaci\'on generalizada de Legendre,
\begin{multline}
\frac{d}{dx}\left[(1-x^2)\frac{dP(x)}{dx}\right]-\left[-l(l+1)+\frac{m^2}{1-x^2}\right]P(x)=\alpha(\alpha-1) x^{\alpha-2}\sum_{j=0}^{\infty}a_jx^j \\
+\alpha x^{\alpha-1}\sum_{j=0}^{\infty}ja_jx^{j-1}+\alpha x^{\alpha-1}\sum_{j=0}^{\infty}ja_jx^{j-1}+x^{\alpha}\sum_{j=0}^{\infty}j(j-1)a_jx^{j-2} \\
-\alpha(\alpha+1) x^{\alpha}\sum_{j=0}^{\infty}a_jx^j-\alpha x^{\alpha+1}\sum_{j=0}^{\infty}ja_jx^{j-1} -(\alpha+2)x^{\alpha+1}\sum_{j=0}^{\infty}ja_jx^{j-1} \\
-x^{\alpha+2}\sum_{j=0}^{\infty}j(j-1)a_jx^{j-2}-\left[-l(l+1)+\frac{m^2}{1-x^2}\right]x^{\alpha}\sum_{j=0}^{\infty}a_jx^j=0,
\end{multline}
donde se verifican las ecuaciones,
\begin{align}
&\left\{\alpha(\alpha-1) x^{\alpha-2}-\left[-l(l+1)+\frac{m^2}{1-x^2}\right]x^{\alpha}\right. \notag \\
&\hspace{2.9cm}-\alpha(\alpha+1) x^{\alpha}\biggr\}\sum_{j=0}^{\infty}a_jx^j=0 \notag \\
&\left[2\alpha x^{\alpha-1}-\alpha x^{\alpha+1}-(\alpha+2)x^{\alpha+1}\right] \notag \\
&\hspace{4.7cm}\cdot\sum_{j=0}^{\infty}ja_jx^{j-1}=0 \notag \\
&\left[x^{\alpha}-x^{\alpha+2}\right]\sum_{j=0}^{\infty}j(j-1)a_jx^{j-2}=0.
\end{align}
que al solucionarlas y al reemplazar los valores en la ecuaci\'on original, (ver \cite{Jackson}, \cite{Goldstein}, \cite{Arfken} y \cite{Jeffreys}) se llega a una funci\'on recursiva para la variable $a_j$ de la forma,
\begin{equation}
a_{j+2}=\frac{(\alpha+j)(\alpha+j+1)-l(l+1)}{(\alpha+j+1)(\alpha+j+2)}a_j,
\end{equation}
lo que da la soluci\'on de las funciones $P_j(x)$ en forma de polinomios, los cuales toman el nombre de polinomios de Legendre, que se compactan mediante la f\'ormula de Rodriguez como,
\begin{equation}
P_j(x)=\frac{1}{2^jl!}\frac{d^l}{dx^l}\left(x^2-1\right)^l.
\end{equation}
Algunos t\'erminos de la serie son:
\begin{align}
P_0(x) &=1 \notag \\
P_1(x) &=x \notag \\
P_2(x) &=\frac{1}{2}\left(3x^2-1\right) \notag \\
P_3(x) &=\frac{1}{2}\left(5x^3-3x\right) \notag \\
P_4(x) &=\frac{1}{8}\left(35x^4-30x^2+3\right) \notag \\
P_5(x) &=\frac{1}{8}\left(63x^5-70x^3+15x\right) \notag \\
P_6(x) &=\frac{1}{16}\left(231x^6-315x^4+105x^2-5\right) \notag \\
P_7(x) &=\frac{1}{16}\left(429x^7-639x^5+315x^3-35x\right) \notag \\
&\hspace{3cm}\vdots
\end{align}
\subsection{Aplicaciones en la Electrodin\'amica}
Una de las aplicaciones m\'as usadas de los Polinomios de Legendre y las funciones de Rodrigues est\'a en la expansi\'on multipolar, que es necesaria en la teor\'{\i}a de radiaci\'on. Los polinomios ayudar a expandir funciones tales como,
\begin{equation}
\frac{1}{\vert\bm{x}-\bm{x'}\vert}=\frac{1}{\sqrt{r^2+r^{'2}-2rr'\cos\gamma}}=\sum^{\infty}_{l=0}\frac{r^{'l}}{r^{l+1}}P_l(\cos\gamma)
\end{equation}
para la restricci\'on multipolar, $r>r'$, la que es usada, por ejemplo, para calcular el potencial de una carga puntual, el cual usando el eje $z$ como eje de simetr\'{\i}a da,
\begin{equation}
\Phi(r,\theta)=\sum^{\infty}_{l=0}\left[A_lr^l+B_lr^{-(l+1)}\right]P_l(\cos\gamma)
\end{equation}
donde $A_l$ y $B_l$ se hallan con las condiciones de frontera del problema.
\newline
La expansi\'on multipolar puede realizarse primero haciendo un expansi\'on de Taylorde una funci\'on arbitraria $\mathcal{F}(\bm{r}-\bm{R})$ alrededor del origen $\bm{r}=\bm{0}$ de la forma,
\begin{equation}
\mathcal{F}(\bm{r}-\bm{R})=\mathcal{F}(\bm{R})+\sum_{i=x,y,z}r_i\mathcal{F}_i(\bm{R})+\frac{1}{2}\sum_{i,j=x,y,z}r_ir_j\mathcal{F}_{ij}(\bm{R})+\ldots
\end{equation}
donde
\begin{equation}
\mathcal{F}_{i}(\bm{R})\equiv\left(\frac{\partial\mathcal{F}(\bm{r}-\bm{R})}{\partial r_i}\right)_{\bm{r=0}}
 \end{equation}
y
\begin{equation}
\mathcal{F}_{ij}(\bm{R})\equiv\left(\frac{\partial^2\mathcal{F}(\bm{r}-\bm{R})}{\partial r_ir_j}\right)_{\bm{r=0}}.
\end{equation}
Si $\mathcal{F}_{i}(\bm{r}-\bm{R})$ satisface la ecuaci\'on de Laplace
\begin{equation}
\left[\bm{\nabla}^2\mathcal{F}(\bm{r}-\bm{R})\right]_{\bm{r=0}}=\sum_{i=x,y,z}\mathcal{F}_{ii}(\bm{R})=0,
\end{equation}
entonces la expansi\'on se escribe tensorialmente de la forma,
\begin{equation}
\sum_{i,j=x,y,z}r_ir_j\mathcal{F}_{ij}(\bm{R})=\frac{1}{3}\sum_{i,j=x,y,z}\left(3r_ir_j-\delta_{ij}r^2\right)\mathcal{F}_{ij}(\bm{R})
\end{equation}
si se considera $\mathcal{F}(\bm{r}-\bm{R})\equiv\frac{1}{\vert\bm{r}-\bm{R}\vert}$, entonces $\mathcal{F}(\bm{R})=\frac{1}{R}$, $\mathcal{F}_i(\bm{R})=\frac{R_i}{R^3}$, $\mathcal{F}_{ij}(\bm{R})=\frac{3R_iR_j-\delta_{ij}R^2}{R^5}$, etc, definiendo, respectivamente, el monopolo, el dipolo y el cuadrupolo por
\begin{align}
q_T &\equiv\sum_{i=1}^{N}q_i \notag \\
P_{i} &\equiv\sum_{\alpha=1}^{N}q_{\alpha}r_{\alpha i} \notag \\
P_{ij} &\equiv\sum_{\alpha=1}^{N}q_{\alpha}\left(3r_{\alpha i}r_{\alpha j}-\delta_{ij}r_{\alpha}^2\right) \notag \\
&\hspace{1.5cm}\vdots
\end{align}
donde se obtiene la expansi\'on multipolar de la funci\'on escalar, denominada potencial total $\Phi(\bm{R})$
\begin{align}
\Phi(\bm{R}) &=\frac{1}{4\pi\varepsilon_0}\sum_{i=1}^Nq_i\mathcal{F}(\bm{r}_i-\bm{R}) \notag \\
&=\frac{q_T}{R}+\frac{1}{R^3}\sum_{i=x,y,z}P_iR_i \notag \\
&\hspace{0.3cm}+\frac{1}{6R^5}\sum_{i,j=x,y,z}Q_{ij}\left(3R_iR_j-\delta_{ij}R^2\right)+\ldots
\end{align}
Si se aproxima a dos cargas, entonces el potencial dipolar el\'ectrico estar\'a dado por,
\begin{equation}
\Phi(\bm{R})=\frac{1}{4\pi\varepsilon_0}\frac{\bm{P}\centerdot\bm{R}}{R^3}.
\end{equation}
En coordenadas esf\'ericas toma la forma (ver referencias \cite{Jackson}, \cite{Goldstein}, \cite{Arfken} y \cite{Jeffreys}),
\begin{equation}
\Phi(\bm{R})=\frac{1}{4\pi\varepsilon_0}\sum_{l=0}^{\infty}\sum_{m=-l}^{l}(-1)^mI_l^{-m}(\bm{R})\centerdot\sum_{i=1}^{N}q_iR_l^m(\bm{r}_i),
\end{equation}
donde $I_l^{-m}(\bm{R})$ se denomina un arm\'onico s\'olido y est\'a definido por,
\begin{equation}
I_l^{-m}(\bm{R})\equiv\sqrt{\frac{4\pi}{2l+1}}\frac{Y_l^m(\bm{\hat{R}})}{R^{l+1}},
\end{equation}
y $R_l^m(\bm{r}_i)$ se denomina arm\'onico s\'olido regular. Definiendo el momento multipolar esf\'erico como $Q_l^m\equiv\sum_{i=1}^Nq_iR_l^m(\bm{r}_i)$ para $-l\leq m\leq-l$, se halla el potencial como,
\begin{align}
\Phi(\bm{R}) &=\frac{1}{4\pi\varepsilon_0}\sum_{l=0}^{\infty}\sum_{m=-l}^{l}(-1)^mI_l^{-m}(\bm{R})Q_l^m \notag \\
&=\frac{1}{4\pi\varepsilon_0}\sum_{l=0}^{\infty}\sqrt{\frac{4\pi}{2l+1}}\frac{1}{R^{l+1}} \notag \\
&\hspace{1.6cm}\cdot\sum_{m=-l}^{l}(-1)^mY_l^m(\bm{\hat{R}})Q_l^m
\end{align}
para $R>r_{\text{max}}$, y donde se observa que la expansi\'on aparece como coeficientes en la expansi\'on $R^{-1}$ del potencial.
\section{Clases de Gauge\label{TiposGauge}}
Algunos tipos de gauge para $\mu=0,1,2,3$ y para $j=1,2,3$ son el gauge de Lorenz, $\partial_{\mu}A^{\mu}=0$; el gauge de Coulomb \'o de radiaci\'on, $\bm{\nabla}\centerdot\bm{A}=\partial_jA^j=0$; el gauge axial, $A_3=0$; el gauge temporal \'o Hamiltoniano, $A_0=0$; el gauge de cono de luz, $n_{\mu}A^{\mu}=0$; el gauge de Poincar\'e, $x_jA_j=0$ y el gauge de Schwinger-Fock, $x_{\mu}A^{\mu}=0$
\section{Campo de un Solenoide Largo y Delgado \label{Magneto}}
El campo de un solenoide largo y delgado se estudia desde el punto de vista del potencial vectorial de la forma,
\begin{equation}
\bm{A}=\frac{1}{c}\int\frac{\bm{J}(\bm{x}')}{\vert\bm{x}-\bm{x}'\vert}d^3\bm{x}'
\end{equation}
en donde se hace la aproximaci\'on $\vert\bm{x}-\bm{x}'\vert^{-1}=\vert\bm{x}\vert^{-1}+\frac{\bm{x}\centerdot\bm{x}'}{\vert\bm{x}\vert^3}+\ldots$ para obtener,
\begin{equation}
\bm{A}(\bm{x})=\frac{1}{c}\int\frac{\bm{J}(\bm{x}')}{\vert\bm{x}\vert}d^3\bm{x}'+\frac{1}{c}\int\frac{(\bm{x}\centerdot\bm{x}')\bm{J}(\bm{x}')}{\vert\bm{x}\vert^3}d^3\bm{x}'+\ldots \label{PotVectA}
\end{equation}
que para un sistema se escribe de la siguiente manera:
\begin{equation}
\bm{A}_i(\bm{x})=\frac{1}{c\vert\bm{x}\vert}\int\bm{J}_i(\bm{x}')d^3\bm{x}'+\frac{\bm{x}}{c\vert\bm{x}\vert^3}\centerdot\int\bm{x}'\bm{J}_i(\bm{x}')d^3\bm{x}'+\ldots
\end{equation}
\textbf{Teorema.}
\par
Si $f(\bm{x}')$ y $g(\bm{x}')$ son dos funciones bi\'en comportadas de $\bm{x}'$. Si $\bm{J}(\bm{x}')$ est\'a localizada y tiene divergencia nula, entonces
\begin{equation}
\int\left(f\bm{J}\centerdot\bm{\nabla}'g+g\bm{J}\centerdot\bm{\nabla}'f\right)d^3x'=0
\end{equation}
que se puede escribir como
\begin{equation}
\int\left[f\bm{\nabla}'\centerdot(g\bm{J})\right]d^3x'=0
\end{equation}
integrando por partes mediante el uso de $u=f$, $du=\bm{\nabla}'fd^3x'$, $dv=\bm{\nabla}'\centerdot\left(g\bm{J}\right)d^3x'$ , $v=g\bm{J}$ y de $U=g$, $dU=\bm{\nabla}'gd^3x'$, $dV=\bm{\nabla}'\centerdot\left(g\bm{J}\right)d^3x'$ y $v=g\bm{J}$ queda,
\begin{equation}
\int\left[f\bm{\nabla}'\centerdot(g\bm{J})\right]d^3x'=-\int (g\bm{J}\centerdot\bm{\nabla}'f)d^3x'=0
\end{equation}
pues $fg\bm{J}$ evaluado en el universo es nulo. Con los valores $f=1$ y $g=x'_i$ entonces se tiene que $\bm{\nabla}'f=0$ y $\bm{\nabla}'g=\bm{\hat{x}}'_i$ y por tanto la integral queda,
\begin{equation}
\int\left(\bm{J}\centerdot\bm{\nabla}'x'_i\right)d^3x'=\int J_i(\bm{x}')d^3x'=0 \label{int1}
\end{equation}
Ahora si se hace la elecci\'on $f=x'_i$ y $g=x'_j$ entonces se tiene que $\bm{\nabla}'f=\bm{\hat{x}}'_i$ y $\bm{\nabla}'g=\bm{\hat{x}}'_j$ y por tanto la integral queda,
\begin{align}
\int &\left(f\bm{J}\centerdot\bm{\nabla}'g+g\bm{J}\centerdot\bm{\nabla}'f\right)d^3x' \notag \\
&=\int\left(x'_i\bm{J}\centerdot\bm{\hat{x}}'_j+x'_j\bm{J}\centerdot\bm{\hat{x}}'_i\right)d^3x' \notag \\
&=\int\left(x'_iJ_j+x'_jJ_i\right)d^3x'
\end{align}
y por tanto,
\begin{equation}
\int\left(x'_iJ_j+x'_jJ_i\right)d^3x'=0  \label{int2}
\end{equation}
Se concluye, debido a la ec. (\ref{int1}), que el primer t\'ermino del potencial vectorial (\ref{PotVectA}) se anula y por tanto el t\'ermino siguiente, usando la ec. (\ref{int2}), se modifica de la siguiente manera,
\begin{align}
\frac{\bm{x}}{c\vert\bm{x}\vert^3} &\centerdot\int J_i(\bm{x}')\bm{x}'d^3x'=\frac{\bm{x}}{c\vert\bm{x}\vert^3}\centerdot\int \bm{x}'J_i(\bm{x}')d^3x' \notag \\
&=\frac{1}{c\vert\bm{x}\vert^3}\sum_j x_j\int x'_jJ_i(\bm{x}')d^3x' \notag \\
&=-\frac{1}{c\vert\bm{x}\vert^3}\sum_j\frac{1}{2} x_j\int (x'_iJ_j-x'_jJ_i)d^3x'
\end{align}
debido a que $x'_iJ_j+x'_jJ_i=0$, donde $x'_iJ_j=-x'_jJ_i$, entonces $2x'_iJ_j=-(x'_jJ_i-x'_iJ_j)$ y por tanto $x'_iJ_j=-\frac{1}{2}(x'_jJ_i-x'_iJ_j)$. Se nota que el resultado est\'a relacionado con el delta de Levi-Civita, por lo que el t\'ermino se modifica a
\begin{align}
-\frac{1}{c\vert\bm{x}\vert^3} &\sum_j\frac{1}{2} x_j\int (x'_iJ_j-x'_jJ_i)d^3x' \notag \\
&=-\frac{1}{2c\vert\bm{x}\vert^3}\sum_{ijk} x_j\varepsilon_{ijk}x_j\int (\bm{x}'\times \bm{J})_kd^3x' \notag \\
&=-\frac{1}{2c\vert\bm{x}\vert^3}\bm{x}\times\int (\bm{x}'\times \bm{J}(\bm{x}')d^3x' \notag \\
&=-\frac{1}{\vert\bm{x}\vert^3}\bm{x}\times\bm{m}=\frac{\bm{m}\times\bm{x}}{\vert\bm{x}\vert^3}=\bm{A}(\bm{x})
\end{align}
donde $\bm{m}$ es el momento magn\'etico y $\bm{\mu}(\bm{x})=\frac{\bm{x}\times\bm{J}(\bm{x})}{2c}$ es la magnetizaci\'on del medio. Entonces la densidad de flujo magn\'etico est\'a definida por,
\begin{equation}
\bm{B}=\bm{\nabla}\times\bm{A}=\bm{\nabla}\times\frac{\bm{m}\times\bm{x}}{\vert\bm{x}\vert^3} \label{MagnF}.
\end{equation}
Usando
\begin{align}
\left[\bm{\nabla}\times\left(\bm{A}\times\bm{B}\right)\right]_i &=\varepsilon_{ijk}\nabla_j\varepsilon_{klm}A_lB_m \notag \\
&=\varepsilon_{ijk}\varepsilon_{klm}\nabla_jA_lB_m \notag \\
&=\left(\delta_{il}\delta_{jm}-\delta_{im}\delta_{jl}\right)\nabla_jA_lB_m
\end{align}
donde usando las propiedades del los delta de Kronecker se obtiene: $\nabla_jA_iB_j-\nabla_jA_jB_i=\left(\nabla_jB_jA_i+B_j\nabla_jA_i\right)-\left(\nabla_jA_jB_i+A_j\nabla_jB_i\right)$, dando como resultado la ecuaci\'on vectorial,
\begin{equation}
\bm{\nabla}\times\left(\bm{A}\times\bm{B}\right)=(\bm{\nabla}\centerdot\bm{B})\bm{A}+(\bm{B}\centerdot\bm{\nabla})\bm{A}-(\bm{\nabla}\centerdot\bm{A})\bm{B}-(\bm{A}\centerdot\bm{\nabla})\bm{B}
\end{equation}
que al aplicarla en la ecuaci\'on (\ref{MagnF}) da el siguiente resultado:
\begin{equation}
\bm{B}=\frac{3\bm{\hat{x}}\left(\bm{x}\centerdot\bm{m}\right)-\bm{m}}{\vert\bm{x}\vert^3}
\end{equation}
Y as\'{\i}, el potencial vectorial es (en el sistema CGS-Gauss),
\begin{equation}
\bm{A}(\bm{x})=\frac{1}{c}\frac{\bm{m}\times\bm{\hat{r}}}{r^2}.
\end{equation}
Si $\bm{m}$ est\'a alineado en la direcci\'on del eje $z$, el potencial en el punto $P(r,\theta,\phi)$ es,
\begin{equation}
\bm{A}(\bm{x})=\frac{1}{c}\frac{\bm{m}\sin\theta}{r^2}\bm{\hat{\phi}}
\end{equation}
y por tanto la densidad de flujo magn\'etico se puede calcular mediante,
\begin{equation}
\bm{B}=\frac{1}{c}\left[\frac{1}{r\sin\theta}\frac{\partial}{\partial\theta}\left(\frac{m\sin\theta}{r^2}\sin\theta\right)\bm{\hat{r}}-\frac{1}{r}\frac{\partial}{\partial r}\left(\frac{m\sin\theta}{r^2}r\right)\bm{\hat{\theta}}\right]
\end{equation}
dando finalmente,
\begin{equation}
\bm{B}=\frac{m}{cr^3}\left(2\cos\theta\bm{\hat{r}}+\sin\theta\bm{\hat{\theta}}\right)
\end{equation}
que es similar al campo el\'ectrico dipolar, con momento dipolar el\'ectrico $\bm{p}$:
\begin{equation}
\bm{E}=\frac{\bm{p}}{r^3}\left(2\cos\theta\bm{\hat{r}}+\sin\theta\bm{\hat{\theta}}\right),
\end{equation}
pero a pesar de dicha afinidad, el potencial escalar no se usa para representar el campo magn\'etico dipolar. Para un solenoide muy delgado con momento magn\'etico por unidad de longitud $\bm{\mathcal{M}}(\bm{x}')$ a lo largo de su longitud, el potencial vectorial en un punto distinto de su eje de simetr\'{\i}a longitudinal es,
\begin{align}
\bm{A}(\bm{x}) &=\frac{1}{c}\int\frac{\bm{\mathcal{M}}(\bm{x}')\times(\bm{x}-\bm{x}')}{\vert\bm{x}-\bm{x}'\vert^3}dL(\bm{x}') \notag \\
&=\frac{\mathcal{M}}{c}\left(\int\frac{\sin\theta'}{r^{'2}}dz'\right)\bm{\hat{\phi}}
\end{align}
donde $r'=r^2+z^{'2}-2zr'\cos\theta$ (Ley de cosenos) y $\sin\theta'=\frac{r}{r'}\sin\theta$ (Ley de senos), y por tanto el potencial vectorial en un punto diferente a lo eje de simetr\'{\i}a queda
\begin{align}
\bm{A}(\bm{x}) &=\frac{1}{c}\int\frac{\bm{\mathcal{M}}(\bm{x}')\times(\bm{x}-\bm{x}')}{\vert\bm{x}-\bm{x}'\vert^3}dL(\bm{x}') \notag \\
&\hspace{-1cm}=\frac{\mathcal{M}r\sin\theta}{c}\left[\int_{-L}^0\frac{\sin\theta'}{(r^2+z^{'2}-2z'r\cos\theta)^{\frac{3}{2}}}dz'\right]\bm{\hat{\phi}} \notag \\
&\hspace{-1cm}=\frac{\mathcal{M\sin\theta}}{cr^2}\left\{\int_{-L}^0\frac{\sin\theta'}{\left[1+\left(\frac{z'}{r}\right)^2-2\frac{z'}{r}\cos\theta\right]^{\frac{3}{2}}}dz'\right\}\bm{\hat{\phi}}
\end{align}
pero $r^{-2}z^{'2}-2r^{-1}\cos\theta+1=\left(r^{-1}z'-\cos\theta\right)^2-\cos^2\theta+1=\left(r^{-1}z'-\cos\theta\right)^2+\sin^2\theta$. Haciendo en cambio de variable $u=r^{-1}z'-\cos\theta$, entonces $du=r^{-1}dz'$, y por tanto,
\begin{equation}
\bm{A}(\bm{x})=\frac{\mathcal{M}}{c}\frac{\sin\theta}{r^2}\left[\int_{-\left(\frac{L}{r}+\cos\theta\right)}^{-\cos\theta}\frac{rdu}{\left(u^2+\sin^2\theta\right)^{\frac{3}{2}}}\right]\bm{\hat{\phi}}
\end{equation}
realizando el reemplazo trigonom\'etrico $u=\sin\theta\sec\alpha$, se resuelve la integral resultando,
\begin{equation}
\bm{A}(\bm{x})=\frac{\mathcal{M}}{c}\frac{1}{r\sin\theta}\left\{-\cos\theta+\frac{\frac{L}{r}+\cos\theta}{\left[\left(\frac{L}{r}+\cos\theta\right)^2+\sin^2\theta\right]^{\frac{1}{2}}}\right\}\bm{\hat{\phi}}
\end{equation}
si se considera la longitud $L$ del solenoide, el \'angulo $\theta$ formado entre el eje $z$ y un vector $r$ que va desde el punto superior del solenoide al punto de medici\'on $P$, y el \'angulo $\theta_2$ formado entre el eje $z$ y un vector $r_2$ que va desde el punto inferior del solenoide al punto de medici\'on $P$, se observa que la ecuaci\'on se reduce a,
\begin{equation}
\bm{A}(\bm{x})=\frac{\mathcal{M}}{c}\frac{1}{r\sin\theta}\left(\cos\theta_2-\cos\theta\right)\bm{\hat{\phi}} \label{SolPotVect}
\end{equation}
dando el valor del potencial vectorial en un punto diferente a lo eje de simetr\'{\i}a. Por tanto la densidad de flujo magn\'etico ser\'a,
\begin{align}
\bm{B}(\bm{x}) &=\bm{\nabla}_{r}\times\bm{A}(\bm{x})+\bm{\nabla}_{r_2}\times\bm{A}(\bm{x}) \notag \\
&=\frac{1}{r\sin\theta}\frac{\partial}{\partial\theta}(A_{\phi}\sin\theta)\bm{\hat{r}}-\frac{1}{r}\frac{\partial}{\partial r}(A_{\phi}r)\bm{\hat{\phi}} \notag \\
&\hspace{1cm}+\frac{1}{r_2\sin\theta_2}\frac{\partial}{\partial\theta_2}(A_{\phi}\sin\theta_2)\bm{\hat{r}}
\end{align}
resultando
\begin{equation}
\bm{B}(\bm{x})=\frac{\mathcal{M}}{cr^2}\bm{\hat{r}}-\frac{\mathcal{M}}{cr_2^2}\bm{\hat{r}}_2. \label{SolMagnField}
\end{equation}
El resultado anterior es una de las razones para la existencia del monopolo magn\'etico, pues tien la forma de dos polos magn\'eticos puntuales con cargas magn\'eticas positiva y negativa, donde la magnitud de \'estas es $q_m=\frac{\mathcal{M}}{c}$, resultando (para los puntos por fuera del eje de simetr\'{\i}a),
\begin{equation}
\bm{B}(\bm{x})=\frac{q_m}{r^2}\bm{\hat{r}}-\frac{q_m}{r_2^2}\bm{\hat{r}}_2.
\end{equation}
con un potencial vectorial magn\'etico dado por,
\begin{equation}
\bm{A}(\bm{x})=\frac{q_m}{r\sin\theta}\left(\cos\theta_2-\cos\theta\right)\bm{\hat{\phi}},
\end{equation}
donde se puede ver una correspondencia con la parte el\'ectrica tanto del potencial escalar $\phi(\bm{x})=\frac{q}{r}-\frac{q}{r_2}$, como de la intensidad de campo el\'ectrico $\bm{E}(\bm{x})=-\bm{\nabla}\phi(\bm{x})\frac{q}{r^2}\bm{\hat{r}}-\frac{q}{r^2_2}\bm{\hat{r}}_2$.
\par
En el caso de que las ecuaciones (\ref{SolPotVect}) y (\ref{SolMagnField}) sigan la restricci\'on $\theta_2=0$ para $L\to\infty$ o sea para $r_2\to\infty$ (con el fin de hacer una aproximaci\'on a un extremo puntual), quedan como
\begin{equation}
\left.
\begin{gathered}
\bm{B}(\bm{x})=\frac{q_m}{r^2}\bm{\hat{r}} \notag \\
\bm{A}(\bm{x})=\frac{q_m}{r\sin\theta}\left(1-\cos\theta\right)\bm{\hat{\phi}}
\end{gathered}
\right\}
\theta\neq\pi
\end{equation}
que no est\'an definidas sobre el eje $z$ negativo, debido a que el solenoide delgado o magneto est\'a sobre dicho eje, que representa la distribuci\'on de corriente fuente. Pero un monopolo real tendr\'{\i}a un campo esf\'erico isotr\'opico $\bm{B}(\bm{x})=\frac{q_m}{r^2}\bm{\hat{r}}$ que no lo mismo que $\bm{B}(\bm{x})=\frac{q_m}{r^2}\bm{\hat{r}}$, para $\theta\neq\pi$. Se observa que usar un solenoide semi-infinito para representar a un monopolo magn\'etico no es correcto, aunque la l\'{\i}nea de singularidad tiene un significado muy claro para el solemoide. Para el monopolo ha sido creado el potencial vectorial,
\begin{equation}
\bm{A}=\frac{q_m(1-\cos\theta)}{r\sin\theta}\bm{\hat{\phi}}=\frac{q_m}{r}\tan\left(\frac{\theta}{2}\right)\bm{\hat{\phi}},\quad\theta\neq\pi
\end{equation}
con el objetivo de reemplazar al solenoide semi-infinito por un monopolo puntual, a pesar de sus inconsistencias f\'{\i}sicas.
\section{Principio de M\'{\i}nima Acci\'on \label{PpioAcc}}
El principio de m\'{\i}nima acci\'on o principio de Hamilton se basa en que para un sistema que evoluciona en el tiempo, la diferencia entre su energ\'{\i}a cin\'etica y su potencial, para la trayectoria mas realista, es m\'{\i}nima. La diferencia entre estas energ\'{\i}as se denomina la funci\'on Lagrangiana, $\mathcal{L}=T-U$, y la acci\'on, $S$, es la evoluci\'on temporal de \'esta Lagrangiana,
\begin{equation}
S=\int_{t_1}^{t_2}\mathcal{L}(q,\dot{q},t)dt
\end{equation}
donde $q,\,\dot{q}$ son las posiciones y velocidades generalizadas del sistema. Seg\'un el teorema fundamental del c\'alculo variacional, la variaci\'on de la acci\'on debe ser m\'{\i}nima y a primer orden en la trayectoria y se debe anular en los puntos extremos, lo que genera una integral extremal. Por tanto,
\begin{equation}
\delta S =\delta\int_{t_1}^{t_2}\mathcal{L}(q,\dot{q},t)dt
\end{equation}
lo que genera variaciones de las coordenadas generalizadas $q(t)\to q(t)+\delta q(t)$. Adicionalmente, debido a la integral extremal se tiene que $\delta q(t_1)=\delta q(t_2)=0$, entonces,
\begin{equation}
\delta S =\delta\int_{t_1}^{t_2}\mathcal{L}(q+\delta q,\dot{q}+\delta\dot{q},t)dt-\delta\int_{t_1}^{t_2}\mathcal{L}(q,\dot{q},t)dt
\end{equation}
como el principio variacional exige que la primera variaci\'on sea cero entonces,
\begin{equation}
\delta S =\int_{t_1}^{t_2}\left(\frac{\partial\mathcal{L}}{\partial q}\delta q+\frac{\partial\mathcal{L}}{\partial\dot{q}}\delta\dot{q}\right)dt=0
\end{equation}
al integrar el segundo t\'ermino por partes se llega a,
\begin{equation}
\delta S =\left[\frac{\partial\mathcal{L}}{\partial\dot{q}}\delta q\right]_{t_1}^{t_2}+\int_{t_1}^{t_2}\left[\frac{\partial\mathcal{L}}{\partial q}-\frac{d}{dt}\left(\frac{\partial\mathcal{L}}{\partial\dot{q}}\right)\right]\delta qdt=0
\end{equation}
pero debido a la condici\'on de extremal se tiene que: $\left[\frac{\partial\mathcal{L}}{\partial\dot{q}}\delta q\right]_{t_1}^{t_2}=0$. Tambi\'en debido a que las variaciones de la Lagrangiana son independientes de $\delta q$ que es distinta a cero, entonces se verifica,
\begin{equation}
\frac{d}{dt}\left(\frac{\partial\mathcal{L}}{\partial\dot{q}}\right)-\frac{\partial\mathcal{L}}{\partial q}=0
\end{equation}
Si hay varios grados de libertad, entonces deben existir diferentes acciones, y por tanto variaciones de \'estas para las $s$ funciones diferentes $q_i(t)$, lo que lleva a
\begin{equation}
\frac{d}{dt}\left(\frac{\partial\mathcal{L}}{\partial\dot{q}_i}\right)-\frac{\partial\mathcal{L}}{\partial q_i}=0 \label{EcLagrange}
\end{equation}
para $i=1,2,3,\ldots,s$, denominadas ecuaciones de Lagrange. De \'estas se observa que si el potencial no es dependiente de la velocidad,
\begin{align}
&\frac{d}{dt}\left\{\frac{\partial [T(\dot{q}_i)-U(q_i)]}{\partial\dot{q}_i}\right\}-\frac{\partial [T(\dot{q}_i)-U(q_i)]}{\partial q_i}=0 \notag \\
&\frac{d}{dt}\left(\frac{\partial T}{\partial\dot{q}_i}\right)=-\frac{\partial U}{\partial q_i}
\end{align}
si $p_i=\frac{\partial\mathcal{L}}{\partial\dot{q}_i}$, es el momento can\'onicamente conjugado, entonces:
\begin{equation}
\frac{dp_i}{dt}=F_i=-\frac{\partial U}{\partial q_i}
\end{equation}
que verifica la din\'amica de Newton.
\newline
Adem\'as si se calcula la variaci\'on temporal de la Lagrangiana (usando suma sobre \'{\i}ndices contraidos), entonces,
\begin{equation}
\frac{d\mathcal{L}}{dt}=\frac{\partial\mathcal{L}}{\partial q_i}\frac{dq_i}{dt}+\frac{\partial\mathcal{L}}{\partial\dot{q}_i}\frac{d\dot{q}_i}{dt}+\frac{\partial\mathcal{L}}{\partial t}
\end{equation}
aplicando las ecuaciones de Lagrange (\ref{EcLagrange})
\begin{align}
\frac{d\mathcal{L}}{dt} &=\frac{d}{dt}\left(\frac{\partial\mathcal{L}}{dq_i}\right)\dot{q}_i+\frac{\partial\mathcal{L}}{\partial\dot{q}_i}\frac{d\dot{q}_i}{dt}+\frac{\partial\mathcal{L}}{\partial t} \notag \\
&=\frac{d}{dt}\left(\dot{q}_i\frac{\partial\mathcal{L}}{dq_i}\right)+\frac{\partial\mathcal{L}}{\partial t}
\end{align}
y por tanto se verifica que
\begin{equation}
\frac{dh}{dt}=-\frac{\partial\mathcal{L}}{\partial t}
\end{equation}
donde $h=\dot{q}_i\frac{\partial\mathcal{L}}{dq_i}-\mathcal{L}$.  \'Esta funci\'on $h(q_i,\dot{q}_i;t)$ se denomina funci\'on energ\'{\i}a y si la Lagrangiana no depende del tiempo, entonces se conserva. \'Esta transformaci\'on es un transformaci\'on de Legendre, y si se incluye el momento conjugado $\dot{p}_i=\frac{\partial\mathcal{L}}{\partial q_i}$, entonces el diferencial de la Lagrangiana es
\begin{equation}
d\mathcal{L}=\dot{p}_idq_i+p_id\dot{q}_i+\frac{\partial\mathcal{L}}{\partial t}dt
\end{equation}
y se define la funci\'on Hamiltoniana mediante una transformaci\'on de Legendre, de la forma
\begin{equation}
\mathcal{H}(q,p,t)=\dot{q}_ip_i-\mathcal{L}(q,\dot{q},t)
\end{equation}
con su diferencial igual a
\begin{align}
d\mathcal{H} &=d\dot{q}_ip_i+\dot{q}_idp_i-\frac{\partial\mathcal{L}}{\partial t} \notag \\
&=\frac{\partial\mathcal{H}}{\partial q_i}dq_i+\frac{\partial\mathcal{H}}{\partial p_i}dp_i+\frac{\partial\mathcal{H}}{\partial t}dt
\end{align}
donde la Hamiltoniana no depende de las velocidades generalizadas y por tanto la transformaci\'on de Legendre produce un cambio del espacio de las configuraciones $(q_i,\dot{q}_i)$ al espacio de las fases $(q_i,p_i)$ \cite{Goldstein}, \cite{Landau2}.

\printindex
\end{document}